\documentclass[preprint, 12pt]{elsarticle}
\usepackage[margin=1in]{geometry}
\usepackage{amsmath,amssymb,bm}
\usepackage{physics}       
\usepackage{graphicx}
\usepackage{booktabs}
\usepackage{array}
\usepackage{multirow}
\usepackage{hyperref}
\hypersetup{hidelinks}
\usepackage{cleveref}
\usepackage{xcolor}
\usepackage{enumitem}
\usepackage{natbib}
\usepackage{caption}
\usepackage{subcaption}
\usepackage{float}
\usepackage{accents}
\usepackage{tabularx}
\newcommand{\cmark}{\ensuremath{\checkmark}}

\begin{document}

\begin{frontmatter}

\title{A Green's-function method for vertical thermal boundary
conductance in anisotropic multilayers}

\author[pitt]{Dihui Wang\corref{cor1}}
\ead{diw25@pitt.edu}

\author[byu]{Troy Munro}

\author[pitt]{Heng Ban}

\cortext[cor1]{Corresponding author}

\affiliation[pitt]{
    organization={Department of Mechanical Engineering and Materials Science, University of Pittsburgh},
    city={Pittsburgh},
    state={PA},
    country={USA}
}

\affiliation[byu]{
    organization={Department of Mechanical Engineering, Brigham Young University},
    city={Provo},
    state={UT},
    country={USA}
}

\begin{abstract}
Vertical thermal interfaces occur in both engineered and natural
materials. Their vertical thermal boundary conductance can
differ from the horizontal counterpart, requiring dedicated
characterization. Yet current thermal metrology
resolves vertical thermal boundary conductance only in
restricted geometries such as two bulk media, for lack of an efficient
forward solution that admits anisotropy, multilayers, and depth-dependent
vertical thermal boundary conductance together. We present a Green's-function boundary
integral equation (GBIE) method that couples transfer-matrix
Green's functions to an interface-only integral equation for
depth-dependent $G_v(z)$, supporting dissimilar orthotropic multilayers
($k_x\neq k_y\neq k_z$) on either side and horizontal conductance
$G_h$. For anisotropic film-on-substrate multilayers with films from
$1~\mu\mathrm{m}$ to $100~\mathrm{nm}$, the GBIE agrees with
three-dimensional finite element method (FEM) predictions
to within one percent mean normalized phase and
amplitude error, while running $29\text{--}210\times$ faster and
reducing peak memory by factors of $120\text{--}450$ in single-core
tests; a JIT-compiled JAX implementation reaches up to $4100\times$
on a matched 16-core comparison. The GBIE further reproduces a
continuous film over a buried interface, representative of a
thermoreflectance measurement, and a finite-depth interface with
depth-dependent $G_v(z)$. The GBIE accommodates
lateral-to-film-thickness ratios above $10^{5}$, where volumetric FEM
can become computationally prohibitive. These results establish an
efficient forward solution for vertical-interface heat transport in
systems ranging from microelectronic device sidewalls to grain
boundaries in polycrystalline solids.
\end{abstract}


\end{frontmatter}
\section{Introduction}
\label{sec:intro}

Vertical thermal interfaces are ubiquitous in engineered and natural
materials. They arise as sidewalls in microelectronic
devices~\citep{Cho2012,Liu2025}, cracks and debonds in composites and nuclear
fuels~\citep{Chowdhury2018CarbonCarbonAerospace,Cho2021MatrixGraphiteOxidation,Zhang2021FCMDebonding},
and grain or phase boundaries in polycrystalline solids~\citep{Sood2018}.
Because these interfaces redirect heat across laterally heterogeneous
structures, their resistance can affect thermal spreading, local temperature
rise, and the detectability of interfacial damage. The vertical thermal
boundary conductance (vTBC) \(G_v\) spans at least six orders of magnitude,
\((10^{3}\text{--}10^{9}~\mathrm{W\,m^{-2}\,K^{-1}})\), and need not equal
the horizontal TBC for the same material pair.
For example, the measured Al/SiO$_2$ sidewall conductance was
approximately 2.4-fold lower than that of planar interfaces
fabricated under comparable conditions \citep{Park2017a}.
A separate Al/SiO$_2$ nanofin study estimated sidewall conductances
approximately 1.1--4.5-fold lower than the planar value \citep{Park2017b}.
Such differences can arise from processing damage,
roughness, voids, or orientation-dependent phonon transport
\citep{Hopkins2011Anisotropy}. Quantifying \(G_v\) is therefore important
for predicting heat flow, assessing interface integrity, and guiding the
thermal design and fabrication of heterogeneous structures.

The most straightforward route to characterizing vTBC is to pair an
established non-contact thermal metrology, such as frequency-domain
thermoreflectance (FDTR)~\citep{Schmidt2009FDTR,Schmidt2008}, lock-in
thermography~\citep{Gonzalez2019,Wang2024SRLIT,Wang2025Tensor,Ma2025Phase,Ma2025Depth},
or photothermal radiometry (PTR)~\citep{Sheikh2025PTR}, with a
forward model that incorporates the vertical interface. Current forward
models for these techniques are mature for horizontal interfaces through the
transfer-matrix method (TMM)%
~\citep{Feldman1999,Cahill2004,Jiang2018,Jiang2022SSTR}. A Fourier transform
in the two in-plane directions reduces the three-dimensional anisotropic heat
equation to a one-dimensional ordinary differential equation along the depth
\(z\). This framework relies on in-plane translational invariance
and therefore cannot accommodate a vertical interface.

Existing vertical-interface models are limited to bulk isotropic
systems. An analytical surface-temperature expression was derived for
an infinite vertical crack in a single-layer half-space~\citep{PechMay2014}
and subsequently extended to two dissimilar media~\citep{Gonzalez2019}.
Both models permit arbitrary surface coordinates but require a triple
integral per observation point. For validation, we use a reduced
expression restricted to observation points on the line through the
beam center perpendicular to the interface.
Neither model accounts for anisotropy, multilayer
structures, or interfaces of finite depth.

Direct three-dimensional FEM removes these geometric restrictions, but
often at substantial computational cost. A discontinuous-Galerkin
model relaxed the infinite-depth assumption~\citep{Celorrio2014}
but targeted low frequencies and low conductances,
$G_v \sim 10^{4}~\mathrm{W\,m^{-2}\,K^{-1}}$, and required minutes per
solve. Reaching higher conductances, $G_v > 10^{6}$, requires much
higher frequencies and correspondingly smaller thermal length scales,
which makes volumetric meshes increasingly expensive. The
low-element-quality warning in our \(100~\mathrm{nm}\)-film COMSOL case
likewise illustrates the difficulty of resolving the film and the much
larger lateral thermal domain simultaneously. An alternative approach
confines heat to one-dimensional flow in a nanograting, requiring
specially designed samples~\citep{Park2017a}.

Boundary-integral methods have also been used to model heat conduction
across imperfect thermal interfaces. They use fundamental solutions to
relate boundary temperatures and normal heat fluxes through integral
equations~\citep{Fahmy2025AnisotropicBEM}. Related methods treat multi-term fractional heat conduction in
heterogeneous media~\citep{FahmyMarin2026FractionalBIE}. For imperfect
interfaces, Ang et al.\ developed a boundary-element formulation for
steady two-dimensional conduction between homogeneous isotropic media
with uniform interfacial conductance~\citep{Ang2004GreenImperfect}.
Yet, no model among those reviewed here combines
multilayers, anisotropy, distinct stacks on the two sides, and a
depth-dependent \(G_v(z)\).
Table~\ref{tab:method_scope} compares the scope of the present formulation
with the reference models.

\begin{table}[htbp]
\centering
\footnotesize
\setlength{\tabcolsep}{2.5pt}
\renewcommand{\arraystretch}{1.25}
\renewcommand{\tabularxcolumn}[1]{m{#1}}

\caption{Scope of the compared heat-transfer formulations.
Checkmarks indicate features included in the cited formulation;
dashes indicate features outside its reported scope.}
\label{tab:method_scope}

\begin{tabularx}{\linewidth}{@{}
    >{\raggedright\arraybackslash}m{0.19\linewidth}
    *{7}{>{\centering\arraybackslash}X}
    >{\raggedright\arraybackslash}m{0.14\linewidth}
@{}}
\toprule
\multirow{2}{*}{Formulation}
& \multicolumn{3}{c}{Geometry and materials}
& \multicolumn{4}{c}{Interface treatment}
& \multirow{2}{*}{Solution type\textsuperscript{b}} \\
\cmidrule(lr){2-4}
\cmidrule(lr){5-8}
& \shortstack{Depth\\layers}
& \shortstack{Lateral\\contrast}
& \shortstack{Ortho-\\tropy}
& $G_h$
& \shortstack{Infinite\\depth}
& \shortstack{Finite\\depth}
& $R_v(z)$
& \\
\midrule

Multilayer TMM~\citep{Feldman1999,Cahill2004,Jiang2018}
& \cmark & --- & \cmark
& \cmark & --- & --- & ---
& Analytical \\

Infinite-crack model~\citep{PechMay2014}
& --- & --- & ---
& --- & \cmark & --- & ---
& Analytical \\

Two-medium model~\citep{Gonzalez2019}
& --- & \cmark & ---
& --- & \cmark & --- & ---
& Analytical \\

Finite-crack DG model~\citep{Celorrio2014}
& --- & --- & ---
& --- & \cmark & \cmark & \cmark\textsuperscript{a}
& Numerical \\

Present GBIE
& \cmark & \cmark & \cmark
& \cmark & \cmark & \cmark & \cmark
& Semi-analytical \\

\bottomrule
\end{tabularx}

\medskip
\begin{minipage}{\linewidth}
\scriptsize
Depth layers denote multilayers along $z$; lateral contrast denotes
dissimilar materials across a vertical interface.
$G_h$ denotes horizontal thermal boundary conductance.
Infinite and finite depth refer to the vertical interface;
$R_v(z)$ denotes depth-dependent interfacial resistance.

\textsuperscript{a}Spatially varying resistance is included in the
DG formulation; the reported examples use uniform crack resistance.

\textsuperscript{b}Analytical: explicit transform/integral solution,
with numerical quadrature as needed; numerical: volumetric DG-FEM
discretization; semi-analytical: analytical layer Green's functions
coupled to numerical depth discretization of the unknown interface flux.
\end{minipage}
\end{table}

The present work addresses these limitations. We develop a semi-analytical
Green's-function boundary integral equation (GBIE) framework posed on the
vertical interface
(Fig.~\ref{fig:sample_sketch}). For each stack, the Green's function is
constructed from two independent transfer-matrix modes: one satisfying the
homogeneous top-surface condition and the other the homogeneous rear
condition. Their Wronskian combination gives the temperature response at any
depth to a source at any other depth while retaining the analytical
multilayer solution in the horizontal directions. The GBIE then couples the
two stacks by enforcing heat-flux conservation and the thermal contact
condition along their shared vertical interface. The formulation accommodates orthotropic anisotropic conductivity
($\mathbf{k}=\mathrm{diag}(k_{x},k_{y},k_{z})$ with
$k_{x}\neq k_{y}\neq k_{z}$), distinct multilayer stacks
on the two sides, horizontal interlayer conductances \(G_h\), and
depth-dependent \(G_v(z)\) profiles on surface-breaking or buried interfaces
of finite or semi-infinite depth.
This work addresses the need for efficient thermal modeling of complex
vertical interfaces in advanced devices and heterogeneous materials.
It extends existing Green's-function models to dissimilar anisotropic
multilayers with buried or finite-depth interfaces and depth-dependent
conductance, while discretizing only the interfacial heat flux.

\section{Theory}\label{sec:theory}

\subsection{Geometry, Governing Equations, and Source}\label{sec:geom}

\begin{figure}[t]
    \centering
    \includegraphics[width=0.80\linewidth]{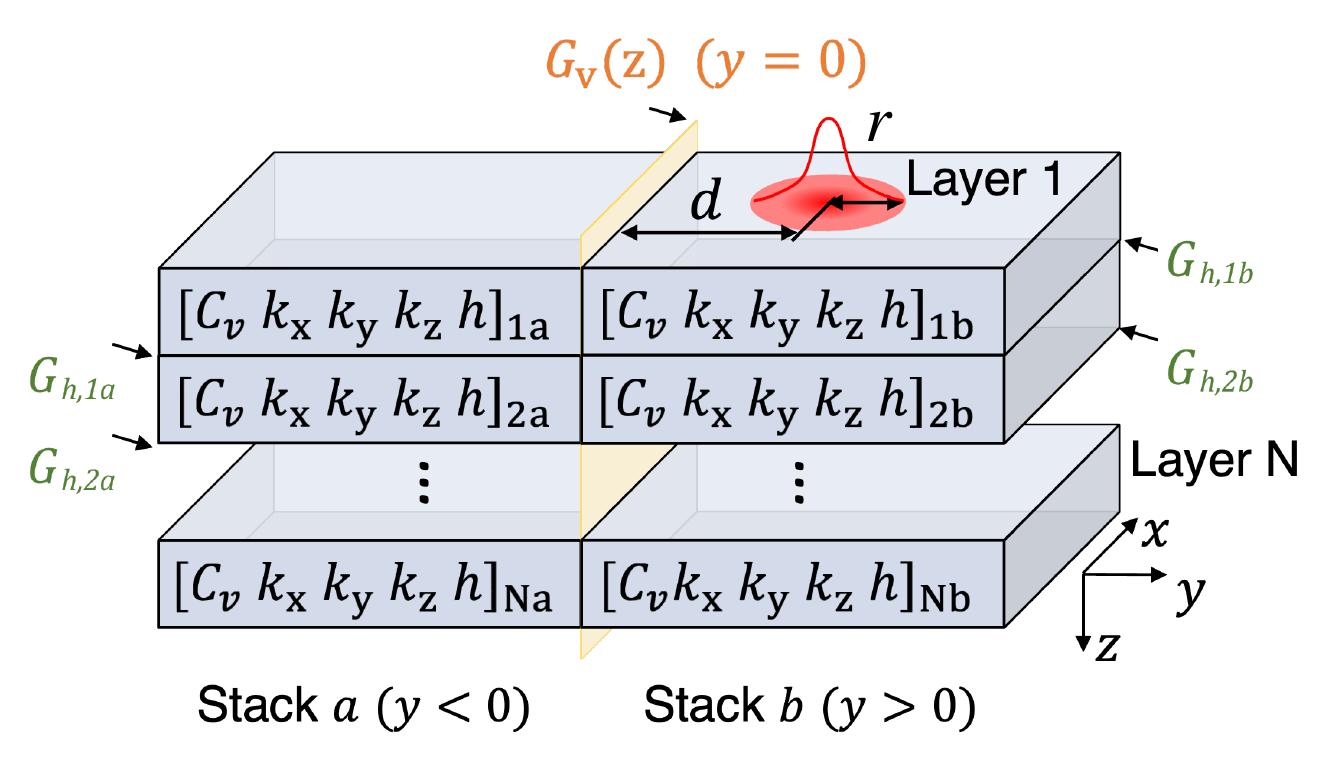}
    \caption{Schematic of the vertical-interface geometry. Stack~$a$
($y<0$) and stack~$b$ ($y>0$) are joined at the vertical interface
$y=0$, where the depth-dependent conductance $G_v(z)$ acts, and the
top surface is at $z=0$. A Gaussian laser of $1/e^2$ radius $r$ is
centered a distance $d$ from the interface; the unknown interfacial
heat flux $Q_y(z)$ is solved along $y=0$.}
    \label{fig:sample_sketch}
\end{figure}

Two multilayer stacks, infinite in $x$ and semi-infinite in $y$, meet at $y=0$
(Fig.~\ref{fig:sample_sketch}): stack~$a$ occupies $y<0$ and stack~$b$
occupies $y>0$, with $z$ increasing into the sample from its top surface
at $z=0$. Layer~$l$ of stack~$j$ has conductivity tensor
$\mathbf{k}_{lj}=\operatorname{diag}(k_{x,lj},k_{y,lj},k_{z,lj})$,
volumetric heat capacity $C_{v,lj}$, thickness $h_{lj}$, and horizontal
TBC $G_{h,lj}$ to the layer below.
We consider orthotropic thermal conductivity with independent principal
components $k_{x}$, $k_{y}$, and $k_{z}$ and zero off-diagonal
components ($k_{xy}=k_{xz}=k_{yz}=0$).
Layer and stack indices are suppressed unless needed for clarity.
Detailed derivations are provided in Supplementary
Section~S1.

Within each homogeneous layer without volumetric heating, the time-domain
heat-diffusion equation is
\begin{equation}
  C_v\frac{\partial T}{\partial t}
  = k_x\frac{\partial^2T}{\partial x^2}
  + k_y\frac{\partial^2T}{\partial y^2}
  + k_z\frac{\partial^2T}{\partial z^2}.
  \label{eq:v2_time_domain_heat}
\end{equation}
Taking the temporal Fourier transform,
$T(x,y,z,t)\overset{\mathcal F_t}{\longleftrightarrow}\Theta(x,y,z;\omega)$,
with kernel $e^{-i\omega t}$ gives $\mathcal F_t[\partial_tT]=i\omega\Theta$, and hence
\begin{equation}
  k_x\,\frac{\partial^2\Theta}{\partial x^2}
  + k_y\,\frac{\partial^2\Theta}{\partial y^2}
  + k_z\,\frac{\partial^2\Theta}{\partial z^2}
  - i\omega\,C_v\,\Theta = 0,
  \qquad \omega=2\pi f.
  \label{eq:v2_governing}
\end{equation}

The $\omega$ dependence is suppressed below. Hats and tildes denote
Fourier transforms in $x$ and $(x,y)$, respectively.

The top-surface boundary condition prescribes the absorbed pump
laser flux, a circular Gaussian,
\begin{equation}
    Q_{\mathrm{pump}}(x,y)
    =\frac{2P_0}{\pi r^2}
    \exp\!\left[-\frac{2(x^2+(y-d)^2)}{r^2}\right],
    \label{eq:v2_gaussian_source}
\end{equation}
where \(P_0\) is the absorbed harmonic-power amplitude, \(r\) the
\(1/e^2\) radius, and the source center is \((0,d)\),
as such \(Q_z(x,y,0)=Q_{\mathrm{pump}}(x,y)\).
The substrate is taken as semi-infinite, with the temperature response
\(\Theta\) decaying to zero as \(z\to\infty\).
At the vertical interface, the heat flux \(Q_y\) is continuous and taken
as positive from stack~\(b\) into stack~\(a\), so
\begin{equation}
    \Theta_b(0^+,z)-\Theta_a(0^-,z)=R_v(z)\,Q_y(z),
    \label{eq:v2_vertical_jump}
\end{equation}
where $R_v(z)=1/G_v(z)$ is the depth-dependent vertical thermal
boundary resistance. The physically resistive segment is
$\Gamma_{R_v}=[z_{\min},z_{\max}]$; outside this interval,
$R_v(z)=0$ represents perfect contact.

\subsection{Spectral Reduction}\label{sec:spectral}

\begin{figure}[htbp]
    \centering
    \includegraphics[width=\linewidth]{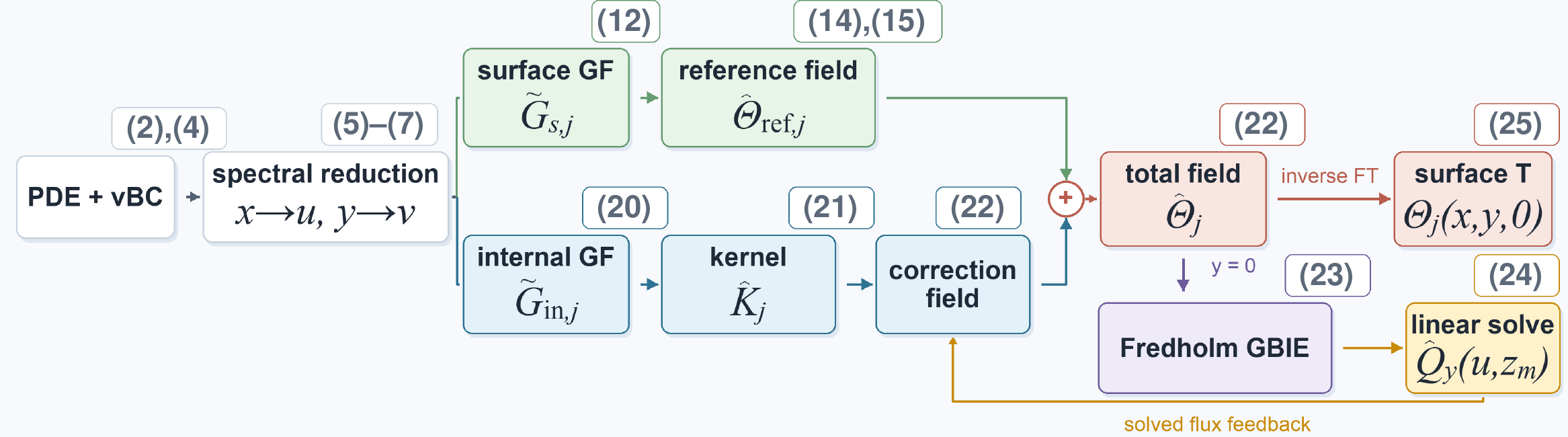}
    \caption{Derivation workflow for the vertical-interface GBIE framework.
    The governing equations and vertical
    boundary condition are first reduced spectrally. The surface Green's
    function constructs the reference field, while the internal Green's
    function supplies the kernel for the interface correction. Imposing
    the vertical-interface condition gives the Fredholm GBIE for
    $\hat{Q}_y$; its solution determines the correction and total surface
    temperature.}
    \label{fig:bie_derivation_workflow}
\end{figure}

Figure~\ref{fig:bie_derivation_workflow} summarizes the derivation workflow.
Fourier-domain solutions for anisotropic multilayers are well established
in thermoreflectance modeling~\citep{Feser2014TensorModel,Tang2021AnisotropicBOFDTR}.
Applying the in-plane Fourier transform to \eqref{eq:v2_governing} on the
laterally extended reference stack, with spatial frequencies $(u,v)$, gives
\begin{equation}
    \frac{d^2\tilde{\Theta}}{dz^2}-\lambda^2\,\tilde{\Theta}=0,
    \qquad
    \lambda(u,v)=
    \left[\frac{k_x(2\pi u)^2+k_y(2\pi v)^2+i\omega C_v}
    {k_z}\right]^{1/2},
    \label{eq:v2_lambda}
\end{equation}
Choosing the square-root branch with $\operatorname{Re}(\lambda)>0$,
the general solution within each homogeneous layer is
$\tilde{\Theta}(z)=c_1\cosh(\lambda z)+c_2\sinh(\lambda z)$.
The coefficients $c_1$ and $c_2$ are determined by the boundary and
interface conditions. This form retains both exponential modes
$e^{\pm\lambda z}$. The detailed derivation is provided in Supplementary
Section~S1.2.

Transforming the Gaussian source separates the symmetric \(x\)-Gaussian from
the \(y\)-profile. Splitting the latter at \(y=0\) gives the source absorbed
by each stack,
\begin{equation}
    \tilde{Q}_{\mathrm{pump},j}(u,v)
    =P_0\exp\!\left(-\frac{\pi^2r^2u^2}{2}\right)F_j(v;d),
    \label{eq:v2_source_transform}
\end{equation}
where \(F_b\) and \(F_a\) are the transforms of the normalized \(y\)-Gaussian
over \(y>0\) and \(y<0\), respectively:
\begin{equation}
    F_b(v;d)=\tfrac{1}{2}\Phi(v;d)\,\operatorname{erfc}[\zeta(v;d)],
    \qquad
    F_a(v;d)=\Phi(v;d)-F_b(v;d),
    \label{eq:v2_half_gaussian}
\end{equation}
where
\(\Phi=\exp(-i2\pi vd-\pi^2r^2v^2/2)\) and
\(\zeta=-\sqrt{2}\,d/r+i\pi rv/\sqrt{2}\). This split retains two-sided
heating when the beam overlaps the interface, which becomes important for
\(d\lesssim r\).
\subsection{Field Decomposition}\label{sec:reference}

The vertical-interface boundary condition~\eqref{eq:v2_vertical_jump}
couples the stacks through the unknown depth-distributed flux
$\hat{Q}_y$. By linearity, the total temperature in stack $j$,
$\hat{\Theta}_j$, is the sum of a reference field,
$\hat{\Theta}_{\mathrm{ref},j}$, and a correction field,
$\hat{\Theta}_{\mathrm{corr},j}$:
\begin{equation}
  \hat{\Theta}_j
  = \hat{\Theta}_{\mathrm{ref},j}
  + \hat{\Theta}_{\mathrm{corr},j}.
  \label{eq:v2_field_decomposition}
\end{equation}
The reference field is obtained by applying the surface Green's
function $\tilde{G}_{s,j}(u,v;z)$, constructed using the standard
transfer-matrix method, to the absorbed laser source. Each reference
problem extends stack $j$ laterally across $y=0$ and applies its
absorbed portion of the source. Accounting for heating in both
stacks is important when the beam is close to the interface
(Supplementary Section~S1.1).
These reference fields satisfy the surface, horizontal-interface,
and rear conditions, but do not satisfy the
vertical-interface conditions together.

The correction fields form the core of the model, coupling stacks
$a$ and $b$ through the vertical-interface conditions.
They introduce no additional top-surface heat flux, since the pump
heating is already included in the reference fields.
The correction field in each stack is constructed using an interface
kernel derived from the internal Green's function
$\tilde{G}_{\mathrm{in},j}(u,v;z,z')$, which gives the temperature
response at depth $z$ to a unit source at depth $z'$.
Integrating this kernel against the flux difference
$\hat{Q}_y-\hat{Q}_{y,\mathrm{ref},j}$ over the interface depth
produces the correction field and ensures continuity of the total
interface flux, whose depth distribution remains unknown.

To determine this flux, we evaluate the total temperature fields
on both sides of the interface and substitute them into the vTBC
temperature-jump condition~\eqref{eq:v2_vertical_jump}.
This yields an equation relating the unknown flux $\hat{Q}_y(u,z)$
to integrals of the same flux over the interface: the
Green's-function boundary integral equation (GBIE).
Solving this equation determines the total interface flux,
from which the correction and total temperature fields follow.

Section~\ref{sec:greens} derives the surface Green's function and
constructs the reference field.
Section~\ref{sec:bie} derives the internal Green's function,
constructs the interface correction, and formulates the GBIE.

\subsection{Surface Green's Function and Reference Field}\label{sec:greens}

The surface Green's function follows from the standard transfer-matrix
method used in layered thermal-diffusion and thermoreflectance
models~\citep{Cahill2004,Schmidt2009FDTR,Schmidt2008,Braun2017TPD,Wang2024SRLIT,Wang2025Tensor}.
Using Fourier's law $\tilde{Q}_z=-k_z\,d\tilde{\Theta}/dz$
together with Eq.~\eqref{eq:v2_lambda}, we express the temperature
and heat flux in each layer $l$ in matrix form, with $z_l$ measured from that layer's
top surface,
\begin{equation}
  \begin{bmatrix}\tilde{\Theta}_l(z_l)\\ \tilde{Q}_{z,l}(z_l)\end{bmatrix}
  =
  \begin{bmatrix}
    \cosh(\lambda_l z_l) &
    -\dfrac{\sinh(\lambda_l z_l)}{k_{z,l}\lambda_l}\\
    -k_{z,l}\lambda_l\sinh(\lambda_l z_l) &
    \cosh(\lambda_l z_l)
  \end{bmatrix}
  \begin{bmatrix}\tilde{\Theta}_l^{\mathrm{top}}\\
  \tilde{Q}_{z,l}^{\mathrm{top}}\end{bmatrix}.
  \label{eq:v2_layer_matrix}
\end{equation}
We denote this layer matrix by $\mathbf{P}_l(z_l)$; its single argument
is the propagation distance from the fixed layer top.
Across a horizontal interface, flux continuity and the temperature jump
give the matrix
$\mathbf{H}_l=\bigl[\begin{smallmatrix}1&-1/G_{h,l}\\0&1\end{smallmatrix}\bigr]$.
For a stack of $N$ layers, their ordered product from the top surface to
the rear surface gives
\begin{equation}
  \begin{aligned}
  \begin{bmatrix}\tilde{\Theta}_N(h_N)\\ \tilde{Q}_{z,N}(h_N)\end{bmatrix}
  &=\mathbf{P}_N(h_N)\mathbf{H}_{N-1}\mathbf{P}_{N-1}(h_{N-1})
    \cdots\mathbf{H}_1\mathbf{P}_1(h_1)
  \begin{bmatrix}\tilde{\Theta}_1^{\mathrm{top}}\\
  \tilde{Q}_{z,1}^{\mathrm{top}}\end{bmatrix}\\
  &\equiv\begin{bmatrix}\tilde{A} & \tilde{B}\\ \tilde{C} & \tilde{D}\end{bmatrix}
  \begin{bmatrix}\tilde{\Theta}_1^{\mathrm{top}}\\
  \tilde{Q}_{z,1}^{\mathrm{top}}\end{bmatrix}.
  \end{aligned}
  \label{eq:v2_total_matrix}
\end{equation}

Here $h_N$ is the thickness of layer $N$. For a substrate thickness
of approximately four or more cross-plane thermal penetration depths,
$h_N\gtrsim4l_{\mathrm{tpd},N}$, where
$l_{\mathrm{tpd},N}=\sqrt{k_{z,N}/(\pi f C_{v,N})}$,
we use the semi-infinite substrate approximation.
This removes the growing exponential, leaving
$\tilde{\Theta}_N(z_N)=\tilde{\Theta}_N^{\mathrm{top}}e^{-\lambda_N z_N}$,
where $z_N$ is measured from the top of layer~$N$.
Fourier's law then gives the rear boundary condition
$\tilde{Q}_z(L)=Y_L\tilde{\Theta}(L)$, where
$L=\sum_{l=1}^{N}h_l$ is the total thickness and
$Y_L=k_{z,N}\lambda_N$.

Propagating a solution of Eq.~\eqref{eq:v2_lambda} upward from
this rear boundary condition defines the \emph{rear mode}, with temperature
$\tilde{\Theta}_j^{\mathrm{R}}(z)$ and heat flux
$\tilde{Q}_{z,j}^{\mathrm{R}}(z)$. The rear condition fixes their ratio
but leaves the amplitude free, since no top-surface flux is prescribed
for this auxiliary solution. We choose
$\tilde{\Theta}_j^{\mathrm{R}}(L)=1$ and
$\tilde{Q}_{z,j}^{\mathrm{R}}(L)=Y_L$; this normalization is independent
of the laser power $P_0$.

Since each layer and contact matrix has unit determinant, the total
transfer matrix satisfies $\tilde A\tilde D-\tilde B\tilde C=1$. The rear-mode
values at the top surface follow by inverting
Eq.~\eqref{eq:v2_total_matrix}:
\begin{equation}
  \begin{aligned}
    \begin{bmatrix}
      \tilde{\Theta}_j^{\mathrm{R}}(0)\\
      \tilde{Q}_{z,j}^{\mathrm{R}}(0)
    \end{bmatrix}
    &=
    \begin{bmatrix}\tilde A&\tilde B\\ \tilde C&\tilde D\end{bmatrix}^{-1}
    \begin{bmatrix}1\\Y_L\end{bmatrix}\\
    &=
    \begin{bmatrix}\tilde D&-\tilde B\\-\tilde C&\tilde A\end{bmatrix}
    \begin{bmatrix}1\\Y_L\end{bmatrix}
    =
    \begin{bmatrix}\tilde D-Y_L\tilde B\\Y_L\tilde A-\tilde C\end{bmatrix}.
  \end{aligned}
  \label{eq:v2_rear_mode_upward}
\end{equation}

The surface Green's function is defined as the ratio of the rear-mode
temperature at depth $z$ to its top-surface heat flux,
\begin{equation}
  \tilde{G}_{s,j}(u,v;z)
  =\frac{\tilde{\Theta}_j^{\mathrm{R}}(z)}
        {\tilde{Q}_{z,j}^{\mathrm{R}}(0)}.
  \label{eq:v2_surface_green}
\end{equation}
At $z=0$, substituting Eq.~\eqref{eq:v2_rear_mode_upward} into
Eq.~\eqref{eq:v2_surface_green} gives
\begin{equation}
  \tilde{G}_{s,j}(u,v;0)
  =-\frac{\tilde{D}-Y_L\tilde{B}}
  {\tilde{C}-Y_L\tilde{A}}.
  \label{eq:v2_surface_green_terminal}
\end{equation}

For an adiabatic rear, $\tilde Q_{z,N}(h_N)=0$. The second row of
Eq.~\eqref{eq:v2_total_matrix} then gives
$\tilde C\tilde\Theta_1^{\mathrm{top}}+
\tilde D\tilde Q_{z,1}^{\mathrm{top}}=0$, yielding the familiar result
$\tilde G_{s,j}(u,v;0)=-\tilde D/\tilde C$, equivalently $Y_L=0$.
For sufficiently thick substrates, the two rear conditions yield
nearly identical surface responses.

Applying the surface Green's function~\eqref{eq:v2_surface_green} to the
known laser source~\eqref{eq:v2_source_transform} and inverting the
transverse Fourier transform gives the reference field,
\begin{equation}
    \hat{\Theta}_{\mathrm{ref},j}(u,y,z)
    =2P_0\exp\!\left(-\frac{\pi^2 r^2 u^2}{2}\right)
    \int_0^\infty \tilde{G}_{s,j}(u,v;z)\,
    \operatorname{Re}\!\left[e^{i2\pi vy}\,F_j(v;d)\right]dv,
    \label{eq:v2_reference_field}
\end{equation}
with the interface flux defined as positive from stack $b$ to
stack $a$ (the negative-$y$ direction), Fourier's law gives
the reference interface flux at $y=0$ as
\begin{equation}
    \hat{Q}_{y,\mathrm{ref},j}(u,z)=k_{y,j}(z)\,
    \left.\frac{\partial\hat{\Theta}_{\mathrm{ref},j}}{\partial y}
    \right|_{y=0}.
    \label{eq:v2_reference_flux}
\end{equation}

\subsection{Internal Green's Function and Correction Field}\label{sec:bie}

For a unit internal source within a layer at $z=z'$, we construct the
internal Green's function from two auxiliary solutions of
Eq.~\eqref{eq:v2_lambda}: the rear mode introduced in
Section~\ref{sec:greens} and a top mode satisfying the homogeneous
top-surface condition.

To write both modes explicitly, define the transfer matrix by
\begin{equation}
  \begin{aligned}
    \begin{bmatrix}
      \tilde{\Theta}_j(z_2)\\ \tilde{Q}_{z,j}(z_2)
    \end{bmatrix}
    &=\mathbf{M}_j(z_2,z_1)
    \begin{bmatrix}
      \tilde{\Theta}_j(z_1)\\ \tilde{Q}_{z,j}(z_1)
    \end{bmatrix}\\
    &=\begin{bmatrix}
      \tilde{A}_j(z_2,z_1) & \tilde{B}_j(z_2,z_1)\\
      \tilde{C}_j(z_2,z_1) & \tilde{D}_j(z_2,z_1)
    \end{bmatrix}
    \begin{bmatrix}
      \tilde{\Theta}_j(z_1)\\ \tilde{Q}_{z,j}(z_1)
    \end{bmatrix}.
  \end{aligned}
  \label{eq:v2_partial_transfer_matrix}
\end{equation}

$z_2$ is the destination depth and $z_1$ is the starting
depth, both in global coordinates. For two points within a
single homogeneous layer, only their signed
separation $\Delta z=z_2-z_1$ matters. Using that layer's $k_z$ and
$\lambda$, the coefficients are explicitly
\begin{equation}
  \begin{aligned}
    \tilde A_j(z_2,z_1)=\tilde D_j(z_2,z_1)
      &=\cosh(\lambda\Delta z),\\
    \tilde B_j(z_2,z_1)&=-\frac{\sinh(\lambda\Delta z)}{k_z\lambda},
    &\qquad \tilde C_j(z_2,z_1)&=-k_z\lambda\sinh(\lambda\Delta z).
  \end{aligned}
  \label{eq:v2_single_layer_coefficients}
\end{equation}
Thus $\mathbf{M}_j(z_2,z_1)$ is the layer matrix evaluated at
$\Delta z$; across multiple layers, its entries depend on both endpoints
through the intervening material layers and contacts.

For the rear mode, following Eq.~\eqref{eq:v2_rear_mode_upward},
we propagate upward from the normalized rear values
$\tilde{\Theta}_j^{\mathrm{R}}(L)=1$ and
$\tilde{Q}_{z,j}^{\mathrm{R}}(L)=Y_L$.
Since $\mathbf{M}_j(L,z)$ maps downward from $z$ to $L$, its inverse maps
upward from $L$ to $z$. Using its unit determinant and
$\mathbf{M}_j(z,L)=\mathbf{M}_j(L,z)^{-1}$ gives
\begin{equation}
  \begin{aligned}
    \begin{bmatrix}
      \tilde{\Theta}_j^{\mathrm{R}}(z)\\
      \tilde{Q}_{z,j}^{\mathrm{R}}(z)
    \end{bmatrix}
    &=\mathbf{M}_j(L,z)^{-1}\begin{bmatrix}1\\Y_L\end{bmatrix}
    =\begin{bmatrix}
      \tilde D_j(L,z)-Y_L\tilde B_j(L,z)\\
      Y_L\tilde A_j(L,z)-\tilde C_j(L,z)
    \end{bmatrix}\\
    &=\mathbf{M}_j(z,L)\begin{bmatrix}1\\Y_L\end{bmatrix}
    =\begin{bmatrix}
      \tilde A_j(z,L)+Y_L\tilde B_j(z,L)\\
      \tilde C_j(z,L)+Y_L\tilde D_j(z,L)
    \end{bmatrix}.
  \end{aligned}
  \label{eq:v2_rear_mode_depth}
\end{equation}

For the top mode, we propagate downward from
$\tilde{\Theta}_j^{\mathrm{T}}(0)=1$ and
$\tilde{Q}_{z,j}^{\mathrm{T}}(0)=0$:
\begin{equation}
  \begin{bmatrix}
    \tilde{\Theta}_j^{\mathrm{T}}(z)\\
    \tilde{Q}_{z,j}^{\mathrm{T}}(z)
  \end{bmatrix}
  =\mathbf{M}_j(z,0)\begin{bmatrix}1\\0\end{bmatrix}
  =\begin{bmatrix}
    \tilde A_j(z,0)&\tilde B_j(z,0)\\
    \tilde C_j(z,0)&\tilde D_j(z,0)
  \end{bmatrix}\begin{bmatrix}1\\0\end{bmatrix}
  =\begin{bmatrix}\tilde A_j(z,0)\\\tilde C_j(z,0)\end{bmatrix}.
  \label{eq:v2_top_mode_depth}
\end{equation}
Here $\mathbf{M}_j(z,0)$ directly maps the top-surface values to depth $z$.
The physical reference field
$\hat{\Theta}_{\mathrm{ref},j}$ in Eq.~\eqref{eq:v2_reference_field}
is obtained by applying the surface Green's function to the pump source.
Each mode satisfies its corresponding endpoint condition; the assembled
Green's function satisfies both.
By the Sturm--Liouville construction~\citep{Cole2011,Stakgold2011},
the top mode applies above the source and the rear mode below it;
imposing temperature continuity and a unit heat-flux jump at $z=z'$
gives the Wronskian form
\begin{equation}
  \begin{aligned}
  \tilde{G}_{\mathrm{in},j}(u,v;z,z')
  &=
  \frac{\tilde{\Theta}_j^{\mathrm{T}}(z_<)\,
        \tilde{\Theta}_j^{\mathrm{R}}(z_>)}{W_j(u,v)},\\
  W_j
  &=
  \tilde{\Theta}_j^{\mathrm{T}}\,\tilde{Q}_{z,j}^{\mathrm{R}}
  -\tilde{Q}_{z,j}^{\mathrm{T}}\,\tilde{\Theta}_j^{\mathrm{R}},
  \qquad
  z_<=\min(z,z'),\quad z_>=\max(z,z').
  \end{aligned}
  \label{eq:v2_depth_green}
\end{equation}

The internal Green's function~\eqref{eq:v2_depth_green} supplies the kernel
relating the interface flux to the temperature it induces, extending the
Green-function treatments of vertical thermal
barriers~\citep{PechMay2014,Gonzalez2019} to anisotropic multilayer stacks:
\begin{equation}
  \hat{K}_j(u,y;z,z')=4\int_0^\infty
  \tilde{G}_{\mathrm{in},j}(u,v;z,z')\cos(2\pi v|y|)\,dv.
  \label{eq:v2_kernel}
\end{equation}
The factor four combines the two-sided Fourier inversion with the factor
two for a flux imposed on the boundary of a half-space. In a distributional
sense the resulting kernel obeys
$k_{y,a}\partial_y\hat K_a|_{0^-}=\delta(z-z')$ and
$k_{y,b}\partial_y\hat K_b|_{0^+}=-\delta(z-z')$.
Consequently, adding the kernel response with signs $s_a=+1$ and $s_b=-1$
supplies exactly the difference between the required total flux and the
reference flux. Integrating that excess-flux response over the computational
coupling support $\Gamma$ gives
\begin{equation}
  \hat{\Theta}_j(u,y,z)=\hat{\Theta}_{\mathrm{ref},j}(u,y,z)
  +s_j\int_{\Gamma}\hat{K}_j(u,y;z,z')
  \bigl[\hat{Q}_y(u,z')-\hat{Q}_{y,\mathrm{ref},j}(u,z')\bigr]\,dz',
  \qquad y\in\Omega_j,
  \label{eq:v2_total_field}
\end{equation}
with $\Omega_a=(-\infty,0)$, $\Omega_b=(0,\infty)$, $s_a=+1$, and
$s_b=-1$. The signs $s_j$ ensure continuity of the total interface
flux, while subtracting $\hat{Q}_{y,\mathrm{ref},j}$ avoids double
counting the reference flux. The integral term defines the correction
field $\hat{\Theta}_{\mathrm{corr},j}$ over the coupling support
$\Gamma=[0,z_{\mathrm{int,max}}]$. Numerical truncation and the
treatment of physical interface endpoints are discussed in
Section~\ref{sec:interface_endpoints}.

At this stage, the reference fields and kernels are known, while the
total interface flux $\hat{Q}_y$ remains unknown.
Evaluating~\eqref{eq:v2_total_field} on both sides of $y=0$ and imposing
the temperature-jump condition~\eqref{eq:v2_vertical_jump} gives,
for each $u$, the Fredholm Green's-function boundary integral equation
(GBIE) for $\hat{Q}_y$:
\begin{equation}
  \begin{aligned}
  &\int_{\Gamma}\bigl[\hat{K}_a(u,0;z,z')+\hat{K}_b(u,0;z,z')\bigr]
  \hat{Q}_y(u,z')\,dz'+R_v(z)\,\hat{Q}_y(u,z)\\
  &\qquad=\Delta\hat{\Theta}_{\mathrm{ref}}(u,z)
  +\int_{\Gamma}\bigl[\hat{K}_a(u,0;z,z')\,
    \hat{Q}_{y,\mathrm{ref},a}(u,z')
  +\hat{K}_b(u,0;z,z')\,
    \hat{Q}_{y,\mathrm{ref},b}(u,z')\bigr]\,dz',
  \end{aligned}
  \label{eq:v2_bie}
\end{equation}
where $\Delta\hat{\Theta}_{\mathrm{ref}}(u,z)=
\hat{\Theta}_{\mathrm{ref},b}(u,0,z)-\hat{\Theta}_{\mathrm{ref},a}(u,0,z)$.
Solving~\eqref{eq:v2_bie} for $\hat{Q}_y(u,z)$ along $\Gamma$ determines
the correction field in~\eqref{eq:v2_total_field}. Adding it to the known
reference field gives the total temperature.

\subsection{Discretization and Surface Reconstruction}\label{sec:discretization}
On \(N_z\) Gauss--Legendre nodes \(\{z_m\}_{m=1}^{N_z}\) with depth weights
\(\{\psi_m^{(z)}\}\), \eqref{eq:v2_bie} becomes
\begin{equation}
    \bigl[(\hat{\mathbf{K}}_a(u)+\hat{\mathbf{K}}_b(u))
    \boldsymbol{\Psi}_z
    +\mathbf{R}_v\bigr]
    \hat{\mathbf{Q}}_y(u)=\hat{\mathbf{b}}(u),
    \qquad
    \boldsymbol{\Psi}_z=\operatorname{diag}(\psi_m^{(z)}),
    \label{eq:v2_bie_matrix}
\end{equation}
with $[\hat{\mathbf{K}}_j(u)]_{mn}=\hat{K}_j(u,0;z_m,z_n)$ for
$m,n=1,\dots,N_z$, $\mathbf{R}_v=\operatorname{diag}[R_v(z_m)]$,
$\hat{\mathbf{Q}}_y(u)$ the vector of nodal values $\hat{Q}_y(u,z_m)$, and
$\hat{\mathbf{b}}(u)$ the quadrature of the right-hand side.

Solving \eqref{eq:v2_bie_matrix} at each $u$ gives the nodal interface flux
$\hat{Q}_y(u,z_m)$. Substituting it into the total-field
equation~\eqref{eq:v2_total_field} at $z=0$ gives the surface field
$\hat{\Theta}_j(u,y,0)$, whose inverse $x$-transform yields the surface
temperature
\begin{equation}
    \Theta_j(x,y,0)=2\int_0^\infty
    \hat{\Theta}_j(u,y,0)\cos(2\pi ux)\,du.
    \label{eq:v2_surface_reconstruction}
\end{equation}
The one-sided integral follows from the \(x\)-symmetry of the centered beam;
an \(x\)-offset replaces the cosine argument by \(2\pi u(x-d_x)\).

\subsection{Numerical Implementation}\label{sec:numerics}

\subsubsection{Quadrature and stable evaluation}\label{sec:quadrature}

The $u$, $v$, and $z$ integrals are evaluated by composite
Gauss--Legendre quadrature. The standard setting
\((N_u,N_v,N_z)=(35,120,25)\), selected from five tested resolution levels,
is refined when convergence requires it. The grids are partitioned at
material boundaries and relevant beam, layer-thickness, and thermal-diffusion
scales. Thus \(N_z\) discretizes only the unknown vertical-interface flux, and no quadrature interval crosses a material
or prescribed \(R_v(z)\) discontinuity. Layer-relative breakpoints also
protect thin films; the interface segment lying within each
nonzero film examined here contains at least five collocation nodes. The
horizontal multilayer response remains analytical through the transfer
matrices.

The \(u\)- and \(v\)-cutoffs are selected separately because the
along-interface Gaussian is rapidly damped while the split transverse source
has a slower spectral tail. Stable evaluation uses a scaled error function,
rear-mode admittance recursion, and normalized top-mode propagation with
logarithmic scale tracking. Stable propagation through thermally thick
layers can also be formulated with scattering matrices~\citep{Li2022ThermalScatteringMatrix}.
Grid construction, the five numerical presets,
input-dependent cutoff selection, diagonal treatment, and spectral-cutoff
and depth-refinement comparisons are detailed in Supplementary Sections~S2--S4 and S7.

\subsubsection{Perfect contact, interface depth, and sample thickness}\label{sec:interface_endpoints}

Equation~\eqref{eq:v2_bie_matrix} is solved directly in resistance form.
Setting $R_v=0$ removes the local resistance term while leaving the
kernel intact, so perfect contact between two dissimilar stacks is
recovered without special treatment. For the $100~\mathrm{nm}$
multilayer benchmark, a near-perfect contact,
$G_v=10^{11}~\mathrm{W\,m^{-2}\,K^{-1}}$, reproduces that limit to
within $0.002\%$ in relative complex $L_2$ norm (Supplementary
Section~S13).

The coupling support $\Gamma$ also includes any perfectly connected
regions, such as a continuous film above a buried interface or a common
support below a finite-depth resistive segment. Their total interface
flux remains an unknown, with $R_v=0$ as defined after
Eq.~\eqref{eq:v2_vertical_jump}. The numerical endpoint
$z_{\mathrm{int,max}}$ truncates the coupling problem and must be checked
independently of the physical endpoint $z_{\max}$ of $\Gamma_{R_v}$.

At a termination $z_t$ on a common support, Eq.~\eqref{eq:v2_vertical_jump} imposes
$\hat\Theta_b=\hat\Theta_a$ for $z>z_t$, together with the horizontal
contact laws of the two stacks. No additional value of $\hat Q_y(z_t)$ is
prescribed. In particular, finite horizontal resistances at the junction
mean that $\hat\Theta_b(z_t^-)-\hat\Theta_a(z_t^-)$ and
$\hat Q_y(z_t^-)$ need not vanish. Material and resistance breakpoints are
quadrature boundaries; the collocation nodes lie inside the intervals.

The formulation also extends to finite-thickness samples by imposing a
rear boundary condition at the physical depth $L$. For each stack, a
laterally uniform linear condition can be written as
\begin{equation}
  \tilde Q_z(L)-Y_L\tilde\Theta(L)=\tilde g_L,
  \label{eq:v2_general_rear_condition}
\end{equation}
where $\tilde g_L$ is the prescribed rear forcing.
Setting $Y_L=0$ gives a prescribed-flux (Neumann) condition;
$Y_L=h_{\mathrm{rear}}$ gives a convective (Robin) condition, with
$h_{\mathrm{rear}}$ the heat-transfer coefficient and $\tilde g_L=0$
for an unmodulated ambient temperature.
The surface and internal Green's functions in
Eqs.~\eqref{eq:v2_surface_green} and~\eqref{eq:v2_depth_green} retain
their forms, using a rear mode that satisfies the homogeneous condition
$\tilde g_L=0$. Any nonzero rear forcing is included through an additional
rear-driven reference field and its interface flux. The correction then
satisfies $\tilde Q_{z,\mathrm{corr},j}(L)
-Y_L\tilde\Theta_{\mathrm{corr},j}(L)=0$; for a Robin condition, this
allows nonzero rear heat flux. In this case, the coupling support is restricted
to the physical interval $\Gamma=[0,L]$.

\subsubsection{Kernel diagonal and cutoff dependence}\label{sec:kernel_diagonal}

The inverse transform in Eq.~\eqref{eq:v2_kernel} produces an integrable
logarithmic self-singularity at \(z=z'\); the \(z\!\lessgtr\!z'\) split
evaluates the Green's function stably but does not remove this singularity.
The diagonal uses the same finite-$v_{\max}$ composite quadrature as
the off-diagonal entries. At an interior point of a homogeneous layer,
putting $\alpha_j=\sqrt{k_{y,j}/k_{z,j}}$ gives the large-$v$ term
\begin{equation}
  \tilde G_{\mathrm{in},j}(u,v;z,z')\sim
  \frac{e^{-2\pi v\alpha_j|z-z'|}}
       {4\pi v\sqrt{k_{y,j}k_{z,j}}},
  \qquad v\to\infty.
  \label{eq:rev_green_asymptotic}
\end{equation}
Consequently,
\begin{equation}
  \hat K_j(u,0;z,z')=
  -\frac{\log(|z-z'|/\ell)}{\pi\sqrt{k_{y,j}k_{z,j}}}+O(1),
  \qquad z'\to z,
  \label{eq:rev_kernel_log}
\end{equation}
where $\ell>0$ is an arbitrary reference length that makes the
logarithm dimensionless; its choice affects only the bounded
remainder~\citep[Sec.~2.1]{Dijkstra2008BEM}.
This interior expansion does not cover a source
exactly at a free surface or material junction. The finite-cutoff diagonal
therefore grows logarithmically with $v_{\max}$. Its nodal value is not an
analytic self-cell integral. The implications for numerical convergence
are discussed in Supplementary Sections~S3 and S12.

\section{Results and validation}\label{sec:validation}

Validation proceeds through three geometries of increasing complexity.
We first compare homogeneous isotropic half-spaces with the published
semi-analytical solution~\citep{Gonzalez2019}. We then examine semi-infinite vertical
interfaces in anisotropic multilayers, including surface-breaking and
buried configurations, followed by a finite-depth interface with
depth-dependent $G_v(z)$ on a connected support.
The multilayer cases are compared with independent three-dimensional FEM
calculations. Error metrics and phase-reference conventions are given in
Supplementary Sections~S7 and~S12.

\subsection{Isotropic bulk}\label{sec:validation_isotropic}
\label{sec:validation_isotropic_accuracy}

\begin{figure}[!ht]
    \centering
    \includegraphics[width=\linewidth]{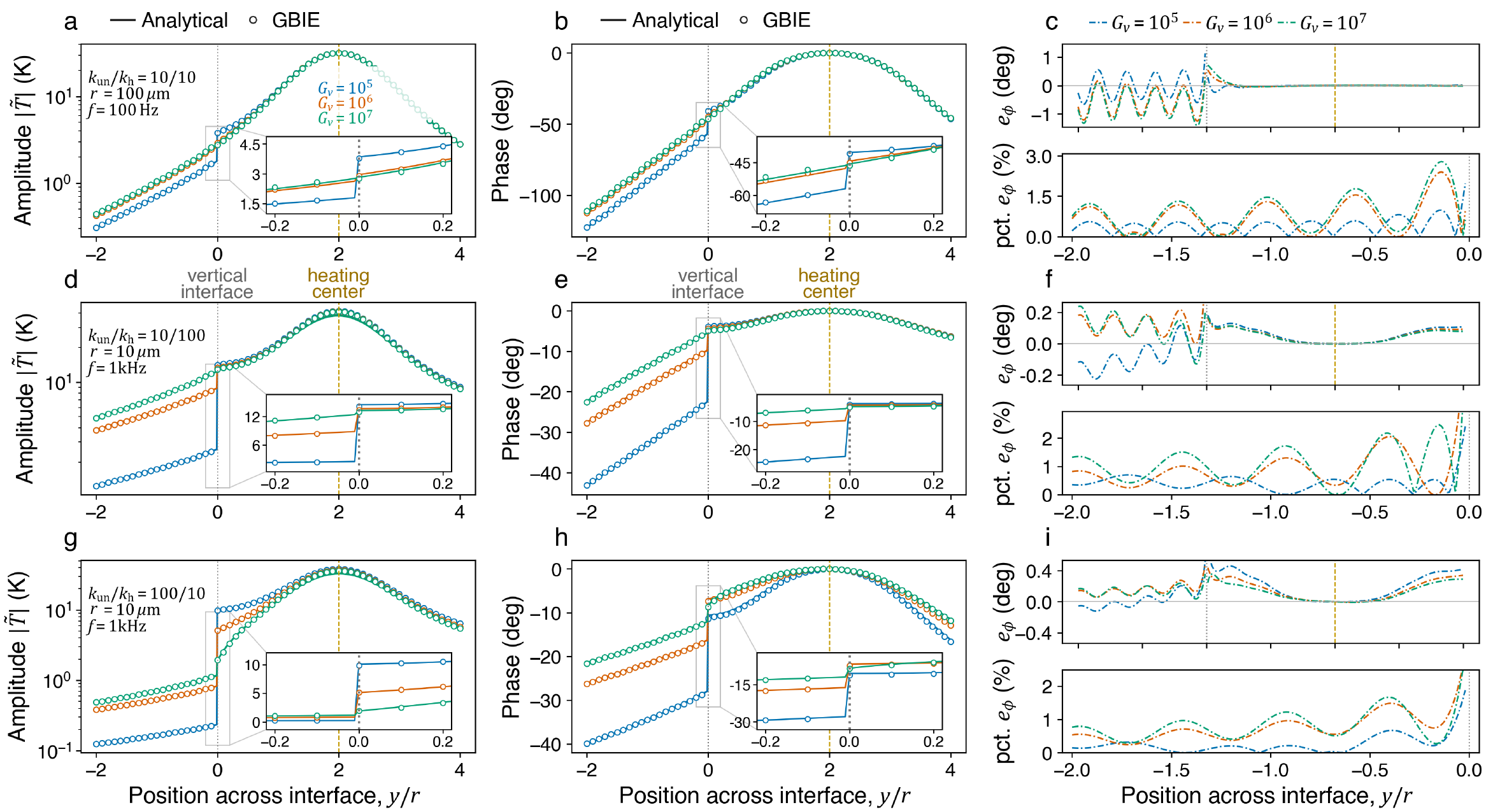}
    \caption{Validation against the semi-analytical isotropic
    model. Rows correspond to
    \((k_{\mathrm{un}},k_{\mathrm{h}})=(10,10)\), \((10,100)\), and
    \((100,10)~\mathrm{W\,m^{-1}\,K^{-1}}\). The first two columns show
    amplitude and phase: solid curves denote the reference and open circles
    denote the GBIE. Colors denote \(G_v=10^5\) (blue), \(10^6\) (orange),
    and \(10^7\) (green), all in \(\mathrm{W\,m^{-2}\,K^{-1}}\). The third
    column shows the phase residual \(e_\phi\) and its pointwise
    percentage magnitude \(e_{\phi,\%}\) as dash-dotted curves in the same
    colors. The dotted and dashed lines mark the
    interface and heating center, respectively; insets enlarge the
    interfacial jump.}
    \label{fig:validation_isotropic}
\end{figure}

We first benchmark the GBIE against the semi-analytical solution
\citep{Gonzalez2019} for an infinite vertical interface between homogeneous
isotropic half-spaces.
Figure~\ref{fig:validation_isotropic} compares surface amplitude and phase
profiles from line scans along $y$ at fixed $x=0$, through the pump center
and across the vertical interface at $y=0$. The nine cases combine
three conductivity pairs $(k_{\rm un},k_{\rm h})=(10,10)$, $(10,100)$,
and $(100,10)$ in $\mathrm{W\,m^{-1}\,K^{-1}}$ and three conductances
$G_v=10^5$, $10^6$, and $10^7~\mathrm{W\,m^{-2}\,K^{-1}}$.
This set tests both directions of conductivity contrast and the matched
case while varying interfacial resistance over two orders of magnitude.
The source is centered at $y/r=2$; the spot size and modulation frequency
are chosen in each case to keep the phase span
$|\phi_{\max}-\phi_{\min}|$ within an experimentally relevant range.

The GBIE reproduces the amplitude and phase profiles across all nine cases,
including the interfacial jump and the changes in profile slope on either
side. Lower $G_v$ increases the jump, as expected for greater resistance
to lateral heat flow. Using the semi-analytical solution as the reference,
the phase residual remains approximately
within $\pm1^\circ$; its percentage magnitude is below $3\%$ and generally
below $1\%$ away from the interface. These percentage residuals are below
the approximately $5\%$ typical experimental noise level.

\label{sec:validation_isotropic_efficiency}

The computational advantage of the GBIE over the analytical model
grows with the number of observation coordinates. For a single
121-point line scan
runtimes are comparable; the GBIE is $1.9$--$4.1\times$ faster when that
line is refined to 241 points and $5.0$--$7.6\times$ faster for five
parallel 121-point scans at different $x$-positions. Expanding its output
from $1\times121$ to $1001\times121$ points increases runtime by roughly $2\%$.
The GBIE solves for interfacial flux once per $u$ node and reuses it at
every coordinate, whereas the
analytical model repeats its dominant source-convolution quadrature at
each coordinate.
Detailed timings appear in Supplementary Sections~S2, S5, and S11.

\subsection{Anisotropic multilayers}
\label{sec:validation_anisotropic_multilayer}
\label{sec:validation_anisotropic_accuracy}

We next examine anisotropic multilayers. No analytical reference exists
for this geometry, so the GBIE is verified against a three-dimensional
FEM model implemented in COMSOL 6.2. The model solves the frequency-domain
heat equation with quadratic Lagrange temperature shape functions.
A hemispherical region of radius $5r$ beneath the pump is locally
refined, with maximum element sizes ranging from $r/10$ to $r/3$
across the mesh study, while the remainder of the domain uses
physics-controlled \emph{Finer} meshing with adaptive refinement
(see Supplementary Section~S6 for detailed numerical settings).
The benchmark consists of a $1~\mu\mathrm{m}$ film over dissimilar
anisotropic substrates, with $r=1~\mu\mathrm{m}$,
$f=100~\mathrm{kHz}$, and the inputs listed in
Table~\ref{tab:validation_multilayer_properties}. Five calculations
vary $G_v$ at fixed $G_h$ or $G_h$ at fixed $G_v$.

\begin{figure}[htbp]
    \centering
    \includegraphics[width=\linewidth]{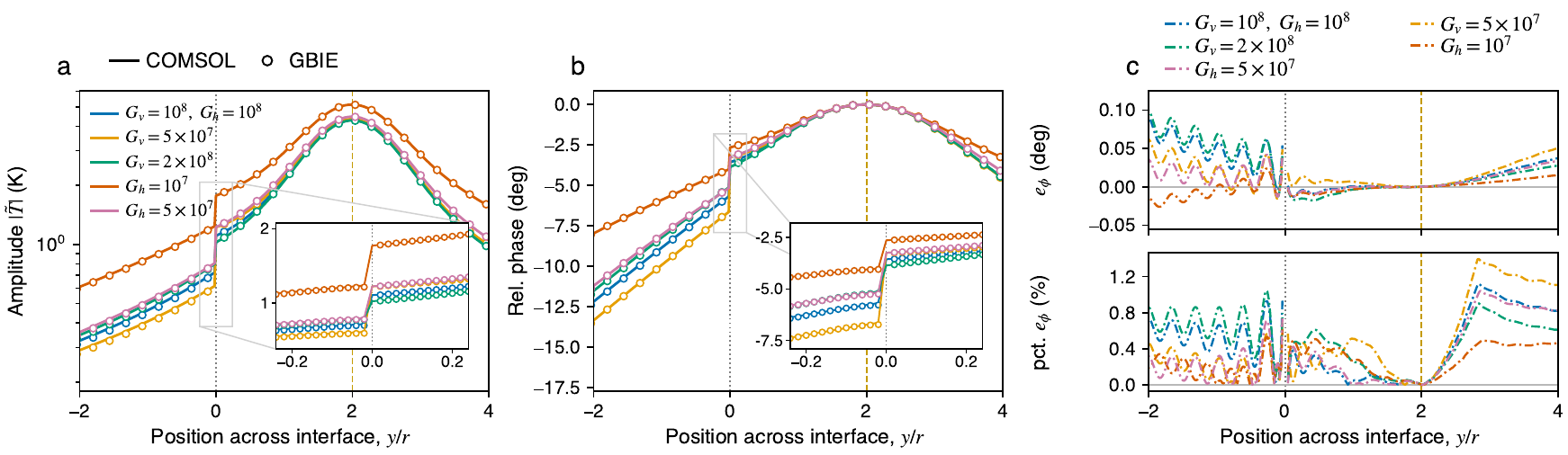}
    \caption{Numerical validation against COMSOL for an anisotropic
    multilayer stack
    with both vertical and horizontal TBCs. Panels show amplitude (a),
    relative phase (b), and the signed phase residual \(e_\phi\) with its
    pointwise percentage magnitude \(e_{\phi,\%}\) (c). COMSOL profiles are
    solid curves, GBIE profiles are open circles, and residuals are
    dash-dotted curves. Colors denote
    \((G_v,G_h)=(10^8,10^8)\) (blue),
    \((5\times10^7,10^8)\) (gold),
    \((2\times10^8,10^8)\) (green),
    \((10^8,10^7)\) (vermilion), and
    \((10^8,5\times10^7)\) (pink), with conductances in
    \(\mathrm{W\,m^{-2}\,K^{-1}}\).}
    \label{fig:validation_multilayer}
\end{figure}

The two conductances produce distinct signatures. Lowering $G_v$ enlarges
the local interface jump. Lowering $G_h$ impedes heat transfer into the
substrates, allowing more heat to spread laterally within the film and
changing the broader surface profile. The GBIE captures both trends and
the interfacial step, with phase residuals approximately
within $\pm0.1^\circ$ (Fig.~\ref{fig:validation_multilayer}) and percentage
magnitude below 1.2 $\%$.

\begin{table}[!t]
    \centering
    \caption{Inputs for the featured anisotropic multilayer cases in the
    results section. Conductivity triples are
    \((k_x,k_y,k_z)\) in \(\mathrm{W\,m^{-1}\,K^{-1}}\), with \(x\)
    parallel to the vertical interface, \(y\) normal to it, and \(z\)
    through the depth. Thicknesses are in \(\mu\mathrm{m}\), volumetric
    heat capacities are in \(\mathrm{J\,m^{-3}\,K^{-1}}\), and
    conductances are in \(\mathrm{W\,m^{-2}\,K^{-1}}\). The common value
    \(C_v=10^6~\mathrm{J\,m^{-3}\,K^{-1}}\) is a deliberate normalization,
    not a material-specific heat capacity. The listed model depth is the
    finite COMSOL truncation; ``semi-infinite'' denotes the corresponding
    GBIE continuation through the terminal substrate.}
    \label{tab:validation_multilayer_properties}
    \begingroup
    \scriptsize
    \setlength{\tabcolsep}{3pt}
    \renewcommand{\arraystretch}{1.14}
    \begin{tabular}{@{}
        >{\raggedright\arraybackslash}p{0.18\linewidth}
        >{\raggedright\arraybackslash}p{0.25\linewidth}
        >{\raggedright\arraybackslash}p{0.25\linewidth}
        >{\raggedright\arraybackslash}p{0.25\linewidth}@{}}
        \toprule
        Input &
        \shortstack[l]{infinite depth interface\\(surface-breaking)} &
        \shortstack[l]{Buried interface\\(surface-breaking/buried)} &
        \shortstack[l]{Finite-depth vertical interface\\on a connected support} \\
        \midrule
        \multicolumn{4}{@{}l}{\textit{Materials and geometry}} \\
        Film thickness \(h_f\) &
        1 & 1 & 2 \\
        Film conductivity &
        \((100,100,100)\) &
        \((100,100,100)\) &
        \((100,100,100)\) \\
        Unheated substrate (\(y<0\)) &
        \((120,100,110)\) &
        \((120,100,110)\) &
        \((55,90,30)\), \(2<z<50\) \\
        Heated substrate (\(y>0\)) &
        \((60,80,90)\) &
        \((60,80,90)\) &
        \((75,60,50)\), \(2<z<50\) \\
        Connected support &
        --- &
        --- &
        \((60,60,60)\), \(50<z\leq300\) \\
        COMSOL model depth &
        100 & 100 & 300 \\
        \(C_v\), all regions &
        \(1.0\times10^6\) &
        \(1.0\times10^6\) &
        \(1.0\times10^6\) \\
        \addlinespace
        \multicolumn{4}{@{}l}{\textit{Vertical and horizontal interfaces}} \\
        Interface onset \(z_{\min}\) &
        0 &
        \(\begin{aligned}
           &0 &&\text{(surface-breaking)},\\[-2pt]
           &1 &&\text{(buried)}
        \end{aligned}\) &
        0 \\
        Interface termination \(z_t\) &
        Through terminal substrate; semi-infinite in GBIE &
        Through terminal substrate; semi-infinite in GBIE &
        \(50\); the two upper regions merge into the connected support for
        \(z>z_t\) \\
        Vertical \(G_v\) &
        \(\{5\times10^7,\,10^8,\,2\times10^8\}\) at
        \(G_h=10^8\) &
        \(10^8\) &
        \(10^8,\ 10^7,\ 2\times10^8\) over
        \(0\leq z<2,\ 2\leq z<5,\ 5\leq z\leq50\), respectively \\
        Horizontal \(G_h\) &
        \(\{10^7,\,5\times10^7,\,10^8\}\) at
        \(G_v=10^8\) &
        \(10^8\) &
        \(10^8\) at the film--upper-region and
        upper-region--support contacts \\
        \addlinespace
        \multicolumn{4}{@{}l}{\textit{Source conditions}} \\
        Gaussian source &
        \multicolumn{3}{l@{}}{
        \(P_0=1~\mathrm{mW}\), \(r=1~\mu\mathrm{m}\),
        \(y_0/r=2\), and \(f=100~\mathrm{kHz}\)} \\
        \midrule
        Figure &
        Fig.~\ref{fig:validation_multilayer} &
        Fig.~\ref{fig:validation_buried}(c--e) &
        Fig.~\ref{fig:validation_buried}(f--h) \\
        \bottomrule
    \end{tabular}
    \endgroup
\end{table}

To guide the choice of lateral domain, we use the estimate
\begin{equation}
 \delta_{\rm lat}=\max_{j,\,i\in\{x,y\}}
 \sqrt{\frac{k_{i,j}}{\pi C_{v,j}f}},\qquad
 L_y\geq L_e+8\delta_{\rm lat},
 \label{eq:lateral_depth}
\end{equation}
where $j$ indexes material regions and $L_e$ is the retained scan width.
This gives $\delta_{\rm lat}\simeq56~\mu\mathrm{m}$ for
$k=100~\mathrm{W\,m^{-1}\,K^{-1}}$, $C_v=10^6~\mathrm{J\,m^{-3}\,K^{-1}}$,
and $f=10~\mathrm{kHz}$. Four penetration depths of padding per side are
a guide for domain sizing; a domain-refinement test is still needed to
quantify the remaining boundary error. The $250$--$500~\mu\mathrm{m}$
domain/property study varies conductivities together with box size, so it
does not isolate domain convergence at fixed properties. The low-frequency
tests in Supplementary Section~S8 show boundary sensitivity but do not
establish negligible finite-box error in every benchmark.

We next compare the computational performance of GBIE and FEM across
different domain sizes and film thicknesses
(Table~\ref{tab:multilayer_computational_cost}). Under matched one-core
execution, the original NumPy/SciPy GBIE is $29$--$210$ times faster
and reduces reported peak memory by factors of $120$--$450$.
Larger domains and thinner films substantially increase the FEM cost,
whereas the GBIE runtime and memory use remain nearly constant.
For the $100~\mathrm{nm}$ film, the one-core original GBIE remains
approximately $90$ times faster than optimized 16-core COMSOL
(Supplementary Section~S10).

The JAX implementation further reduces the computational cost. JAX is a
Python array library that vectorizes numerical programs over batch axes
and compiles them just-in-time (JIT) through XLA; vectorizing the
$(u,v)$ quadrature replaces the per-node loop over transfer-matrix
propagations with batched array operations, and compilation removes
per-node interpreter overhead.
The JAX implementation has two timing regimes: fresh-start timing
and post-compilation timing. Fresh-start timing describes the cost
of an isolated calculation, including startup, compilation, and
output export. The complete workflow takes $2.02$ and
$1.80~\mathrm{s}$ on four and sixteen cores, respectively,
representing approximately $260$- and $290$-fold reductions
relative to the corresponding COMSOL workflows. After compilation,
a sixteen-core full forward evaluation takes $0.128~\mathrm{s}$,
approximately $4100$ times faster than the complete $521.6~\mathrm{s}$
COMSOL workflow at the same core count. This repeated-evaluation time
excludes JAX startup, compilation, and output export; it is compared
with the complete COMSOL workflow, not a repeated solve using an
initialized COMSOL model. Timings at all three core counts are given
in Table~\ref{tab:multilayer_computational_cost}.
This post-compilation regime governs parameter sweeps
and inverse fitting, where a material-property update reuses the
compiled program whenever array shapes, precision, and compilation
options are unchanged
(see detailed implementation comparisons, timing definitions,
and benchmark procedures in Supplementary
Sections~S10,
S14,
and~S15).

\begin{table}[htbp]
    \centering
    \captionsetup{justification=raggedright,singlelinecheck=false}
    \caption{COMSOL and GBIE efficiency comparison}
    \label{tab:multilayer_computational_cost}
    \begingroup
    \scriptsize
    \setlength{\tabcolsep}{3pt}
    \resizebox{\linewidth}{!}{%
    \begin{tabular}{@{}lrrrrrrr@{}}
        \toprule
        Case & Domain & \(h_f\) & Cores & Wall time &
        Time ratio & Peak memory\textsuperscript{c} & Memory ratio \\
        & (\(\mu\mathrm{m}\)) & (\(\mu\mathrm{m}\)) & COMSOL/GBIE &
        COMSOL/GBIE (s) & & COMSOL/GBIE (GiB) & \\
        \midrule
        1 & 250 & 1.00 & \(1/1\) & \(264.0/8.6\)  & \(31\times\)  & \(9.3/0.078\)  & \(120\times\) \\
        2 & 250 & 1.00 & \(1/1\) & \(250.0/8.8\)  & \(29\times\)  & \(9.4/0.074\)  & \(130\times\) \\
        3 & 300 & 1.00 & \(1/1\) & \(432.0/7.2\)  & \(60\times\)  & \(12.9/0.074\) & \(170\times\) \\
        4 & 400 & 1.00 & \(1/1\) & \(681.0/7.9\)  & \(86\times\)  & \(20.3/0.075\) & \(270\times\) \\
        5 & 500 & 1.00 & \(1/1\) & \(1022.0/6.9\) & \(150\times\) & \(30.4/0.074\) & \(410\times\) \\
        \midrule
        500-nm series & 300 & 0.50 & \(1/1\) & \(1099.0/6.1\) &
        \(180\times\) & \(32.9/0.076\) & \(440\times\) \\
        200-nm series & 300 & 0.20 & \(1/1\) & \(1201.0/5.9\) &
        \(200\times\) & \(33.3/0.075\) & \(450\times\) \\
        100-nm series & 300 & 0.10 & \(1/1\) & \(1244.0/5.9\) &
        \(210\times\) & \(33.1/0.075\) & \(440\times\) \\
        100-nm JAX startup\textsuperscript{a} & 300 & 0.10 & \(1/1\) & \(1244.0/3.04\) &
        \(410\times\) & \(33.1/0.399\) & \(83\times\) \\
        100-nm JAX compiled\textsuperscript{b} & 300 & 0.10 & \(1/1\) & \(1244.0/0.162\) & \(7700\times\)\textsuperscript{b} & --- & --- \\
        \midrule
        100-nm optimized & 300 & 0.10 & \(4/1\) & \(533.7/5.9\) &
        \(91\times\) & \(34.8/0.075\) & \(460\times\) \\
        100-nm JAX startup\textsuperscript{a} & 300 & 0.10 & \(4/4\) & \(533.7/2.02\) &
        \(260\times\) & \(34.8/0.381\) & \(91\times\) \\
        100-nm JAX compiled\textsuperscript{b} & 300 & 0.10 & \(4/4\) & \(533.7/0.146\) & \(3600\times\)\textsuperscript{b} & --- & --- \\
        \midrule
        100-nm optimized & 300 & 0.10 & \(16/1\) & \(521.6/5.9\) &
        \(89\times\) & \(36.2/0.075\) & \(480\times\) \\
        100-nm JAX startup\textsuperscript{a} & 300 & 0.10 & \(16/16\) & \(521.6/1.80\) &
        \(290\times\) & \(36.2/0.408\) & \(89\times\) \\
        100-nm JAX compiled\textsuperscript{b} & 300 & 0.10 & \(16/16\) & \(521.6/0.128\) & \(4100\times\)\textsuperscript{b} & --- & --- \\
        \bottomrule
    \end{tabular}%
    }
    \par\vspace{2pt}
    \begin{minipage}{\linewidth}
    \scriptsize
    \textsuperscript{a}\,\textbf{JAX startup:} Median of three fresh
    processes, including imports, compilation, synchronized calculation,
    validation, and output.
    \par\smallskip\noindent
    \textsuperscript{b}\,\textbf{JAX compiled:} Repeated full forward
    evaluation after compilation, with synchronized output and no reused
    solution or factorization.
    \par\smallskip\noindent
    \textsuperscript{c}\,\textbf{Peak memory:} COMSOL-reported peak
    physical memory (a median for optimized runs); JAX uses the largest
    process peak RSS across three fresh runs, including compilation.
    The reporters and aggregation differ. Per-evaluation memory was not
    measured separately and is shown as a dash.
    \par\smallskip\noindent
    All runs used AMD EPYC 9374F processors. Thin-film comparisons
    matched physical nodes within each case; the 100-nm COMSOL and JAX
    series used the same node in independent runs. Further timing and
    memory definitions appear in Supplementary Sections~S10
    and~S15.
    \end{minipage}
    \endgroup
\end{table}

We also test whether the multilayer solution approaches the bulk limit
as the film vanishes. The film enters the Green's function through its
transfer matrix; only the interfacial flux is depth-discretized.
With $G_h\to\infty$, reducing the film to $1~\mathrm{nm}$ gives P95
differences from the zero-film GBIE reference of $0.004^\circ$ in phase
and $0.08\%$ in amplitude (Supplementary Sections~S9 and S11).
This limit tests geometric consistency within the continuum model;
applying local Fourier transport to a real $1~\mathrm{nm}$ film requires
separate physical justification.

\subsection{Buried and finite-depth interfaces}
\label{sec:validation_buried_multilayer}

\begin{figure}[htbp]
    \centering
    \makebox[\linewidth][c]{%
        \includegraphics[
            width=0.98\linewidth,
            height=0.68\textheight,
            keepaspectratio
        ]{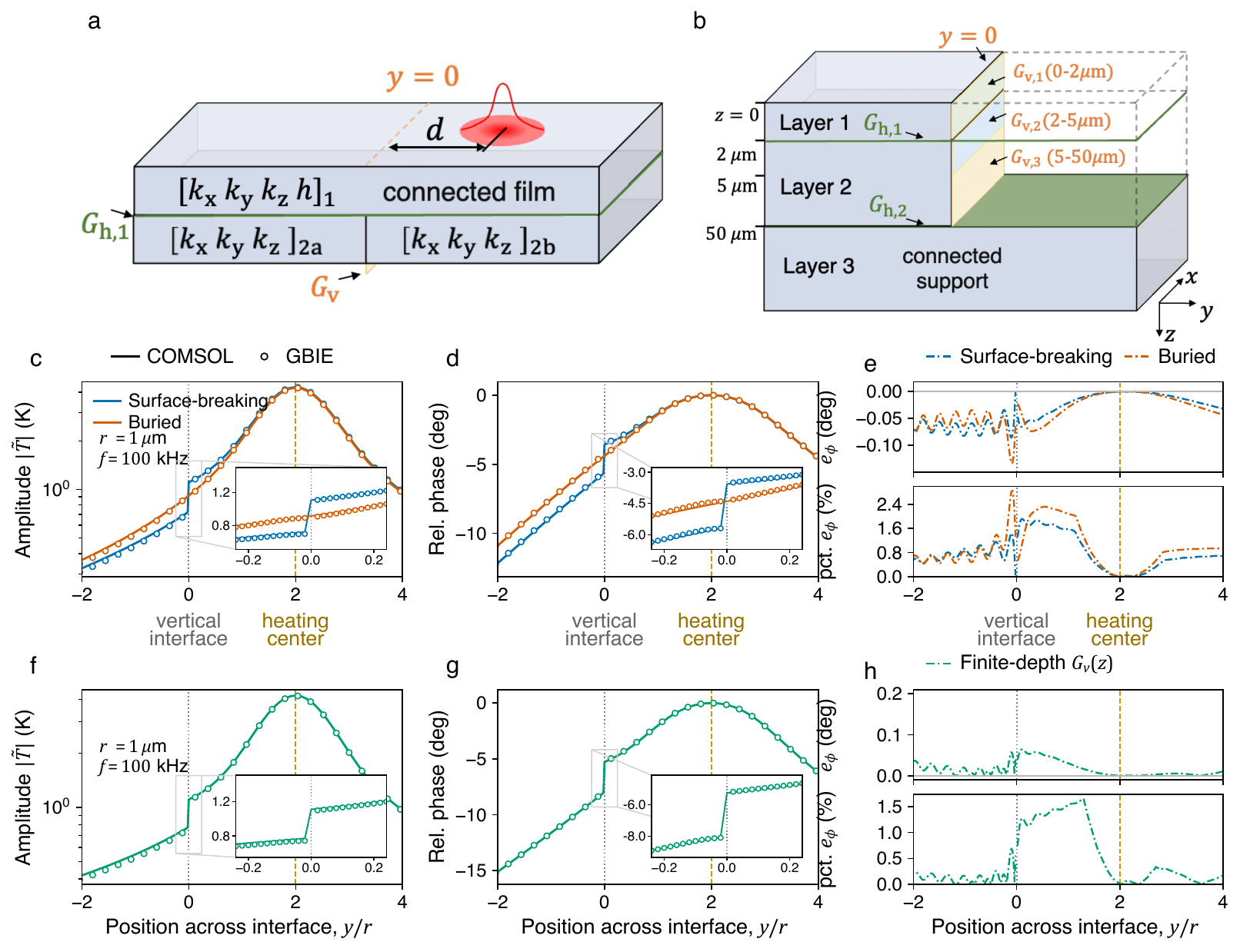}%
    }
      \caption{Geometries and COMSOL verification of buried and
    finite-depth interfaces. Panels (a) and (b) show the buried
    semi-infinite and finite-depth connected-support geometries.
    Panels (c--e) compare surface-breaking (blue) and buried (orange)
    semi-infinite interfaces; panels (f--h) show the finite-depth
    three-band \(G_v(z)\) interface (green), terminated by a connected
    support at \(z=50~\mu\mathrm{m}\). Each group shows amplitude
    (c,f), relative phase (d,g), and the signed phase residual
    \(e_\phi\) with its percentage magnitude \(e_{\phi,\%}\) (e,h).
    Solid curves are COMSOL and open circles are GBIE in the amplitude
    and phase panels; residuals use dash-dotted curves in the same case
    colors. Conductances, depth intervals, and material inputs are in
    Table~\ref{tab:validation_multilayer_properties}.}
    \label{fig:validation_buried}
\end{figure}

We next test the GBIE against COMSOL for a vertical interface buried
beneath a continuous film, as encountered in thermoreflectance
measurements. Figure~\ref{fig:validation_buried}a,c--e shows close
agreement in surface amplitude and phase for both the surface-breaking
and buried configurations with $h_f=1~\mu\mathrm{m}$. The P95 phase
residuals remain within $0.079^\circ$, with a maximum residual of
$0.13^\circ$. For the buried interface, lateral heat spreading through
the film strongly suppresses the surface signature of the interface,
even though the film's cross-plane thermal penetration depth,
$l_{\mathrm{tpd}}=\sqrt{k_z/(\pi f C_v)}\approx18~\mu\mathrm{m}$ at
$100~\mathrm{kHz}$, greatly exceeds its thickness. This loss of
sensitivity can hinder the extraction of vertical TBC from
transducer-based thermoreflectance measurements, even when the
transducer is thermally thin.

Finally, we assess the model for a depth-dependent interface that
terminates on a common support at $z_t=50~\mu\mathrm{m}$
(Fig.~\ref{fig:validation_buried}b,f--h). We prescribe three conductance
bands: $G_v=10^8$, $10^7$, and $2\times10^8~\mathrm{W\,m^{-2}\,K^{-1}}$
over depths of $0$--$2$, $2$--$5$, and $5$--$50~\mu\mathrm{m}$,
respectively, with perfect thermal continuity below the termination.
The remaining inputs are listed in
Table~\ref{tab:validation_multilayer_properties}.
At the selected numerical settings, the GBIE agrees with COMSOL
to a P95 phase residual of $0.055^\circ$ and a relative amplitude
error of $0.70\%$, demonstrating agreement for both discontinuous
depth-dependent conductance and a finite-depth termination.
Supplementary Sections~S6.4 and S12 report the cutoff comparisons
and depth discretization.

\section{Discussion}
\label{sec:discussion}
Several physical assumptions were made in the modeling. Heating
is represented as surface absorption, as appropriate for a metal
transducer \citep{Cahill2004,Schmidt2009FDTR}; optically penetrative
samples require a depth-dependent volumetric source
\citep{Wang2016Transducerless,Warkander2022Transducerless}. Radiative
and convective losses are neglected. Either can be incorporated as a
finite surface admittance in place of the adiabatic top-surface
condition, or at the rear through the terminal transfer-matrix
condition~\citep{Jiang2022SSTR}. The formulation further assumes linear
Fourier diffusion and local interfacial conductances, \(G_v(z)\) and
\(G_h\). It therefore predicts effective interfacial resistances and excludes
nonlocal or quasi-ballistic regimes.

The default numerical settings give close agreement with the reference
surface profiles in the reported benchmarks but are not universally
optimal. More extreme properties, such as conductivity contrasts of
$100{:}1$ or anisotropy ratios of $k_x/k_y=10^3$, may require
further adjustment of the spectral cutoffs and quadrature density.
To balance accuracy and efficiency for new material configurations,
a practical approach is to check these settings against domain- and
mesh-converged FEM solutions at representative conditions spanning the
expected property range. Acceleration with JAX and
additional computational resources can make finer numerical settings
more affordable, enabling stricter convergence checks and better
control of numerical error.

The low evaluation cost makes systematic
surveys of measurement conditions practical across vertical-interface
geometries.  For
surface-breaking grain boundaries with conductances on the order of
$10^8$--$10^9~\mathrm{W\,m^{-2}\,K^{-1}}$,
small heating spots combined with high modulation
frequencies may improve spatial resolution and sensitivity.
For buried interfaces, lower frequencies and varying
pump-to-interface offsets could increase the probing depth and help
identify conditions that reduce masking by lateral heat spreading
in the film. The optimal choices depend on material properties,
interface depth, and transducer geometry. Future studies combining
these parameter sweeps with noise-weighted sensitivity and
identifiability analyses could provide practical measurement
guidelines for a broad range of applications involving vertical
thermal interfaces.

Finally, the model assumes orthotropic conductivity with
\(k_{xy}=k_{xz}=k_{yz}=0\). Admitting nonzero off-diagonal components,
following the full-tensor heat-transport
formulation~\citep{Wang2026TensorIdentifiability}, would provide a basis
for simultaneous determination of the conductivity tensor and vTBC in
fully anisotropic materials. Pairing this extension with noise-weighted sensitivity
and identifiability analysis would convert the scaling arguments above
into quantitative measurement guidelines.

\section{Conclusion}
\label{sec:conclusion}

We developed a semi-analytical GBIE framework for vertical-interface
heat transport between dissimilar orthotropic multilayers, incorporating
horizontal TBC and depth-dependent vertical conductance while
discretizing only the interface flux. The method reproduces reference
surface temperature profiles for surface-breaking, buried, and
finite-depth interfaces. In matched single-core benchmarks, the model
runs $29$--$210$ times faster than FEM and reduces reported peak memory by
factors of $120$--$450$. After compilation, a sixteen-core JAX evaluation
of the $100~\mathrm{nm}$ benchmark takes $0.128~\mathrm{s}$, approximately
$4100$ times faster than the complete COMSOL workflow at the same core
count, enabling efficient repeated forward evaluations. The
buried-interface results show how lateral heat spreading can mask
interface sensitivity even beneath a film thinner than the thermal
penetration depth. The framework thus provides an efficient forward
model for vertical thermal transport in systems ranging from
microelectronic device sidewalls to grain boundaries in polycrystalline
solids.

\section*{CRediT authorship contribution statement}

\noindent\textbf{Dihui Wang:} Conceptualization, Data curation, Formal analysis,
Investigation, Methodology, Resources, Software, Validation, Visualization,
Writing---original draft, Writing---review and editing.

\noindent\textbf{Troy Munro:} Conceptualization, Funding acquisition, Project
administration, Supervision, Writing---review and editing.

\noindent\textbf{Heng Ban:} Conceptualization, Funding acquisition, Project
administration, Supervision, Writing---review and editing.

\section*{Declaration of competing interests}

The authors have no conflicts to disclose.

\section*{Acknowledgements}

This research was performed using funding received from the U.S. Department
of Energy, Office of Nuclear Energy's Nuclear Energy University Program,
under Award No.~DE-NE0009490 and through Brigham Young University under Award
No.~23-0606 (prime sponsor: U.S. Department of Energy).

\section*{Code and data availability}

The implementation, input metadata, processed validation profiles, and
resource logs underlying this study are retained by the corresponding author.
The accompanying Supplementary Information identifies the principal numerical
settings and per-case metrics. A versioned public archive containing the
code, raw profiles, exact COMSOL inputs, and execution scripts is required
for independent reproduction; no public release or archival identifier is
claimed in this revision.

\setlength{\bibsep}{2pt}
\bibliographystyle{unsrtnat}
\bibliography{bibliography/reference}

\end{document}


\maketitle
\clearpage

\begingroup
\footnotesize
\setlength{\cftbeforesecskip}{0.10em}
\tableofcontents
\endgroup
\clearpage

\section{Green-Function Theory for a Vertical Interface}
\label{sec:supp_theory_vertical_interface}

This section gives the derivation details for the Green's-function boundary
integral equation (GBIE) framework introduced in the main text.

\subsection{Fourier Representation and Split Gaussian Source}
\label{sec:supp_source_split}

For the in-plane coordinates we use the cycle-wavenumber Fourier convention
\begin{equation}
    \begin{aligned}
    \tilde{f}(u,v,z)
    &=
    \int_{-\infty}^{\infty}\int_{-\infty}^{\infty}
    f(x,y,z)e^{-i2\pi(ux+vy)}\,dx\,dy,\\
    f(x,y,z)
    &=
    \int_{-\infty}^{\infty}\int_{-\infty}^{\infty}
    \tilde{f}(u,v,z)e^{i2\pi(ux+vy)}\,du\,dv .
    \end{aligned}
    \label{eq:supp_fourier_pair}
\end{equation}
The absorbed laser source has a Gaussian distribution. Writing its surface
flux as $Q_{\mathrm{pump}}(x,y)=A\exp[-2(x^2+(y-d)^2)/r^2]$, the absorbed
harmonic-power amplitude $P_0$ fixes the prefactor through
\begin{equation}
    P_0=\int_{\mathbb R^2}Q_{\mathrm{pump}}\,dx\,dy
    =\frac{A\pi r^2}{2},
    \qquad A=\frac{2P_0}{\pi r^2}.
    \label{eq:supp_gaussian_power_normalization}
\end{equation}
Here $r$ is the $1/e^2$ intensity radius and $(0,d)$ is the source center.

The normalized Gaussian factors in \(x\) and \(y\) are
\begin{equation}
    g_x(x)=\frac{\sqrt{2}}{\sqrt{\pi}r}\exp\left(-\frac{2x^2}{r^2}\right),
    \qquad
    g_y(y;d)=\frac{\sqrt{2}}{\sqrt{\pi}r}
    \exp\left[-\frac{2(y-d)^2}{r^2}\right].
    \label{eq:supp_gaussian_factors}
\end{equation}
We remove the common temporal factor
\(2\pi\delta(\omega-\omega_0)\). The remaining phasor heat-flux amplitude is
\(P_0g_x(x)g_y(y;d)\). Its Fourier transform in \(x\) is
\begin{equation}
    \int_{-\infty}^{\infty}g_x(x)e^{-i2\pi ux}\,dx
    =
    \exp\left(-\frac{\pi^2r^2u^2}{2}\right).
    \label{eq:supp_gx_transform}
\end{equation}
When the beam overlaps the vertical interface, both stacks absorb laser
power. At $d=0$ the symmetric Gaussian deposits half of $P_0$ on each side;
assigning the full source to one stack would miss this two-sided heating.
We therefore split the \(y\)-Gaussian into its \(y>0\) and \(y<0\) contributions,
\begin{equation}
    F_+(v;d)=\int_{0}^{\infty}g_y(y;d)e^{-i2\pi vy}\,dy,
    \qquad
    F_-(v;d)=\int_{-\infty}^{0}g_y(y;d)e^{-i2\pi vy}\,dy .
    \label{eq:supp_half_transforms_def}
\end{equation}
Completing the square gives
\begin{equation}
    F_+(v;d)=\frac{1}{2}\Phi(v;d)\operatorname{erfc}[\zeta(v;d)],
    \qquad
    F_-(v;d)=\Phi(v;d)-F_+(v;d),
    \label{eq:supp_half_transforms}
\end{equation}
with
\begin{equation}
    \Phi(v;d)=\exp\left(-i2\pi vd-\frac{\pi^2r^2v^2}{2}\right),
    \qquad
    \zeta(v;d)=-\frac{\sqrt{2}d}{r}+\frac{i\pi rv}{\sqrt{2}} .
    \label{eq:supp_phi_zeta}
\end{equation}
Thus the side-resolved spectral source used in the main text is
\begin{equation}
    \tilde{Q}_{\mathrm{pump},j}(u,v)
    =
    P_0
    \exp\left(-\frac{\pi^2r^2u^2}{2}\right)
    F_j(v;d),
    \qquad
    F_a=F_-,\quad F_b=F_+,\quad j\in\{a,b\}.
    \label{eq:supp_source_amplitude}
\end{equation}
Here and throughout the supplement, stack \(a\) occupies \(y<0\) and stack
\(b\) occupies \(y>0\), matching the main-text notation.

\subsection{Layer Equation and Transfer Matrices}
\label{sec:supp_transfer_matrix}

Applying the Fourier convention in Eq.~\eqref{eq:supp_fourier_pair} to the
laterally extended reference stack, integration by parts twice in each
in-plane coordinate gives
\begin{equation}
    \mathcal F[\partial_x^2\Theta]=-(2\pi u)^2\tilde\Theta,
    \qquad
    \mathcal F[\partial_y^2\Theta]=-(2\pi v)^2\tilde\Theta.
    \label{eq:supp_fourier_second_derivatives}
\end{equation}
The boundary terms vanish because the temperature perturbation and its
first derivatives decay at lateral infinity. Substituting these identities
into the frequency-domain heat equation in layer $l$ of stack
$j\in\{a,b\}$ gives
\begin{equation}
    k_{z,lj}\frac{d^2\tilde{\Theta}_{lj}}{dz^2}
    -
    \left[
    k_{x,lj}(2\pi u)^2
    +k_{y,lj}(2\pi v)^2
    +i\omega_0 C_{v,lj}
    \right]\tilde{\Theta}_{lj}
    =0 .
    \label{eq:supp_layer_ode}
\end{equation}
Dividing by $k_{z,lj}$ gives
\begin{equation}
    \frac{d^2\tilde{\Theta}_{lj}}{dz^2}
    -\lambda_{lj}^2\tilde{\Theta}_{lj}=0,
    \qquad
    \lambda_{lj}
    =
    \left[
    \frac{
    k_{x,lj}(2\pi u)^2
    +k_{y,lj}(2\pi v)^2
    +i\omega_0 C_{v,lj}}
    {k_{z,lj}}
    \right]^{1/2}.
    \label{eq:supp_lambda}
\end{equation}
We choose the square-root branch with $\operatorname{Re}(\lambda)>0$.
Within each homogeneous layer the general solution is
$\tilde{\Theta}(z)=c_1\cosh(\lambda z)+c_2\sinh(\lambda z)$,
with coefficients set by the boundary and interface conditions.
Both exponential modes $e^{\pm\lambda z}$ are retained. Introducing the
state vector
\begin{equation}
    \mathbf{s}_{lj}(z)=
    \begin{bmatrix}
    \tilde{\Theta}_{lj}(z)\\
    \tilde{Q}_{z,lj}(z)
    \end{bmatrix},
    \qquad
    \tilde{Q}_{z,lj}=-k_{z,lj}
    \frac{d\tilde{\Theta}_{lj}}{dz},
    \label{eq:supp_state_vector}
\end{equation}
and eliminating $c_1$ and $c_2$ gives the temperature and flux at depth $z$
measured from the top of the layer,
\begin{equation}
    \begin{bmatrix}\tilde{\Theta}_{lj}(z)\\ \tilde{Q}_{z,lj}(z)\end{bmatrix}
    =
    \begin{bmatrix}
    \cosh(\lambda z) &
    -\sinh(\lambda z)/(k_z\lambda)\\
    -k_z\lambda\sinh(\lambda z) &
    \cosh(\lambda z)
    \end{bmatrix}
    \begin{bmatrix}\tilde{\Theta}_{lj}^{\mathrm{top}}\\
    \tilde{Q}_{z,lj}^{\mathrm{top}}\end{bmatrix},
    \label{eq:supp_layer_matrix}
\end{equation}
where the layer and stack subscripts on \(k_z\) and \(\lambda\) are suppressed
inside the matrix. A horizontal thermal boundary resistance
\(R_{h,lj}=1/G_{h,lj}\) between layers \(l\) and \(l+1\) of stack \(j\)
imposes flux continuity and a temperature jump,
\begin{equation}
    \tilde{\Theta}_{l+1,j}^{\mathrm{top}}
    =
    \tilde{\Theta}_{lj}^{\mathrm{bot}}
    -R_{h,lj}\tilde{Q}_{z,lj}^{\mathrm{bot}},
    \qquad
    \tilde{Q}_{z,l+1,j}^{\mathrm{top}}
    =
    \tilde{Q}_{z,lj}^{\mathrm{bot}},
    \label{eq:supp_horizontal_jump}
\end{equation}
or, equivalently,
\begin{equation}
    \mathbf{s}_{l+1,j}^{\mathrm{top}}
    =
    \begin{bmatrix}
    1 & -R_{h,lj}\\
    0 & 1
    \end{bmatrix}
    \mathbf{s}_{lj}^{\mathrm{bot}} .
    \label{eq:supp_horizontal_jump_matrix}
\end{equation}

\subsection{Depth Green Function}
\label{sec:supp_depth_green}

For each \((u,v)\), define two auxiliary homogeneous modes of the depth
equation. Each mode satisfies one endpoint condition and leaves its amplitude
free; the assembled Green function satisfies both endpoint conditions.
The top mode satisfies the homogeneous top-surface condition,
\begin{equation}
    \mathbf{s}_j^{\mathrm{T}}(z)=
    \begin{bmatrix}
    \tilde{\Theta}_j^{\mathrm{T}}(z)\\
    \tilde{Q}_{z,j}^{\mathrm{T}}(z)
    \end{bmatrix},
    \qquad
    \tilde{Q}_{z,j}^{\mathrm{T}}(0)=0,
    \label{eq:supp_left_state}
\end{equation}
and the rear state satisfies the selected homogeneous rear condition,
\begin{equation}
    \mathbf{s}_j^{\mathrm{R}}(z)=
    \begin{bmatrix}
    \tilde{\Theta}_j^{\mathrm{R}}(z)\\
    \tilde{Q}_{z,j}^{\mathrm{R}}(z)
    \end{bmatrix},
    \qquad
    \tilde{Q}_{z,j}^{\mathrm{R}}(L)=0
    \quad \text{for an adiabatic finite rear}.
    \label{eq:supp_right_state}
\end{equation}
For a substrate (layer $N$) of thickness $h_N$, the rear surface is at
global depth $L=\sum_{l=1}^{N}h_l$. When the substrate is sufficiently thick
compared with its thermal diffusion length, we use the semi-infinite decay
condition. With $z_N$ measured from the top of layer~$N$ and
$\operatorname{Re}(\lambda_{Nj})>0$, decay at infinity removes the growing
exponential, so
$\tilde{\Theta}_{Nj}^{\mathrm{R}}(z_N)
=\tilde{\Theta}_{Nj}^{\mathrm{R}}(0)e^{-\lambda_{Nj}z_N}$.
Using $\tilde{Q}_{z,Nj}^{\mathrm{R}}=-k_{z,Nj}\,d\tilde{\Theta}_{Nj}^{\mathrm{R}}/dz_N$
at $z_N=h_N$ gives, in global coordinates,
$\tilde{Q}_{z,j}^{\mathrm{R}}(L)
=k_{z,Nj}\lambda_{Nj}\tilde{\Theta}_j^{\mathrm{R}}(L)$.
The top and rear modes may be normalized independently by
$\tilde\Theta_j^{\mathrm T}(0)=1$ and
$\tilde\Theta_j^{\mathrm R}(L)=1$, respectively. These unit temperatures
fix only auxiliary mode scales and are independent of the laser power;
the physical reference temperature follows after applying the surface
Green function to the prescribed heating.
The depth coordinates below are global. Let $\mathbf M_j(L,z)$ denote the
ordered transfer matrix carrying the temperature--flux state downward
from $z$ to $L$, with entries $\tilde A_j(L,z)$, $\tilde B_j(L,z)$,
$\tilde C_j(L,z)$, and $\tilde D_j(L,z)$. Each layer and contact matrix
has unit determinant. Writing the homogeneous rear condition as
$\tilde Q_{z,j}^{\mathrm R}(L)=Y_L\tilde\Theta_j^{\mathrm R}(L)$, the
unit-normalized rear mode is therefore obtained by inverse propagation:
\begin{equation}
    \mathbf s_j^{\mathrm R}(z)
    =\mathbf M_j(L,z)^{-1}
      \begin{bmatrix}1\\Y_L\end{bmatrix}
    =\begin{bmatrix}
      \tilde D_j(L,z)-Y_L\tilde B_j(L,z)\\
      -\tilde C_j(L,z)+Y_L\tilde A_j(L,z)
     \end{bmatrix}.
    \label{eq:supp_rear_mode_upward}
\end{equation}
The internal Green function is continuous at \(z=z'\), and its depth flux has a
unit jump,
\begin{equation}
    \tilde{G}_{\mathrm{in},j}(z'^-,z')
    =\tilde{G}_{\mathrm{in},j}(z'^+,z'),
    \qquad
    \tilde{Q}_z(z'^+)-\tilde{Q}_z(z'^-)=1 .
    \label{eq:supp_green_jump}
\end{equation}
Writing the Green function as a multiple of the top mode above the source and
the rear mode below the source, these two conditions give
\begin{equation}
    \tilde{G}_{\mathrm{in},j}(u,v;z,z')
    =
    \frac{
    \tilde{\Theta}_j^{\mathrm{T}}(z_<)
    \tilde{\Theta}_j^{\mathrm{R}}(z_>)}
    {
    \tilde{\Theta}_j^{\mathrm{T}}(z')
    \tilde{Q}_{z,j}^{\mathrm{R}}(z')
    -
    \tilde{Q}_{z,j}^{\mathrm{T}}(z')
    \tilde{\Theta}_j^{\mathrm{R}}(z')},
    \qquad
    z_< = \min(z,z'),\quad z_> = \max(z,z') .
    \label{eq:supp_depth_green_formula}
\end{equation}
The denominator is the Wronskian written in the temperature--flux state
variables. The resulting Green function is independent of the arbitrary
normalization of the top and rear homogeneous modes.

For a prescribed surface heat flux \(Q_z(0)=\tilde{Q}_s\), only the rear
state is required. Scaling it so that its top flux equals \(\tilde{Q}_s\)
gives
\begin{equation}
    \tilde{\Theta}_j(u,v;z)=
    \tilde{Q}_s
    \frac{\tilde{\Theta}_j^{\mathrm{R}}(z)}
    {\tilde{Q}_{z,j}^{\mathrm{R}}(0)}
    \equiv
    \tilde{Q}_s\tilde{G}_{s,j}(u,v;z).
    \label{eq:supp_surface_green}
\end{equation}

An adiabatic rear is an independent choice: its zero-flux condition
$\tilde Q_{z,N}(h_N)=0$ applied to the second row of the total transfer
matrix gives $\tilde C\tilde\Theta_1^{\mathrm{top}}+
\tilde D\tilde Q_{z,1}^{\mathrm{top}}=0$. Thus the surface response is
$\tilde G_s(0)=-\tilde D/\tilde C$, equivalently $Y_L=0$, without using
the decaying-mode assumption.

For a finite sample, a laterally uniform linear rear condition can instead
be written as $\tilde Q_z(L)-Y_L\tilde\Theta(L)=\tilde g_L$.
The Green functions use the homogeneous rear mode with $\tilde g_L=0$;
$Y_L=0$ gives a prescribed-flux condition and $Y_L=h_{\mathrm{rear}}$
gives a convective condition. Nonzero prescribed rear forcing contributes
an additional rear-driven reference field and its interface-normal flux.
The correction obeys the homogeneous rear condition, which allows nonzero
rear heat flux for a convective boundary. With these Green functions and
reference fields, the GBIE retains its form on the physical coupling
support $\Gamma=[0,L]$.

\subsection{Reference Field and Direct Interface Flux}
\label{sec:supp_reference_field}

We separate laser heating from vertical-interface coupling by writing the
temperature as a reference field plus a correction field. Each stack is
extended formally across all \(y\), and its reference field satisfies the
surface, horizontal-interface, and rear conditions under its absorbed
portion of the Gaussian source (Section~\ref{sec:supp_source_split}).
The correction supplies the excess interfacial flux and enforces the
vertical temperature-jump and flux-continuity conditions. The reference
field is
\begin{equation}
    \hat{\Theta}_{\mathrm{ref},j}(u,y,z)
    =
    2P_0
    \exp\left(-\frac{\pi^2r^2u^2}{2}\right)
    \int_0^\infty
    \tilde{G}_{s,j}(u,v;z)
    \operatorname{Re}\left[e^{i2\pi vy}F_j(v;d)\right]\,dv .
    \label{eq:supp_reference_field}
\end{equation}
The factor of two follows from combining positive and negative \(v\) using the
symmetry of the real-space Gaussian. Since the interfacial heat flux is defined
positive from stack \(b\) into stack \(a\), i.e., in the negative
\(y\)-direction,
the interface-normal flux already present in the reference field is
\begin{equation}
    \hat{Q}_{y,\mathrm{ref},j}(u,z)
    =
    k_{y,l(z)j}
    \left.
    \frac{\partial \hat{\Theta}_{\mathrm{ref},j}}{\partial y}
    \right|_{y=0}.
    \label{eq:supp_direct_flux}
\end{equation}
This direct flux is not the final interfacial flux; it is the flux associated
with the fictitious extended-medium reference solution.

\subsection{Boundary-Integral Equation}
\label{sec:supp_bie_derivation}

The vertical interface requires
\begin{equation}
    \hat{\Theta}_b(u,0,z)-\hat{\Theta}_a(u,0,z)
    =
    R_v(z)\hat{Q}_y(u,z),
    \label{eq:supp_vertical_jump}
\end{equation}
where $\hat{Q}_y(u,z)$ is the total interfacial flux.
The resistance profile is prescribed as $R_v(z)=1/G_v(z)$ on the
physically resistive segment $\Gamma_{R_v}=[z_{\min},z_{\max}]$ and
$R_v(z)=0$ on perfectly connected portions of the coupling plane outside
this segment. These connected regions have continuous temperature but
can carry nonzero interface flux.
The even-\(y\) kernel generated by an interface-normal flux distribution is
\begin{equation}
    \hat{K}_j(u,y;z,z')
    =
    4\int_0^\infty
    \tilde{G}_{\mathrm{in},j}(u,v;z,z')
    \cos(2\pi v|y|)\,dv .
    \label{eq:supp_interface_kernel}
\end{equation}
The correction field must supply only the excess flux beyond the reference
field. Thus
\begin{equation}
    \hat{\Theta}_a(u,y,z)
    =
    \hat{\Theta}_{\mathrm{ref},a}(u,y,z)
    +
    \int_{\Gamma}\hat{K}_a(u,|y|;z,z')
    [\hat{Q}_y(u,z')-\hat{Q}_{y,\mathrm{ref},a}(u,z')]\,dz',
    \qquad y<0,
    \label{eq:supp_total_field_1}
\end{equation}
and
\begin{equation}
    \hat{\Theta}_b(u,y,z)
    =
    \hat{\Theta}_{\mathrm{ref},b}(u,y,z)
    -
    \int_{\Gamma}\hat{K}_b(u,|y|;z,z')
    [\hat{Q}_y(u,z')-\hat{Q}_{y,\mathrm{ref},b}(u,z')]\,dz',
    \qquad y>0.
    \label{eq:supp_total_field_2}
\end{equation}
Substituting Eqs.~\eqref{eq:supp_total_field_1} and
\eqref{eq:supp_total_field_2} into Eq.~\eqref{eq:supp_vertical_jump} gives
\begin{equation}
    \begin{aligned}
    &\int_{\Gamma}
    [\hat{K}_a(u,0;z,z')+\hat{K}_b(u,0;z,z')]
    \hat{Q}_y(u,z')\,dz'
    +R_v(z)\hat{Q}_y(u,z)\\
    &\qquad =
    \Delta\hat{\Theta}_{\mathrm{ref}}(u,z)
    +
    \int_{\Gamma}
    [\hat{K}_b(u,0;z,z')\hat{Q}_{y,\mathrm{ref},b}(u,z')
    +\hat{K}_a(u,0;z,z')\hat{Q}_{y,\mathrm{ref},a}(u,z')]\,dz',
    \end{aligned}
    \label{eq:supp_bie}
\end{equation}
where
\begin{equation}
    \Delta\hat{\Theta}_{\mathrm{ref}}(u,z)
    =
    \hat{\Theta}_{\mathrm{ref},b}(u,0,z)
    -
    \hat{\Theta}_{\mathrm{ref},a}(u,0,z).
    \label{eq:supp_direct_jump}
\end{equation}
Equation~\eqref{eq:supp_bie} is a linear Fredholm equation for the total
interfacial heat flux \(\hat{Q}_y(u,z)\).

\subsection{Collocation and Surface Reconstruction}
\label{sec:supp_discretization}

Let $z_m$ and $\psi_m^{(z)}$ denote quadrature nodes and weights on the computational
coupling plane $\Gamma=[0,z_{\mathrm{int,max}}]$. The prescribed resistance
profile $R_v(z)$ is evaluated on this support.
On perfectly connected portions of $\Gamma$ outside $\Gamma_{R_v}$, $R_v=0$ enforces continuity in a continuous film or
common substrate, while the lateral flux remains an unknown. No zero-flux
endpoint condition is prescribed. At a termination on finite horizontal
contacts, the upper-side temperature traces need not be equal; the common
support below obeys temperature continuity. The numerical truncation depth
must be assessed separately from the physical contact depth. Collocation of
Eq.~\eqref{eq:supp_bie} at the same nodes gives
\begin{equation}
    [(\hat{\mathbf{K}}_a+\hat{\mathbf{K}}_b)\boldsymbol{\Psi}_z+\mathbf{R}_v]
    \hat{\mathbf{Q}}_y
    =
    \hat{\mathbf{b}},
    \label{eq:supp_bie_matrix}
\end{equation}
with
\begin{equation}
    (\hat{\mathbf{K}}_j)_{mn}=\hat{K}_j(u,0;z_m,z_n),
    \qquad
    \boldsymbol{\Psi}_z=\operatorname{diag}(\psi_n^{(z)}),
    \qquad
    \mathbf{R}_v=\operatorname{diag}[R_v(z_m)].
    \label{eq:supp_matrix_entries}
\end{equation}
The right-hand side is
\begin{equation}
    \hat{b}_m
    =
    \Delta\hat{\Theta}_{\mathrm{ref}}(u,z_m)
    +
    \sum_n
    \left[
    \hat{K}_b(u,0;z_m,z_n)\hat{Q}_{y,\mathrm{ref},b}(u,z_n)
    +
    \hat{K}_a(u,0;z_m,z_n)\hat{Q}_{y,\mathrm{ref},a}(u,z_n)
    \right]\psi_n^{(z)} .
    \label{eq:supp_rhs_entries}
\end{equation}
After solving Eq.~\eqref{eq:supp_bie_matrix} for each \(u\), the surface
temperature on stack \(j\) is reconstructed from
Eq.~\eqref{eq:supp_total_field_1} or
Eq.~\eqref{eq:supp_total_field_2} evaluated at \(z=0\), followed by the
inverse cosine transform in \(x\):
\begin{equation}
    \Theta_j(x,y,0;\omega_0)
    =
    2\int_0^\infty
    \hat{\Theta}_j(u,y,0;\omega_0)\cos(2\pi ux)\,du .
    \label{eq:supp_surface_reconstruction}
\end{equation}
The lock-in amplitude and phase are then
\begin{equation}
    A_j=|\Theta_j|,
    \qquad
    \varphi_j=\arg(\Theta_j).
    \label{eq:supp_amp_phase}
\end{equation}

\section{Supplementary Numerical Considerations}
\label{sec:supp_numerical_considerations}

This section records the numerical choices used for the multilayer GBIE
solver.  The discussion
complements the numerical implementation in Section~2.7 of the main text by
separating the spectral cutoff choice, the composite quadrature construction,
and the stabilization steps used in the code.

The derivation in Supplementary Section~S1 uses cycle wavenumbers \(u\) and
\(v\), with Fourier factors \(e^{-i2\pi(ux+vy)}\). The implementation uses the
corresponding angular wavenumbers
\begin{equation}
    \xi=2\pi u,\qquad \eta=2\pi v .
    \label{eq:supp_angular_wavenumber_mapping}
\end{equation}
Accordingly, the dimensionless source multipliers reported as \(U/V\) set
\(\xi_{\max}r=U\) and \(\eta_{\max}r=V\). This mapping removes an otherwise
implicit factor of \(2\pi\) between the theory and code notation.

\subsection{Composite Spectral and Depth Grids}
\label{sec:supp_composite_grids}

For each non-negative \(x\)-wavenumber \(\xi\), the solver evaluates the
side-resolved reference fields and the GBIE kernel by quadrature over the
non-negative transverse wavenumber \(\eta\) and the interface depth \(z\). The
three integrations use composite Gauss--Legendre rules.  The \(z\)-grid is
split at layer interfaces, at the limits of any finite-depth vertical
interface, and at thermal diffusion-length break points.  This prevents a
single quadrature interval from averaging discontinuous material properties or
a discontinuous vertical resistance \(R_v(z)\).  The spectral grids are split
at the optical beam scale and the thermal-wave scale,
\begin{equation}
    q_{\max}
    =
    \max_{l,j,\alpha}
    \sqrt{\frac{\omega_0 C_{v,lj}}{k_{\alpha,lj}}},
    \qquad
    \alpha\in\{x,y,z\},
    \label{eq:supp_numerics_qmax}
\end{equation}
so that the low-wavenumber thermal response and the high-wavenumber source
tail are both resolved.

For the reported metadata-driven calculations, the physical inputs are
evaluated first and the workflow then passes the spectral endpoints explicitly:
\begin{equation}
    \xi_{\max}
    =
    \max\left(\frac{U}{r},\,\alpha_u q_x\right),
    \qquad
    \eta_{\max}
    =
    \max\left(\frac{V}{r},\,\alpha_v q_y\right),
    \label{eq:supp_numerics_cutoffs}
\end{equation}
where \(q_x\) and \(q_y\) are the largest directional thermal wavenumbers in
the supplied stacks.  The reported workflow uses
\(\alpha_u=4\) and \(\alpha_v=6\).  The source multipliers \(U\) and \(V\)
are chosen from the \(rq_{\max}\)-based rule and convergence audit documented
in Supplementary Section~S4.

\subsection{Gear Settings Used for the COMSOL--GBIE Comparisons}
\label{sec:supp_gear_settings}

For the COMSOL--GBIE comparison figures, the practical numerical control is
specified by a small ``gear'' table rather than by varying every quadrature
order independently.  The standard production setting is Gear~2,
\begin{equation}
    (N_u,N_v,N_z)=(35,120,25),
    \label{eq:supp_gear2_standard}
\end{equation}
with \(q_u=4q_{\max}\), \(q_v=6q_{\max}\), and a depth target of
\(5\mu_z\).  Gear~2 fixes the node counts but does not hard-code the spectral
endpoints: \(U/V\) is assigned after evaluating the supplied material,
frequency, and beam-radius inputs.  For homogeneous cases the category rule in
Supplementary Section~S4 supplies the initial pair.  For multilayer or
anisotropic cases, that pair is checked against a more resolved GBIE solution
because \(rq_{\max}\) alone does not encode every layer-thickness and
anisotropy scale.  In the expanded five-execution COMSOL audit,
\(U/V=10/20\) gives the lowest execution-weighted mean
robust-span-normalized phase error among
the tested \(10/10\), \(10/20\), and \(20/20\) pairs; the complete profile,
runtime, and resource comparison is reported in Supplementary Section~S7.

\subsection{Asymmetric Source Cutoffs}
\label{sec:supp_asymmetric_cutoffs}

The remaining cutoff and quadrature sweeps are diagnostic checks supporting
the gear choices above, rather than a separate recommendation to tune every
parameter independently.

The large difference between the two source cutoffs comes from the source
transform rather than from the thermal Green function.  In the implementation
the normalized full Gaussian along \(x\) contributes the factor
\(\exp[-(r\xi)^2/8]\), which is negligible by \(\xi r \simeq 12\). In contrast,
the Gaussian along \(y\) is split at the vertical interface. With the
implementation convention
\(F(\eta)=\int g(y)\exp(-i\eta y)\,dy\), the \(y>0\) half-source is evaluated as
\begin{equation}
    F_+(\eta;d)
    =
    \frac{1}{2}
    \exp\left(-\frac{2d^2}{r^2}\right)
    \operatorname{erfcx}
    \left(
        -\frac{\sqrt{2}d}{r}
        +\frac{i \eta r}{2\sqrt{2}}
    \right),
    \label{eq:supp_numerics_half_source_stable}
\end{equation}
and
\(F_-(\eta;d)=\exp[-i\eta d-(r\eta)^2/8]-F_+(\eta;d)\). This form is algebraically
equivalent to the ordinary complementary-error-function expression, but avoids
overflow and cancellation when \(\eta_{\max}r\) is large. Because splitting a
smooth Gaussian by a step at \(y=0\) introduces a jump in the truncated source,
\(F_\pm\) has an algebraic high-\(\eta\) tail. The tail amplitude is proportional
to the source value at the interface and therefore grows as \(d/r\) decreases.

\begin{figure}[htbp]
    \centering
    \includegraphics[width=0.78\linewidth]{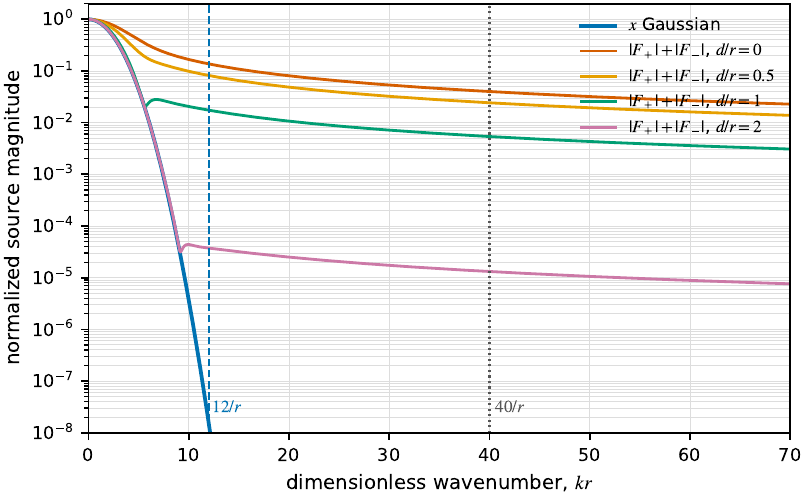}
    \caption{Normalized source-spectrum magnitudes.  The full Gaussian
    factor in the \(x\) direction is exponentially suppressed by
    \(\xi r\approx 12\), while the split half-Gaussian source in the \(y\)
    direction retains a slow algebraic tail.  The tail is strongest when the
    source center approaches the interface (\(d/r\to 0\)), motivating a
    larger transverse cutoff than the full-Gaussian \(x\) cutoff.}
    \label{fig:supp_source_spectral_decay}
\end{figure}

\begin{figure}[htbp]
    \centering
    \includegraphics[width=0.95\linewidth]{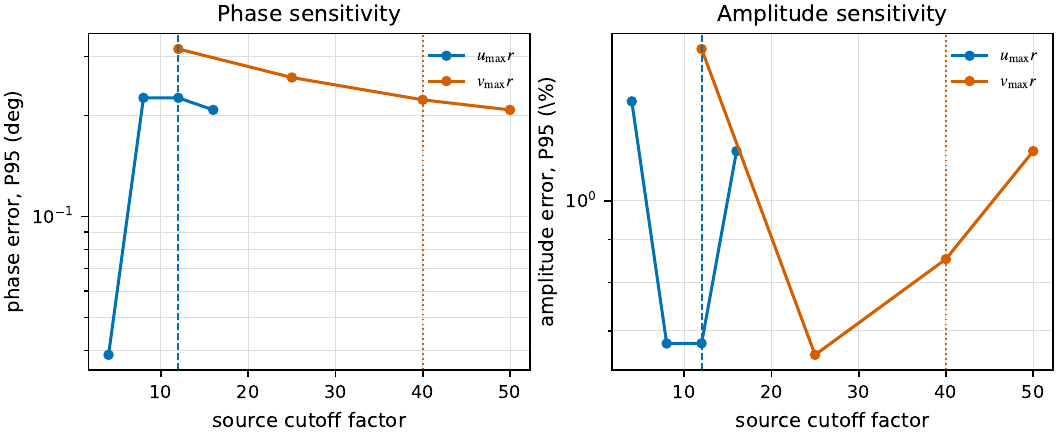}
    \caption{Full-field sensitivity to source cutoff factors for the
    representative anisotropic multilayer case.  The plotted values are P95
    differences from a refined GBIE reference.  The \(v\)-direction cutoff is
    the more restrictive control because the half-Gaussian source has an
    algebraic tail after splitting at \(y=0\).}
    \label{fig:supp_cutoff_sensitivity}
\end{figure}

Figure~\ref{fig:supp_cutoff_sensitivity} shows the full-field effect of the
cutoffs for the anisotropic multilayer case with \(G_v=G_h=10^8~
\mathrm{W\,m^{-2}\,K^{-1}}\), \(d/r=2\), and observation points
\(-2\le y/r\le 4\).  Errors are reported relative to a refined GBIE reference
with \((N_u,N_v,N_z)=(56,180,56)\), \(\xi_{\max}r=16\), and
\(\eta_{\max}r=50\). This benchmark has a phase span of approximately
\(11.3^\circ\).  The full-field sensitivity is modest because the source is two
beam radii from the interface, but under-resolving the transverse cutoff still
produces the largest cutoff error. Increasing \(\eta_{\max}r\) from 12 to 40
reduces the P95 phase difference from \(0.316^\circ\) to \(0.223^\circ\) and
keeps the P95 amplitude difference below \(1\%\).  The default
\(\xi_{\max}r=12\) is conservative for the Gaussian \(x\)-integral; further
increases mostly expose small composite-quadrature differences rather than a
systematic truncation error.

\subsection{Cost of Reconstructing Two-Dimensional Maps}
\label{sec:supp_map_reconstruction_cost}

The GBIE solve is performed independently for each \(u\)-node, but it is not
repeated for each observation point in \(x\).  For a fixed set of \(y\) points,
the expensive operations are the \(v,z\) Green-function assembly, solution of
the depth-discretized interface equation, and evaluation of the one-dimensional
surface response \(\hat{\Theta}^u(y,0)\).  Once this quantity is available, a
two-dimensional map is reconstructed by the outer product
\begin{equation}
    \Theta(x_m,y_n,0)
    \leftarrow
    \Theta(x_m,y_n,0)
    +
    \psi_\xi\,\cos[\xi(x_m-d_x)]\,\hat{\Theta}^{\xi}(y_n,0),
    \label{eq:supp_map_outer_product}
\end{equation}
which is a dense multiplication over the requested \(x\)- and \(y\)-points.  A
map with \(m\) points in \(x\) and \(n\) points in \(y\) therefore has the same
kernel-build cost as one \(y\)-line with the same \(n\), plus only an
\(\mathcal{O}(N_u m n)\) reconstruction cost and storage for the output map.

This was tested with the same representative anisotropic multilayer case and
working settings \((N_u,N_v,N_z)=(48,160,48)\), \(\xi_{\max}r=12\), and
\(\eta_{\max}r=40\). The \(y\)-grid was fixed at \(n=121\), and \(m\) was varied
from a single \(x=0\) line to a \(1001\times121\) map.  The center row of each
map matched the separately computed \(x=0\) line to machine precision.  The
median line runtime was \(10.09~\mathrm{s}\); increasing to \(m=101\), 401, and
1001 raised the runtime by factors of only 1.010, 1.013, and 1.020,
respectively.  Thus, for the map sizes used here, the cost of requesting the
two-dimensional map is negligible compared with the GBIE kernel solve.  The
dominant cost still scales with \(N_u\), \(N_v\), \(N_z\), and the number of
distinct \(y\)-points; extremely large maps would eventually become limited by
the final outer product and memory traffic.

\begin{figure}[htbp]
    \centering
    \includegraphics[width=0.86\linewidth]{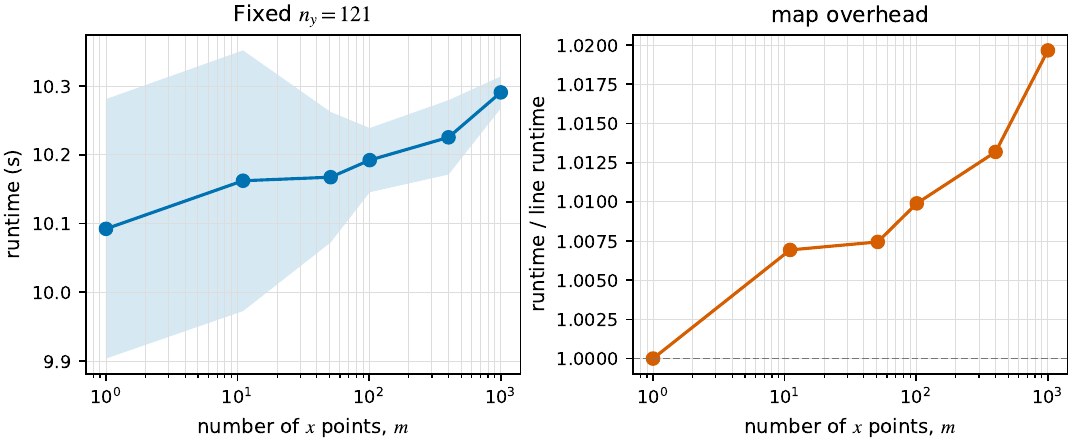}
    \caption{Measured cost of requesting a two-dimensional GBIE surface map
    instead of one \(y\)-line.  The \(y\)-grid is fixed at \(n=121\), while the
    number of \(x\)-points is varied.  The shaded band shows the min--max range
    from two repeats.  Even the \(1001\times121\) map is only about \(2\%\)
    slower than the single-line calculation because the same GBIE kernel and
    interfacial solution serve all map coordinates.}
    \label{fig:supp_map_cost_benchmark}
\end{figure}

\begin{table}[htbp]
    \centering
    \caption{Representative errors from a refined diagnostic study. These
    \((48,160,48)\) settings are convergence checks rather than the Gear~2
    production default. Cutoff rows are compared with the high-cutoff reference
    \((\xi_{\max}r,\eta_{\max}r)=(16,50)\). The quadrature row is compared with a
    same-cutoff refined reference \((N_u,N_v,N_z)=(56,180,56)\).  Timings are
    single-run wall times for the local Python implementation and are intended
    only as relative costs.}
    \label{tab:supp_recommended_numerics}
    \begin{tabular}{lcccc}
        \toprule
        Control checked & Working value & P95 phase error & P95 amplitude error & Runtime \\
        \midrule
        \(\xi_{\max}r\) & 12 & \(0.226^\circ\) & \(0.677\%\) & \(10.1~\mathrm{s}\) \\
        \(\eta_{\max}r\) & 40 & \(0.223^\circ\) & \(0.853\%\) & \(10.0~\mathrm{s}\) \\
        \((N_u,N_v,N_z)\) & \((48,160,48)\) & \(0.225^\circ\) & \(0.672\%\) & \(10.2~\mathrm{s}\) \\
        \bottomrule
    \end{tabular}
\end{table}

\subsection{Stabilization of the Depth Green Function}
\label{sec:supp_depth_stabilization}

The depth Green function is built from top and rear homogeneous modes in
each multilayer stack.  Directly propagating the rear mode upward with
hyperbolic transfer matrices can mix the decaying and growing solutions when
\(\operatorname{Re}(\lambda z)\) is large.  The implementation avoids this by
propagating the local admittance \(Y=Q_z/\Theta\) upward from the terminal substrate
condition.  For a layer segment of thickness \(h\), characteristic admittance
\(Y_c=k_z\lambda\), and bottom admittance \(Y_b\), the stable update is
\begin{equation}
    Y_t
    =
    \frac{Y_b+Y_c\tanh(\lambda h)}
    {1+(Y_b/Y_c)\tanh(\lambda h)} .
    \label{eq:supp_admittance_update}
\end{equation}
At a horizontal thermal resistance \(R_h=1/G_h\), the corresponding update is
\begin{equation}
    Y_{\mathrm{above}}
    =
    \frac{Y_{\mathrm{below}}}
    {1+R_hY_{\mathrm{below}}}.
    \label{eq:supp_interface_admittance_update}
\end{equation}
This recursion applies the semi-infinite decay condition without explicitly
carrying the exponentially growing solution \(e^{+\lambda z}\).

The top-satisfying mode is stabilized symmetrically during downward
propagation. Each layer-transfer result is normalized to order unity and its
removed magnitude is accumulated as a logarithmic scale. In the Wronskian
Green-function formula, only differences of these top-mode scales and the
corresponding rear-mode scales remain, so arbitrary normalizations cancel
analytically. Thus neither homogeneous mode is represented by a capped raw
exponential.

For each \(\xi\), the final vertical-interface operator has the form
\begin{equation}
    \mathbf{A}_{\xi}
    =
    \left[\mathbf{K}_{a,\xi}(0)+\mathbf{K}_{b,\xi}(0)\right]\boldsymbol{\Psi}_z
    +\operatorname{diag}\!\left[R_v(z_n)\right],
    \label{eq:supp_discrete_operator_stability}
\end{equation}
where \(\boldsymbol{\Psi}_z\) contains the depth quadrature weights. The continuum
kernel has an integrable logarithmic self-singularity at \(z=z'\). In the
code, every diagonal entry is evaluated with the same finite
\(\eta_{\max}\) composite Gauss--Legendre sum as the off-diagonal entries; no
entry is deleted or interpolated. The finite spectral limit makes the
assembled diagonal finite but cutoff-dependent. The local \(R_v\) term
gives a uniformly second-kind equation only if resistance is bounded
strictly away from zero on the entire coupling support. It does not define
the diagonal quadrature rule. Large conductance and connected segments
require joint checks in $N_z$, $N_v$, and $\eta_{\max}$. The depth study in
Section~S3 holds the transverse cutoff fixed and does not establish this
joint limit.

\section{Finite-Cutoff Kernel Diagonal and Depth-Quadrature Convergence}
\label{sec:supp_nz_diagonal_convergence}

This section clarifies two implementation details that are coupled in the
discrete boundary-integral equation: the finite-spectral-cutoff treatment of
the kernel diagonal and the distribution of the \(N_z\) depth nodes.  It also
tests the default \(N_z=25\) against denser choices for five anisotropic
multilayer COMSOL executions representing four unique profiles.

\subsection{Depth-node distribution}
\label{sec:supp_nz_distribution}

\(N_z\) is the total requested number of quadrature and collocation nodes for
the unknown vertical-interface flux \(\hat Q_y(\xi,z)\).  It is not a
three-dimensional finite-element mesh, nor is it the number of nodes assigned
to every layer.  Temperature propagation through each horizontal layer is
still evaluated analytically by the transfer matrices.  The depth nodes are
used only to approximate the remaining interface integral in
\cref{eq:supp_discrete_operator_stability}.

The interface integral begins at the surface, \(z=0\), and ends at an
automatically selected \(z_{\mathrm{int,max}}\).  The current metadata-driven
workflow forms a field depth from the beam radius \(r\) and the largest
cross-plane penetration depth,
\begin{equation}
  \mu_{z,\ell}
  =
  \sqrt{\frac{2k_{z,\ell}}{\omega C_{v,\ell}}},
  \qquad
  z_{\mathrm{field}}
  =
  \max\!\left(6r,\,5\max_{\ell}\mu_{z,\ell}\right),
  \label{eq:supp_nz_depth_scale}
\end{equation}
and places \(z_{\mathrm{int,max}}\) at or beyond this depth while respecting
the finite layer stack.  For the five \(10~\si{\kilo\hertz}\) executions,
\(z_{\mathrm{int,max}}\) ranges from
\(141.05\) to \(295.86~\si{\micro\metre}\).

The nodes are not uniform over this long interval.  Every physical layer
boundary is inserted as an exact interval boundary.  Additional breakpoints
are placed at
\begin{equation}
  r/20,\ r/10,\ r/5,\ r/2,\ r,\ 2r,\ 5r
    \label{eq:supp_beam_scale_depth_breakpoints}
\end{equation}
and at selected multiples of the smallest and largest \(\mu_{z,\ell}\).
In addition, every finite layer with \(h_\ell\le10r\) receives layer-relative
breaks at
\begin{equation}
  z_{\ell,\mathrm{top}}+
  \{h_\ell/20,\ h_\ell/10,\ h_\ell/5,\ h_\ell/2\},
    \label{eq:supp_layer_relative_depth_breakpoints}
\end{equation}
followed by its exact bottom boundary.  This makes ultrathin-layer resolution
independent of whether \(h_\ell\) happens to equal the beam radius.
An independent Gauss--Legendre rule is then applied in every resulting
subinterval.  All five present executions have 20 subintervals.  The allocator
first attempts to assign at least four nodes per interval.  If the requested
\(N_z\) is too small for that assignment, it places one node in every interval
and distributes the remaining nodes in proportion to interval length.
Consequently, \(N_z=25\) gives one node in each of the 20 intervals plus five
additional nodes in the longer deep intervals, whereas \(N_z=80\) gives four
nodes in every interval.

More explicitly, let \(b_s\) and \(b_{s+1}\) be two consecutive protected
breakpoints and let \(\{x_p,\psi_p^{(\mathrm{GL})}\}_{p=1}^{n_s}\) be the standard
Gauss--Legendre rule on \([-1,1]\).  The collocation nodes and integration
weights in that interval are
\begin{equation}
  z_{s,p}
  =
  \frac{b_s+b_{s+1}}{2}
  +
  \frac{b_{s+1}-b_s}{2}x_p,
  \qquad
  \psi_{s,p}^{(z)}
  =
  \frac{b_{s+1}-b_s}{2}\psi_p^{(\mathrm{GL})} .
  \label{eq:supp_mapped_z_gauss_rule}
\end{equation}
When \(n_s=1\), \(x_1=0\), so the single node is exactly the interval
midpoint and its weight is the interval width.  The requested total is first
raised to the number of protected intervals if necessary.  The allocator then
uses four nodes per interval when \(N_z\ge4N_{\mathrm{seg}}\); otherwise it
uses one per interval and apportions the remaining nodes by interval length,
with residual nodes assigned to the largest fractional allocations.  The
actual depth order reported by the solver is
\(N_z^{\mathrm{actual}}=\sum_s n_s\).

The \(1~\si{\micro\metre}\) film is explicitly protected by breakpoints at
\begin{equation}
  z=0,\ 0.05,\ 0.10,\ 0.20,\ 0.50,\ 1.00~\si{\micro\metre}.
    \label{eq:supp_one_micron_film_breakpoints}
\end{equation}
Thus \(N_z=25\) still places five film nodes, at
\(z=0.025,\ 0.075,\ 0.15,\ 0.35,\) and
\(0.75~\si{\micro\metre}\), and no quadrature interval crosses the
film/substrate boundary.  The same construction gives protected breaks at
\(0,0.05,0.10,0.20,0.50,\) and \(1.00~\si{\nano\metre}\) for a
\(1~\si{\nano\metre}\) film, so \(N_z=25\) again gives five film nodes,
at \(0.025,0.075,0.15,0.35,\) and \(0.75~\si{\nano\metre}\).
The same layer-relative construction is used throughout the corrected
thin-film COMSOL comparison series. At \(N_z=25\), the
\(500\), \(200\), and \(100~\si{\nano\metre}\) films contain six, six,
and five in-film nodes, respectively; the two thicker films receive one
additional interval from a beam-scale breakpoint. Thus the interface segment within every nonzero film
examined here, from \(1~\si{\nano\metre}\) to
\(1~\si{\micro\metre}\), contains at least five protected depth intervals.
Increasing \(N_z\) refines these same protected intervals; it does not change
the analytical layer representation.

\subsection{Finite-cutoff evaluation of the diagonal}
\label{sec:supp_nz_diagonal}

At equal source and observation depths, the continuous inverse-transverse
kernel has a weak logarithmic self-singularity as the transverse spectral
limit tends to infinity.  The production implementation evaluates the depth
Green function directly at \(z_m=z_n\) and uses the same finite
\(\eta_{\max}\) composite Gauss--Legendre rule on and off the diagonal:
\begin{equation}
  [\mathbf K_{j,\xi}]_{mm}^{(\eta_{\max})}
  =
  \frac{2}{\pi}
  \sum_{p=1}^{N_v}
  \psi_{\eta,p}\,
  \widetilde G_{\mathrm{in},j}
  (\xi,\eta_p;z_m,z_m).
  \label{eq:supp_finite_cutoff_diagonal}
\end{equation}
No diagonal entry is deleted or interpolated, and the code does not apply a
principal value, analytic singularity subtraction, or a separate diagonal
cell average.  The finite spectral limit makes
\cref{eq:supp_finite_cutoff_diagonal} finite but cutoff-dependent.  For the
present \(r=1~\si{\micro\metre}\), \(V=20\) calculations,
\(\eta_{\max}=2.0\times10^7~\si{\per\metre}\), corresponding to the nominal
regularization length \(1/\eta_{\max}=0.05~\si{\micro\metre}\).  The local resistance term is a physical contact law, not a singular
quadrature prescription. A uniformly second-kind equation requires the
resistance to remain bounded away from zero on the entire coupling support;
that condition does not hold on perfectly connected segments. No
condition-number sweep is supplied by this depth-refinement comparison.

\subsection{Depth-order comparison with COMSOL}
\label{sec:supp_nz_comsol_comparison}

The five repaired metadata/result pairs were recalculated with
\((N_u,N_v)=(35,120)\), \(U/V=10/20\), and
\(N_z=25,80,\) and \(160\).  All calculations used one numerical thread, and
all five physical models contain the \(1~\si{\micro\metre}\) film described
above.  The phase metric is the robust-span-normalized RMSE in
\cref{eq:supp_comsol_bie_phase_nrmse}; the amplitude metric is the relative
\(L_2\) error in \cref{eq:supp_comsol_bie_amplitude_l2}.

\begin{table}[htbp]
  \centering
  \caption{Effect of \(N_z\) on agreement with COMSOL.  The two
  \(250~\si{\micro\metre}\) rows are independent executions with identical
  physical profiles.  Every phase error remains below \(1.26\%\), and every
  amplitude error remains below \(0.73\%\).}
  \label{tab:supp_nz_comsol_comparison}
  \small
  \begin{tabular}{lrrrrrr}
    \toprule
    & \multicolumn{3}{c}{Phase RMSE / robust span (\%)} &
      \multicolumn{3}{c}{Amplitude \(L_2\) error (\%)} \\
    Job & \(N_z=25\) & \(N_z=80\) & \(N_z=160\) &
          \(N_z=25\) & \(N_z=80\) & \(N_z=160\) \\
    \midrule
    1 & 0.253 & 0.796 & 0.817 & 0.454 & 0.197 & 0.206 \\
    2 & 0.253 & 0.796 & 0.817 & 0.454 & 0.197 & 0.206 \\
    3 & 0.576 & 1.238 & 1.259 & 0.538 & 0.161 & 0.166 \\
    4 & 0.594 & 1.237 & 1.248 & 0.648 & 0.153 & 0.158 \\
    5 & 0.721 & 1.202 & 1.220 & 0.728 & 0.178 & 0.180 \\
    \bottomrule
  \end{tabular}
\end{table}

\begin{figure}[htbp]
  \centering
  \includegraphics[width=\linewidth]{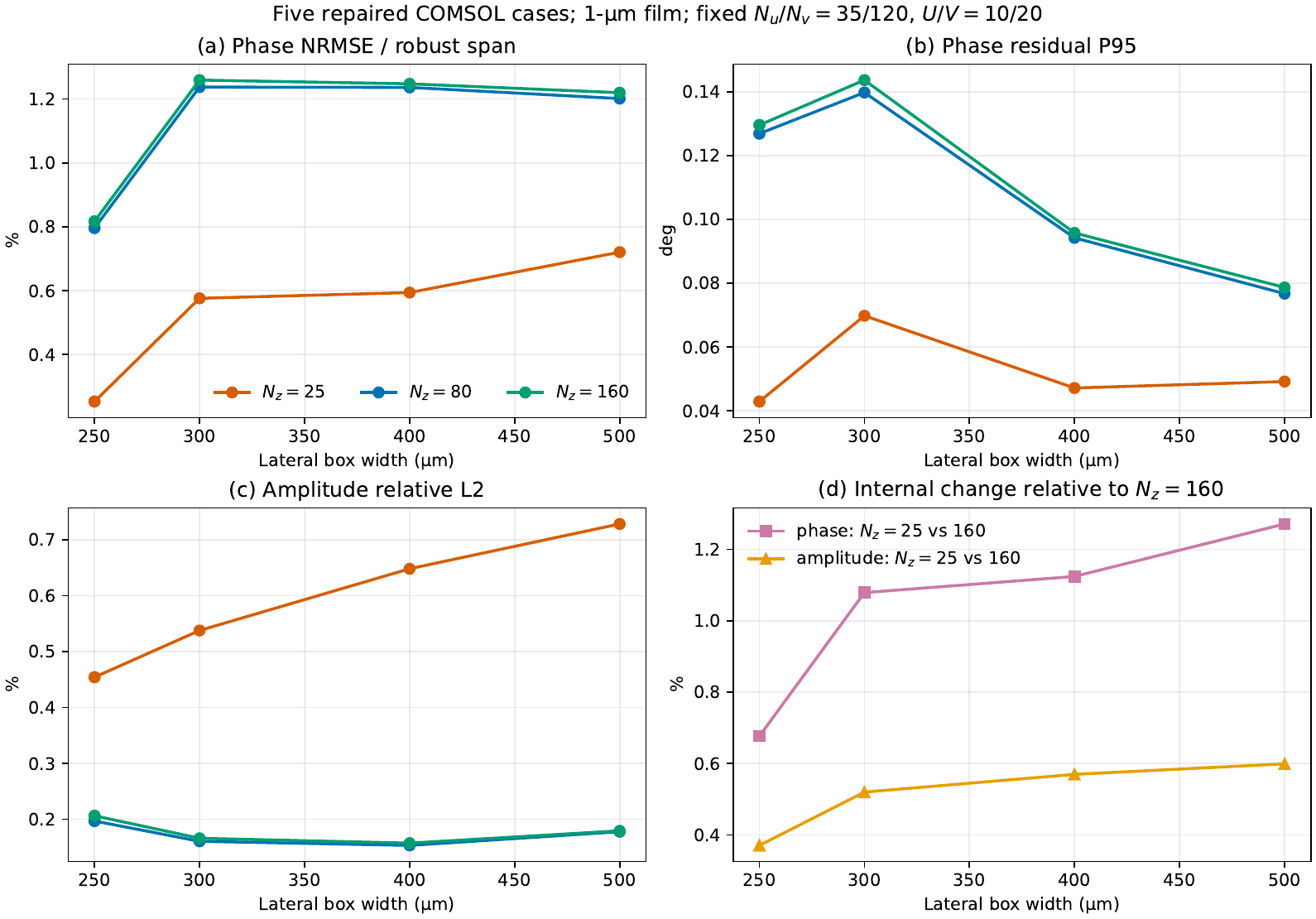}
  \caption{COMSOL agreement for all five \(1~\si{\micro\metre}\)-film executions
  as the depth quadrature is refined.  The practical accuracy classification
  is unchanged: \(N_z=25\), \(80\), and \(160\) all remain below the
  \(2\%\) phase criterion.  The amplitude residual decreases substantially on
  the dense branch.}
  \label{fig:supp_nz_all_cases_comsol}
\end{figure}

The \(N_z=80\) and \(160\) GBIE solutions are nearly identical.  Over the five
executions, their mutual phase difference is only
\(0.014\%\)--\(0.026\%\) of the COMSOL robust span, their P95 absolute phase
difference is \(0.0015^\circ\)--\(0.0044^\circ\), and their amplitude
difference is \(0.0147\%\)--\(0.0172\%\).  By contrast, \(N_z=25\) differs
from \(N_z=160\) by \(0.677\%\)--\(1.271\%\) of phase span and
\(0.370\%\)--\(0.599\%\) in amplitude.  Therefore \(N_z=25\) is not
strictly identical to the depth-converged branch.

Nevertheless, \(N_z=25\) is adequate for the stated fast standard-setting comparison
criterion: its COMSOL phase error is
\(0.253\%\)--\(0.721\%\), its P95 absolute phase residual is
\(0.043^\circ\)--\(0.070^\circ\), and its amplitude error is
\(0.454\%\)--\(0.728\%\).  All remain below the adopted \(2\%\) phase and
\(1\%\) amplitude thresholds for all five \(1~\si{\micro\metre}\)-film
executions. The smaller
coarse-grid phase score should not be interpreted as superior internal
accuracy; comparison with the stable \(N_z\ge80\) branch shows that it partly
results from cancellation with the remaining COMSOL--GBIE modeling and
discretization differences.

\begin{figure}[htbp]
  \centering
  \includegraphics[width=\linewidth]{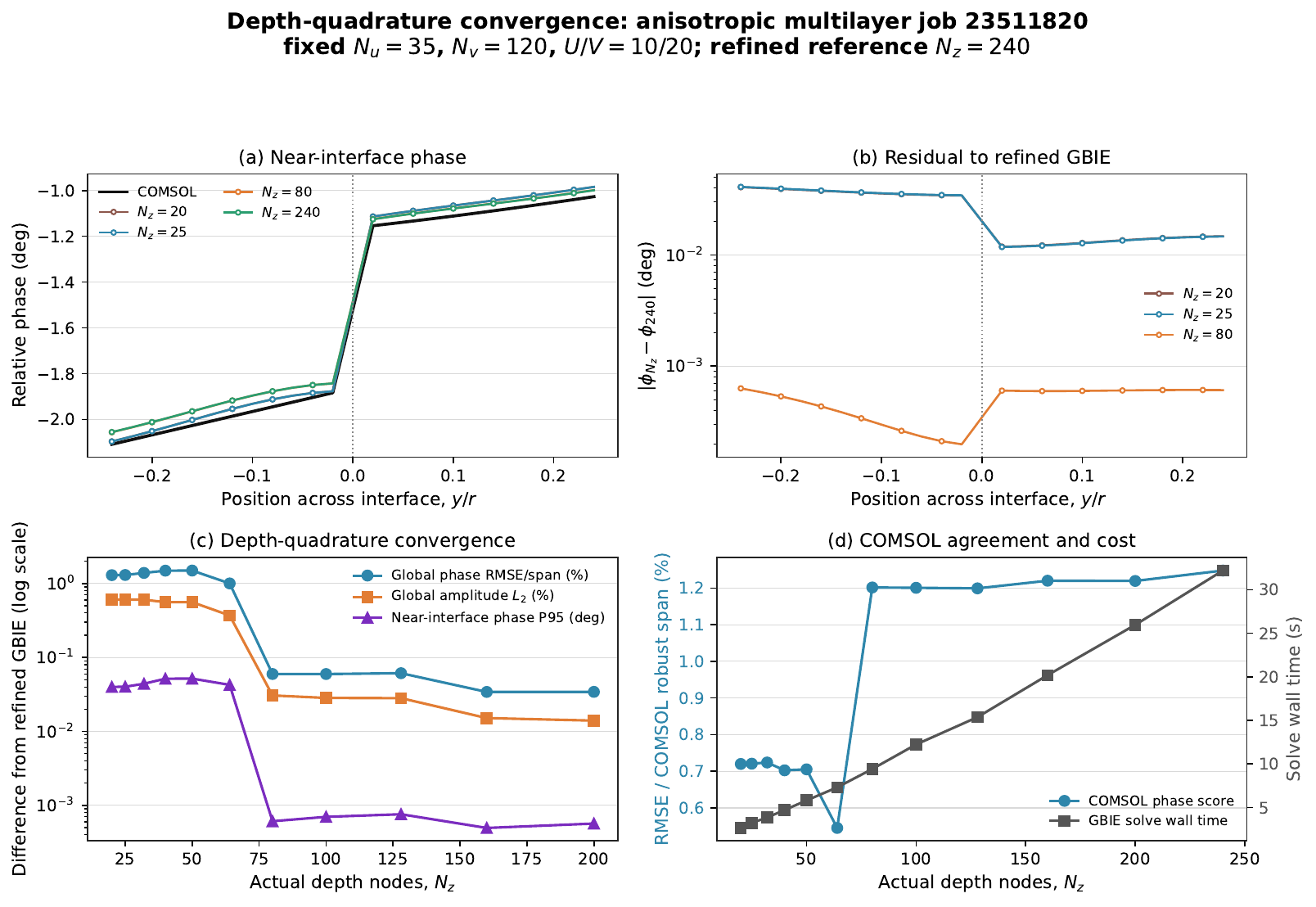}
  \caption{Detailed depth-convergence study for the highest-conductivity
  \(500~\si{\micro\metre}\) case.  \(N_z=240\) is used only as the internal
  GBIE reference.  The transition at \(N_z=80\) occurs because this is the
  first requested grid that assigns four Gauss nodes to each of the 20
  protected depth intervals.  The dense branch is stable, while all shown
  settings preserve the same sub-\(2\%\) external COMSOL phase classification.}
  \label{fig:supp_nz_detailed_convergence}
\end{figure}

Accordingly, the manuscript retains \(N_z=25\) as the fast standard default for
the reported COMSOL comparison and source-cutoff screening.  A calculation
intended to serve as an internally converged GBIE reference should use
\(N_z\ge80\) for this geometry, and a new geometry class should repeat the
coupled \(N_z\), \(N_v\), and spectral-cutoff check.

\clearpage

\section{Selection and Evaluation of the GBIE Source-Cutoff Multipliers}
\label{sec:supp_uv_source_factor}

This section documents how the dimensionless source-cutoff multipliers
\(U\) and \(V\) were selected for the Green's-function boundary integral equation (GBIE)
solver.  It also records the numerical-case audit, the error measure used to
rank candidate multipliers, the resulting category rule, and the twenty
featured examples used for detailed visual inspection.  Every candidate
solution discussed here is a full GBIE solution.  The analytical Eq.~(10)
solution is used only as a numerical reference for the homogeneous
semi-infinite configurations for which it is applicable.

\subsection{Meaning of the multipliers}
\label{sec:supp_uv_definition}

Here $\xi=2\pi u$ and $\eta=2\pi v$ are the angular wavenumbers
used by the implementation. The cycle-wavenumber cutoffs in Section~S1
are $u_{\max}=\xi_{\max}/(2\pi)$ and
$v_{\max}=\eta_{\max}/(2\pi)$.
For the Gaussian \(1/e^2\) radius \(r\), the source-controlled spectral limits are
\begin{equation}
    \xi_{\mathrm{src}}=\frac{U}{r},
    \qquad
    \eta_{\mathrm{src}}=\frac{V}{r}.
    \label{eq:supp_uv_source_limits}
\end{equation}
The limits supplied to the GBIE calculation retain both the source and thermal
scales,
\begin{equation}
    \xi_{\max}=\max\!\left(\frac{U}{r},\alpha_u q_x\right),
    \qquad
    \eta_{\max}=\max\!\left(\frac{V}{r},\alpha_v q_y\right),
    \label{eq:supp_uv_combined_limits}
\end{equation}
where \(q_x\) and \(q_y\) are the largest directional thermal wavenumbers
among the materials included in the calculation.  Thus, increasing \(U\) or
\(V\) enlarges the represented source spectrum only when the source term is
the active cutoff.  It does not by itself increase the quadrature order, and a
larger cutoff can be counterproductive if the existing nodes are spread over a
wider interval.  The cutoff choice must therefore be interpreted together
with the numerical gear.

The initial rule is organized by
\begin{equation}
    rq_{\max}=r\max_{j,\alpha}
    \sqrt{\frac{\omega C_j}{k_{\alpha,j}}},
    \qquad \alpha\in\{x,y,z\},
    \label{eq:supp_uv_aqmax}
\end{equation}
using Category I for \(rq_{\max}<0.10\), Category II for
\(0.10\leq rq_{\max}<0.60\), Category III for
\(0.60\leq rq_{\max}<1.00\), and Category IV for
\(rq_{\max}\geq1.00\).

\subsection{Numerical-case audit and study design}
\label{sec:supp_uv_study_design}

The archive contains \(2700\) saved candidate-result rows from the original
factorial sweep.  It is reasonable to describe this as roughly three thousand
numerical solver evaluations, but it is \emph{not} correct to describe it as
three thousand independent physical cases: the rows repeat cutoff and gear
evaluations for only \(108\) distinct structured physical conditions.  We
subsequently generated \(60\) genuinely random, \(rq_{\max}\)-stratified
physical conditions, with twenty cases in each of Categories I--III.  Nine
full-GBIE candidates,
\begin{equation}
    U,V\in\{5,10,20\},
    \label{eq:supp_uv_candidate_grid}
\end{equation}
were evaluated for every physical condition.  Therefore the common final
reevaluation contains \(168\) distinct physical conditions and
\(168\times9=1512\) candidate GBIE evaluations.  Counting the archived sweep
and the added random-case calculations gives more than three thousand solver
evaluations overall, while the statistically relevant physical-condition
count remains \(168\).  The twenty examples in
\cref{sec:supp_uv_featured_examples} are an illustrative subset and are not an
additional independent cohort.

The structured set spans conductivity, spot radius, modulation frequency, and
vertical conductance combinations.  The random set samples the two bulk
conductivities log-uniformly from \(10\) to
\(100~\si{\watt\per\metre\per\kelvin}\), frequency log-uniformly from
\(10^3\) to \(10^5~\si{\hertz}\), and conductance log-uniformly from
\(10^5\) to \(10^7~\si{\watt\per\square\metre\per\kelvin}\).  The spot
radius is obtained from a randomly selected target \(rq_{\max}\) and retained
when it lies between \(1\) and \(10~\si{\micro\metre}\).

\subsection{Primary error measure}
\label{sec:supp_uv_error_measure}

The earlier rule emphasized a worst-case absolute phase P95 value.  That
criterion can be controlled by a single atypical condition and does not weight
small-phase-span measurements appropriately.  The reevaluation instead uses
the mean pointwise phase-percentage error.  After unwrapping the phase and
setting the phase at the beam center to zero for both solutions, the score for
case \(i\) and candidate \((U,V)\) is
\begin{equation}
    E_i(U,V)=
    \frac{100}{N_i}
    \sum_{y/r<-0.2}
    \frac{\left|\phi_{\mathrm{GBIE}}(y;U,V)
    -\phi_{\mathrm{ref}}(y)\right|}
    {\max\!\left(\left|\phi_{\mathrm{ref}}(y)\right|,1^\circ\right)}.
    \label{eq:supp_uv_phase_percentage}
\end{equation}
The one-degree denominator floor controls the relative error near a zero
crossing without discarding low-span cases.  A category-level candidate is
ranked by the unweighted mean of \(E_i\) over the physical cases in that
category.  Bootstrap winner probabilities use \(5000\) resamples of the
physical cases, rather than resampling individual spatial points or repeated
candidate rows.

\subsection{Category-level results and adopted rule}
\label{sec:supp_uv_results}

\Cref{tab:supp_uv_category_results} summarizes the combined \(168\)-condition
result.  Category II changes to \(U/V=10/10\), reducing the category mean
phase-percentage error by \(30.2\%\) relative to the previous \(20/20\)
choice.  Category III changes to \(5/20\), reducing the mean by \(12.6\%\)
relative to \(20/20\).  For Category I, the interface-coupling split is
retained: \(20/5\) is used at weak coupling, while \(20/10\) is used when
\(Gw/k_{\min}\geq0.09\).  The bootstrap winner probabilities are \(0.558\),
\(0.964\), and \(0.994\) for the adopted Category I, II, and III rules,
respectively.

\begin{table}[htbp]
    \centering
    \caption{Combined structured-plus-random source-factor result.  The score
    is the category mean of the per-case mean phase-percentage error in
    \cref{eq:supp_uv_phase_percentage}.}
    \label{tab:supp_uv_category_results}
    \resizebox{\linewidth}{!}{%
    \begin{tabular}{lcccc}
        \toprule
        Category & Distinct cases & Adopted \(U/V\) & Mean error & Interpretation \\
        \midrule
        I & 50 & \(20/5\), or \(20/10\) at strong coupling & \(1.509\%\) & Conditional rule retained \\
        II & 41 & \(10/10\) & \(0.791\%\) & \(30.2\%\) below previous rule \\
        III & 35 & \(5/20\) & \(0.787\%\) & \(12.6\%\) below previous rule \\
        IV & 42 & \(20/20\) only as a starting point & \(96.0\%\) & Explicit convergence required \\
        \bottomrule
    \end{tabular}%
    }
\end{table}

Category IV is deliberately excluded from an unconditional fixed-factor rule.
Its mean is dominated by unresolved outliers, and changing \(U/V\) within the
Gear-2 grid does not control them.  These cases use Gear 5 and a GBIE-to-GBIE
convergence sweep.  The same safeguard is used for conductivity contrast
above ten and for multilayer or anisotropic configurations outside the
homogeneous calibration domain.

The twenty featured examples provide an additional, manually inspectable
check.  Averaged within this subset, the adopted rule gives mean errors of
\(0.670\%\), \(0.419\%\), and \(0.567\%\) in Categories I, II, and III,
respectively.  A different candidate can be locally better for an individual
case; the adopted rule minimizes the category average and is not claimed to be
the pointwise optimum for every physical condition.

\clearpage
\subsection{Twenty featured examples}
\label{sec:supp_uv_featured_examples}

In each three-case figure, the top, middle, and bottom cases use panels
(a)--(d), (e)--(h), and (i)--(l), respectively; the final two-case figure uses
panels (a)--(d) and (e)--(h).  Within each case row, the first two panels
compare the analytical reference with the current GBIE rule and the best
plotted alternative when it differs.  The last two panels compare the error
profiles for the current rule and the requested higher-cutoff alternatives
\(10/10\), \(10/20\), \(20/10\), and \(20/20\).  The percentage values in
the captions are the per-case means from \cref{eq:supp_uv_phase_percentage}.

\begin{figure}[htbp]
    \centering
    \includegraphics[width=\linewidth]{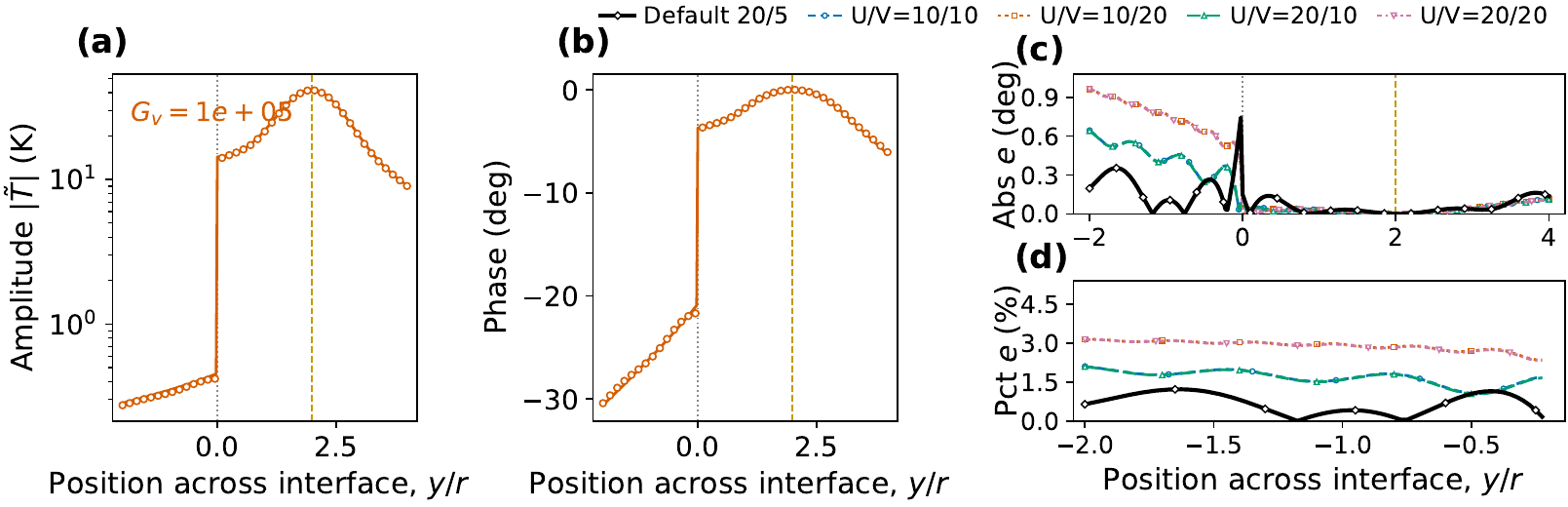}\par\vspace{-0.35em}
    \includegraphics[width=\linewidth]{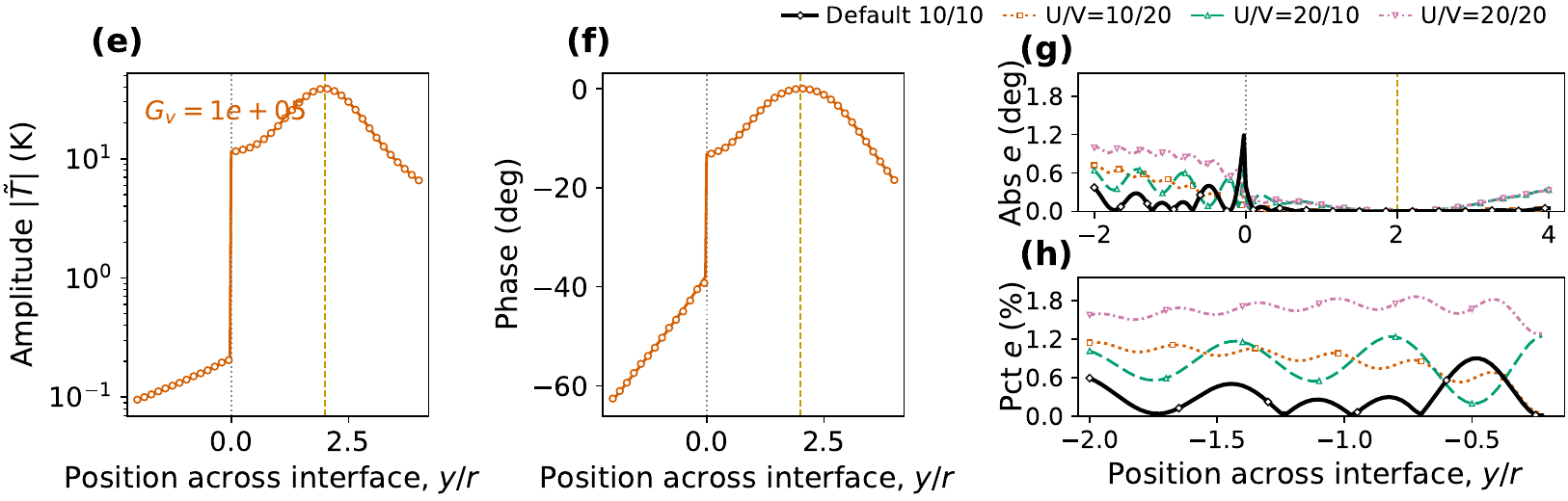}\par\vspace{-0.35em}
    \includegraphics[width=\linewidth]{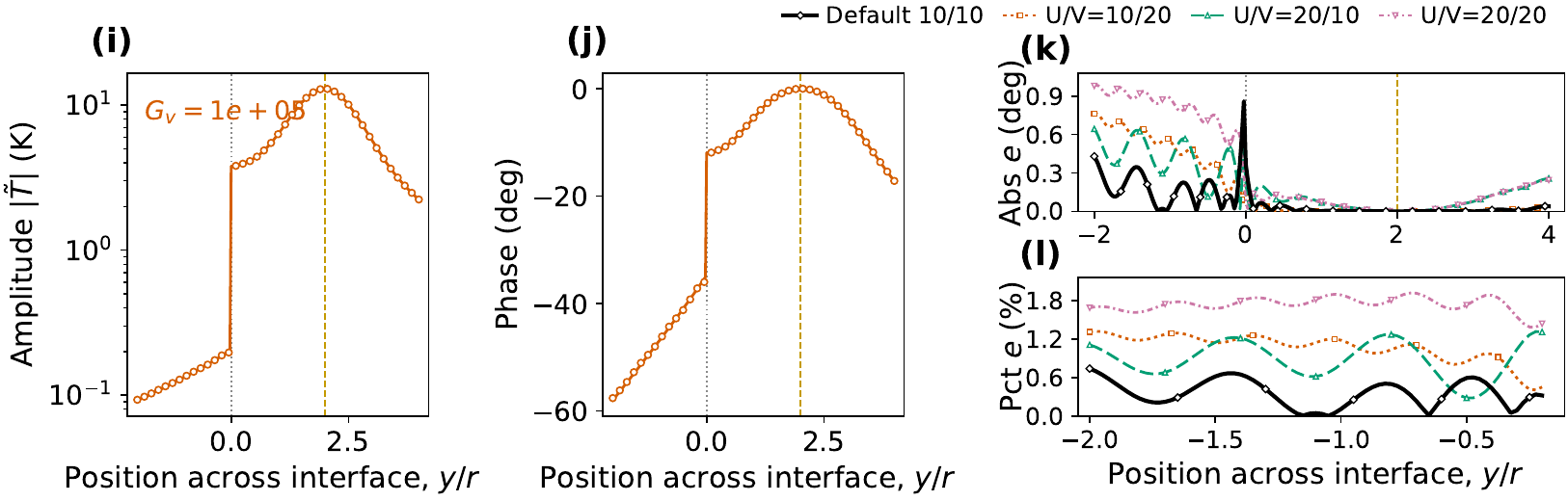}
    \caption{Featured cases 01--03, shown as panels (a)--(d), (e)--(h), and
    (i)--(l), respectively.  Case 01:
    Category I, \(rq_{\max}=0.079\), \(k_{\mathrm{un}}/k_{\mathrm{h}}=10/10\),
    \(r=1~\si{\micro\metre}\), \(f=10~\si{\kilo\hertz}\), and
    \(G=10^5~\si{\watt\per\square\metre\per\kelvin}\); \(20/5\) is best
    with \(E_i=0.683\%\).  Cases 02 and 03 are Category II with matched
    \(10/10~\si{\watt\per\metre\per\kelvin}\) materials and
    \((r,f)=(1~\si{\micro\metre},100~\si{\kilo\hertz})\) and
    \((3~\si{\micro\metre},10~\si{\kilo\hertz})\), respectively; \(10/10\)
    is best with \(E_i=0.310\%\) and \(0.354\%\).}
    \label{fig:supp_uv_cases01_03}
\end{figure}

\begin{figure}[p]
    \centering
    \includegraphics[width=\linewidth]{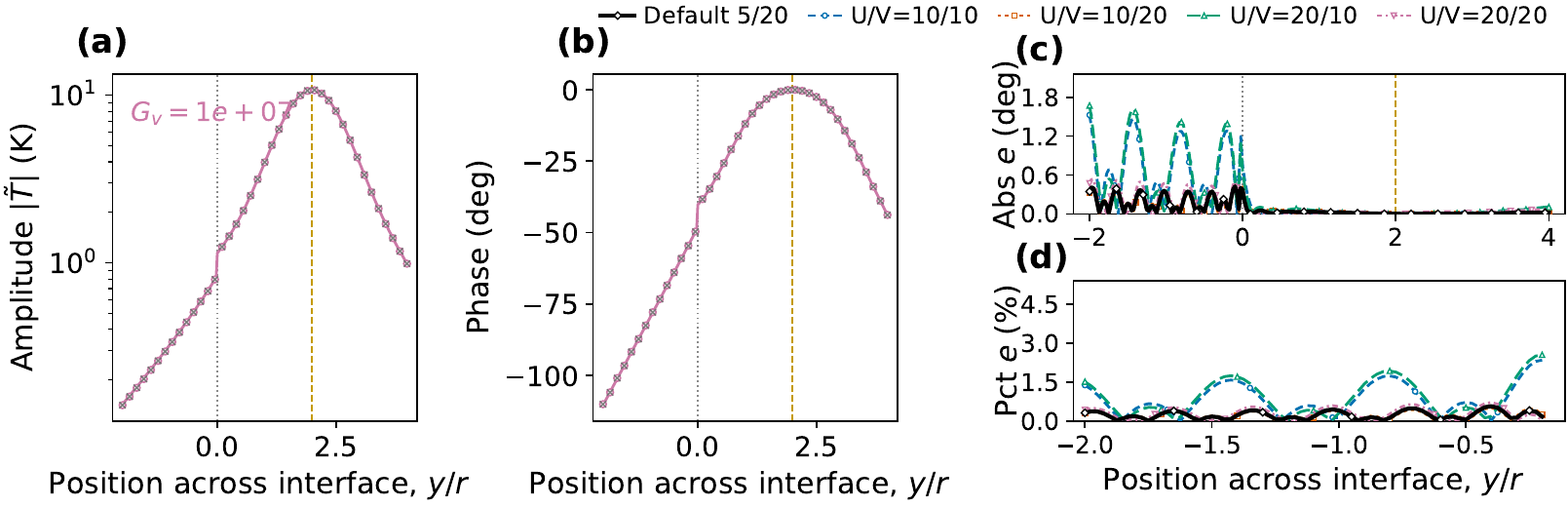}\par\vspace{-0.35em}
    \includegraphics[width=\linewidth]{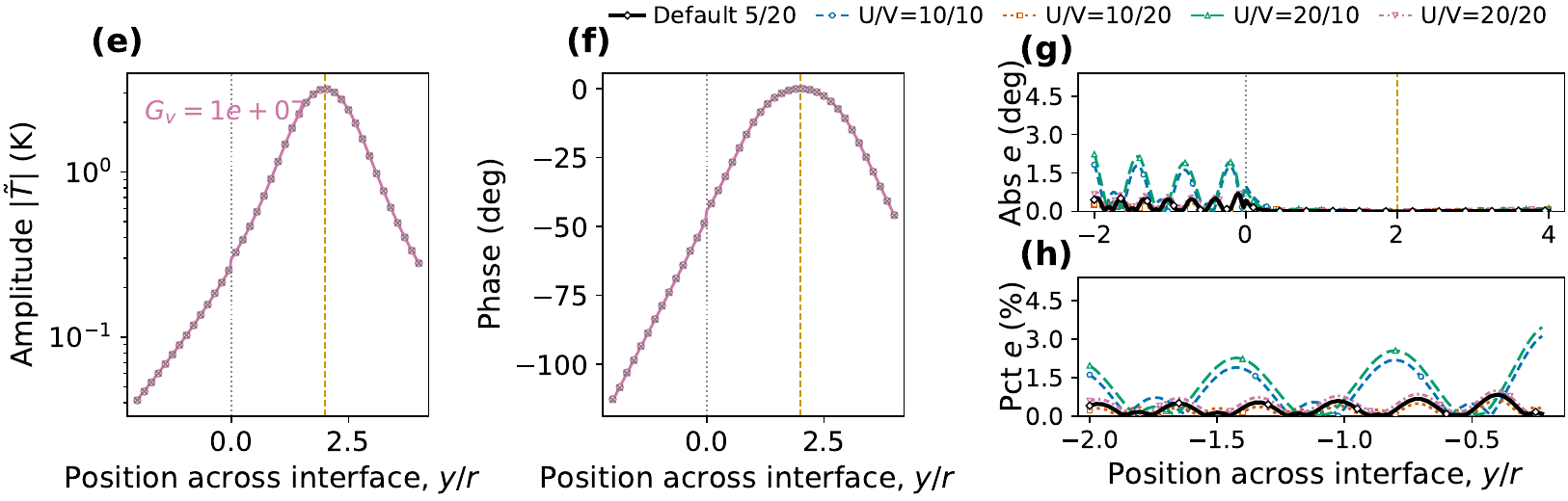}\par\vspace{-0.35em}
    \includegraphics[width=\linewidth]{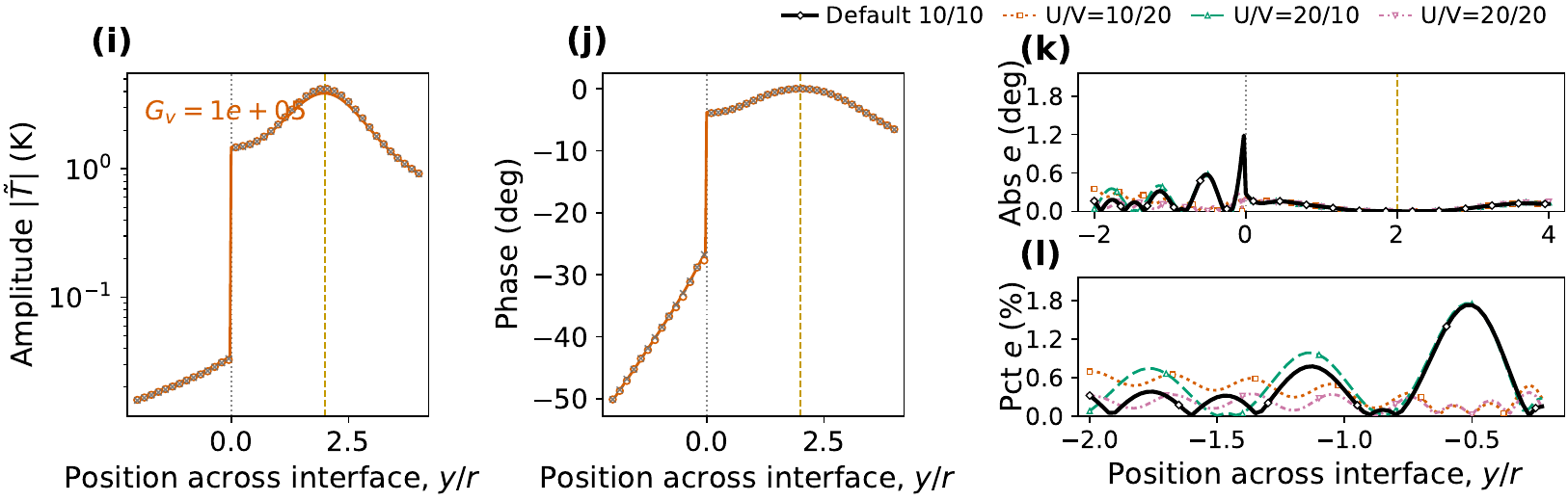}
    \caption{Featured cases 04--06, shown as panels (a)--(d), (e)--(h), and
    (i)--(l), respectively.  Cases 04 and 05:
    Category III, matched \(10/10~\si{\watt\per\metre\per\kelvin}\)
    materials, with \((rq_{\max},r,f)=(0.752,3~\si{\micro\metre},
    100~\si{\kilo\hertz})\) and \((0.793,10~\si{\micro\metre},
    10~\si{\kilo\hertz})\); \(5/20\) gives \(E_i=0.230\%\) and
    \(0.292\%\), while \(10/20\) gives \(0.222\%\) and \(0.240\%\).
    Case 06: Category II, \(rq_{\max}=0.251\),
    \(k_{\mathrm{un}}/k_{\mathrm{h}}=10/100\),
    \(r=1~\si{\micro\metre}\), and \(f=100~\si{\kilo\hertz}\);
    \(10/10\) gives \(0.484\%\), while \(20/20\) gives \(0.199\%\).}
    \label{fig:supp_uv_cases04_06}
\end{figure}

\begin{figure}[p]
    \centering
    \includegraphics[width=\linewidth]{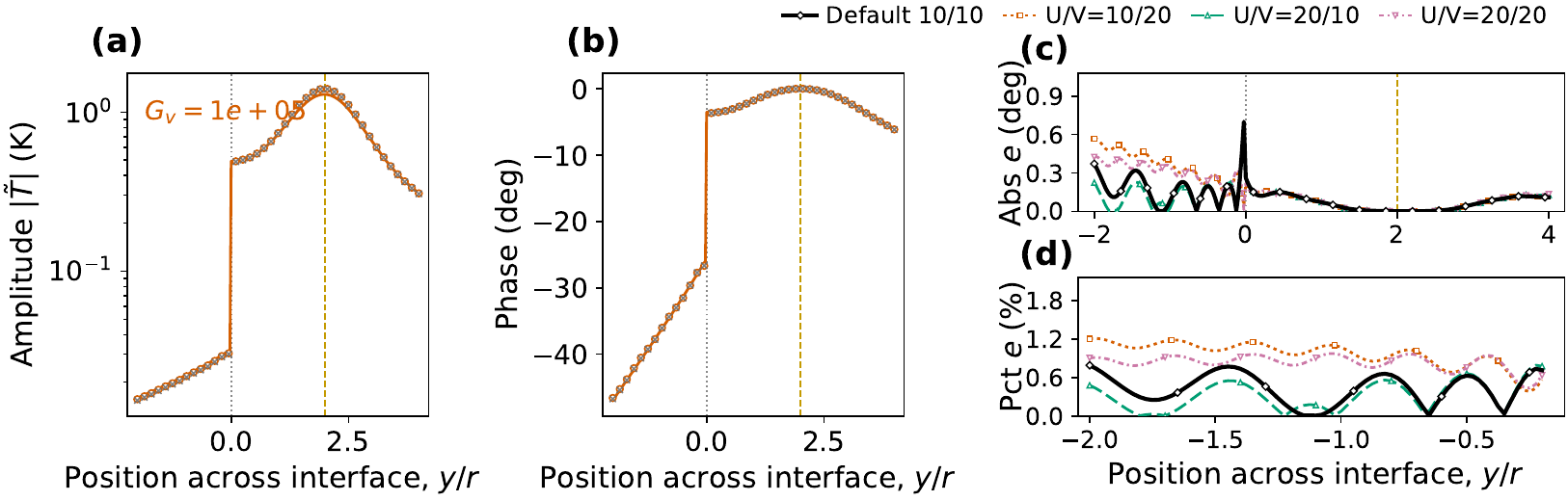}\par\vspace{-0.35em}
    \includegraphics[width=\linewidth]{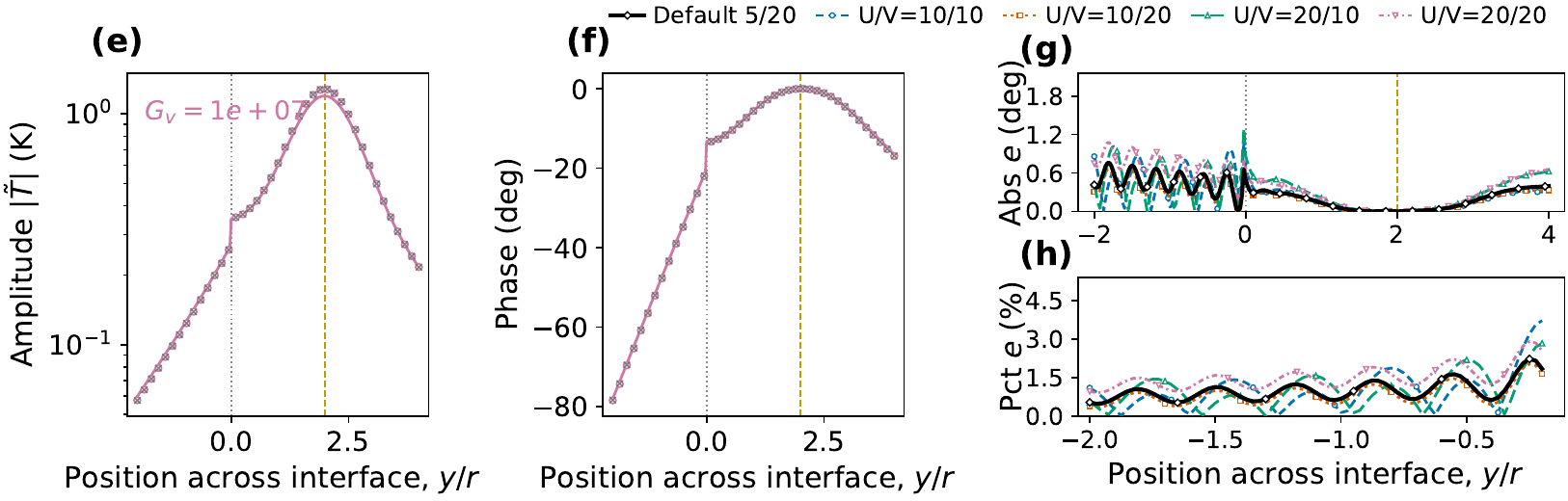}\par\vspace{-0.35em}
    \includegraphics[width=\linewidth]{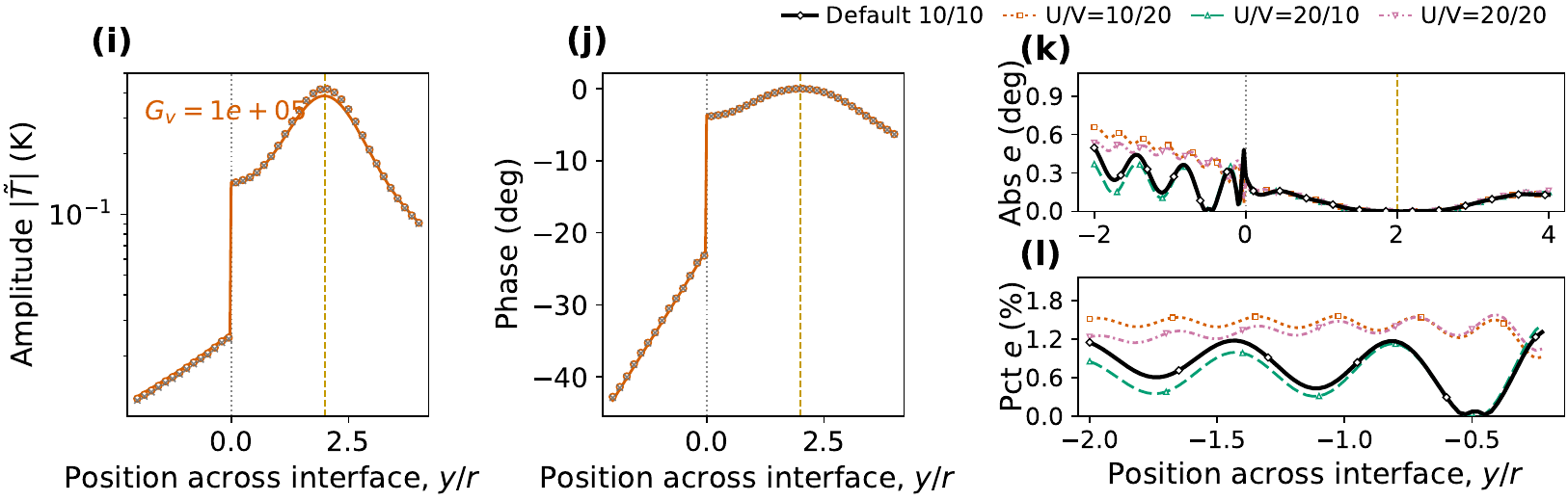}
    \caption{Featured cases 07--09, shown as panels (a)--(d), (e)--(h), and
    (i)--(l), respectively.  All use
    \(k_{\mathrm{un}}/k_{\mathrm{h}}=10/100\).  Case 07: Category II,
    \(rq_{\max}=0.238\), \(r=3~\si{\micro\metre}\), and
    \(f=10~\si{\kilo\hertz}\); \(10/10\) gives \(0.429\%\), while \(20/10\)
    gives \(0.317\%\).  Case 08: Category III, \(rq_{\max}=0.752\),
    \(r=3~\si{\micro\metre}\), and \(f=100~\si{\kilo\hertz}\); \(5/20\)
    gives \(0.999\%\), while \(10/20\) gives \(0.843\%\).  Case 09:
    Category II, \(rq_{\max}=0.251\), \(r=10~\si{\micro\metre}\), and
    \(f=1~\si{\kilo\hertz}\); \(10/10\) gives \(0.754\%\), while \(20/10\)
    gives \(0.641\%\).}
    \label{fig:supp_uv_cases07_09}
\end{figure}

\begin{figure}[p]
    \centering
    \includegraphics[width=\linewidth]{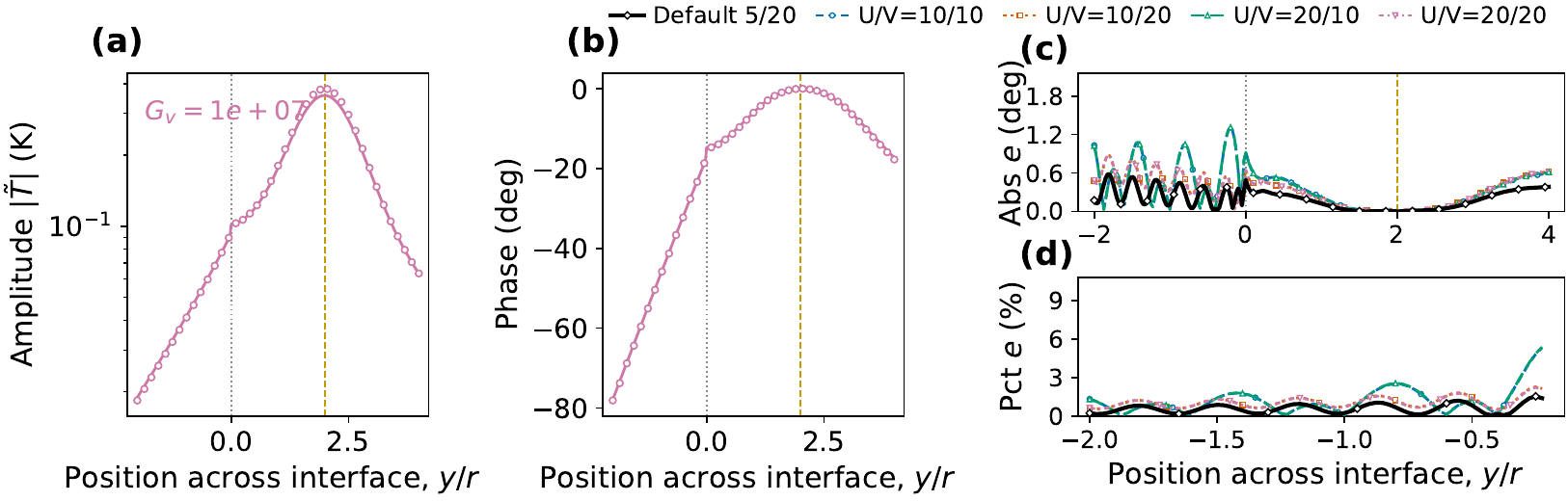}\par\vspace{-0.35em}
    \includegraphics[width=\linewidth]{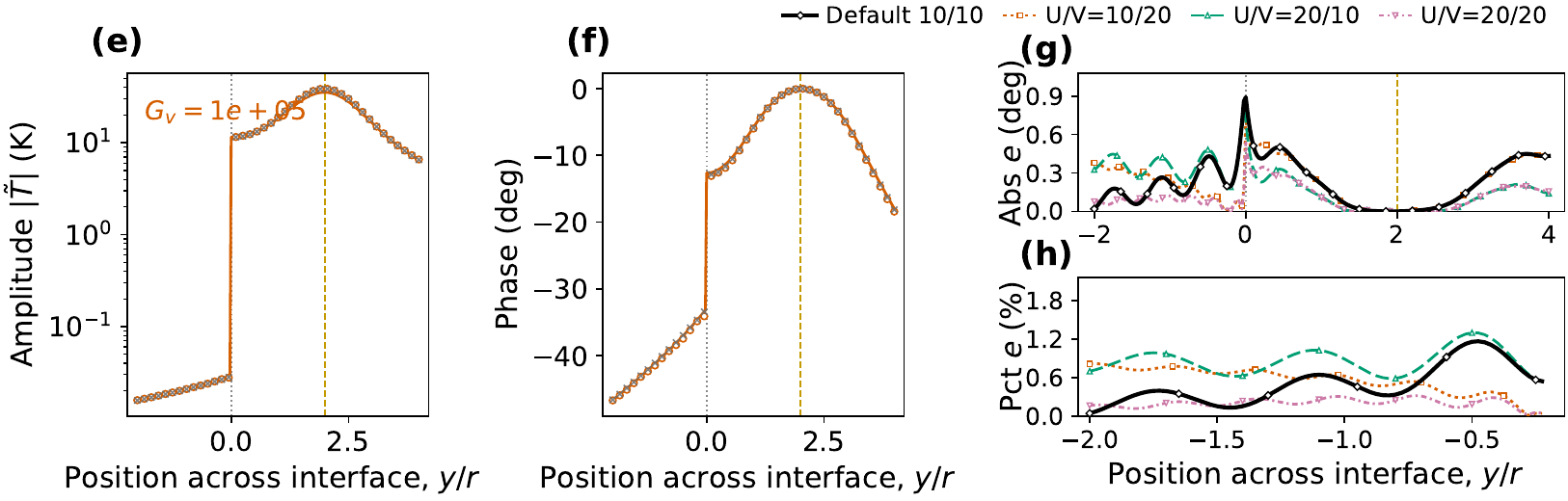}\par\vspace{-0.35em}
    \includegraphics[width=\linewidth]{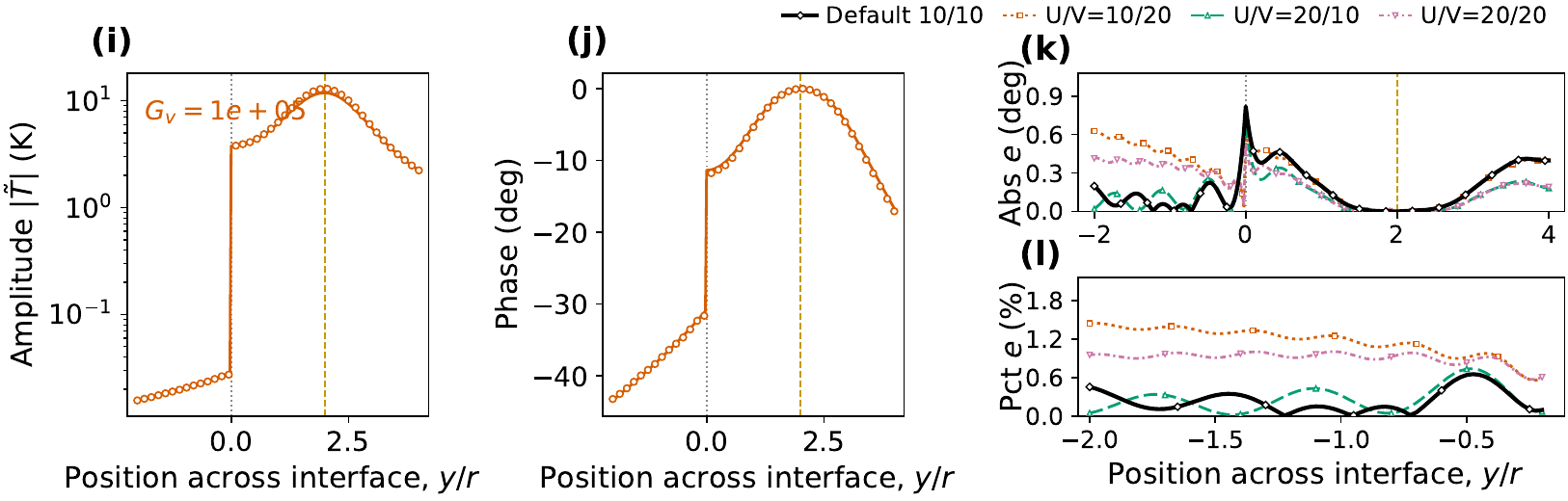}
    \caption{Featured cases 10--12, shown as panels (a)--(d), (e)--(h), and
    (i)--(l), respectively.  Case 10:
    Category III, \(rq_{\max}=0.793\),
    \(k_{\mathrm{un}}/k_{\mathrm{h}}=10/100\),
    \(r=10~\si{\micro\metre}\), and \(f=10~\si{\kilo\hertz}\); \(5/20\)
    is best with \(E_i=0.577\%\).  Cases 11 and 12 are Category II with
    \(k_{\mathrm{un}}/k_{\mathrm{h}}=100/10\) and
    \((rq_{\max},r,f)=(0.251,1~\si{\micro\metre},100~\si{\kilo\hertz})\)
    and \((0.238,3~\si{\micro\metre},10~\si{\kilo\hertz})\), respectively.
    For case 11, \(10/10\) gives \(0.494\%\) and \(20/20\) gives
    \(0.202\%\); for case 12, \(10/10\) is best with \(0.235\%\).}
    \label{fig:supp_uv_cases10_12}
\end{figure}

\begin{figure}[p]
    \centering
    \includegraphics[width=\linewidth]{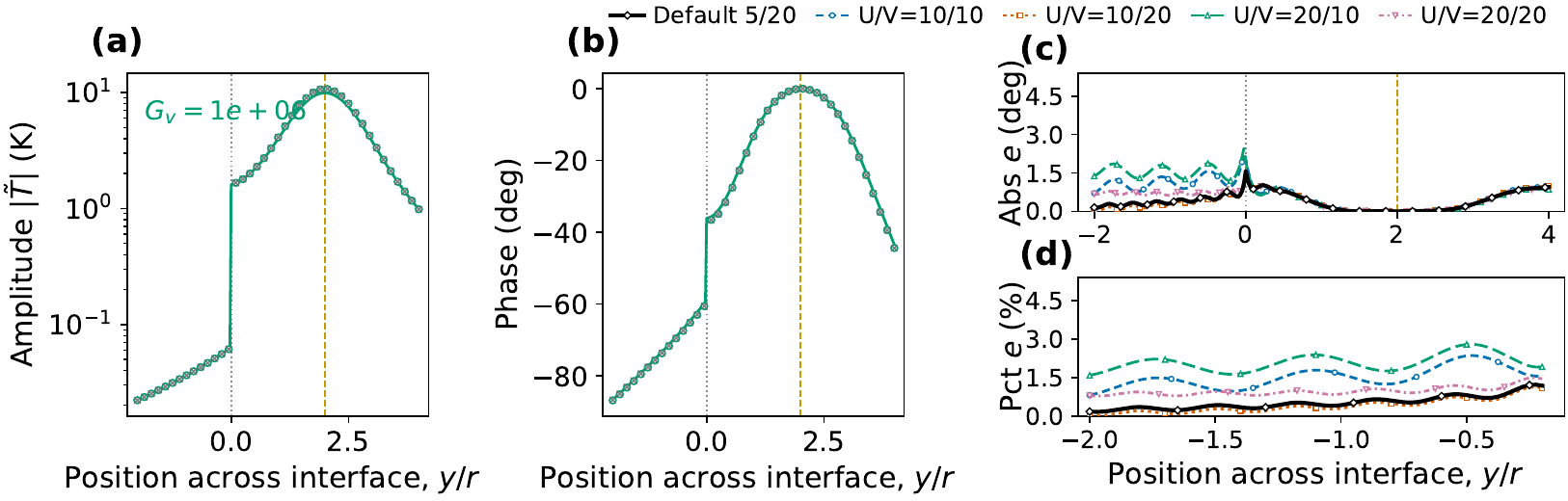}\par\vspace{-0.35em}
    \includegraphics[width=\linewidth]{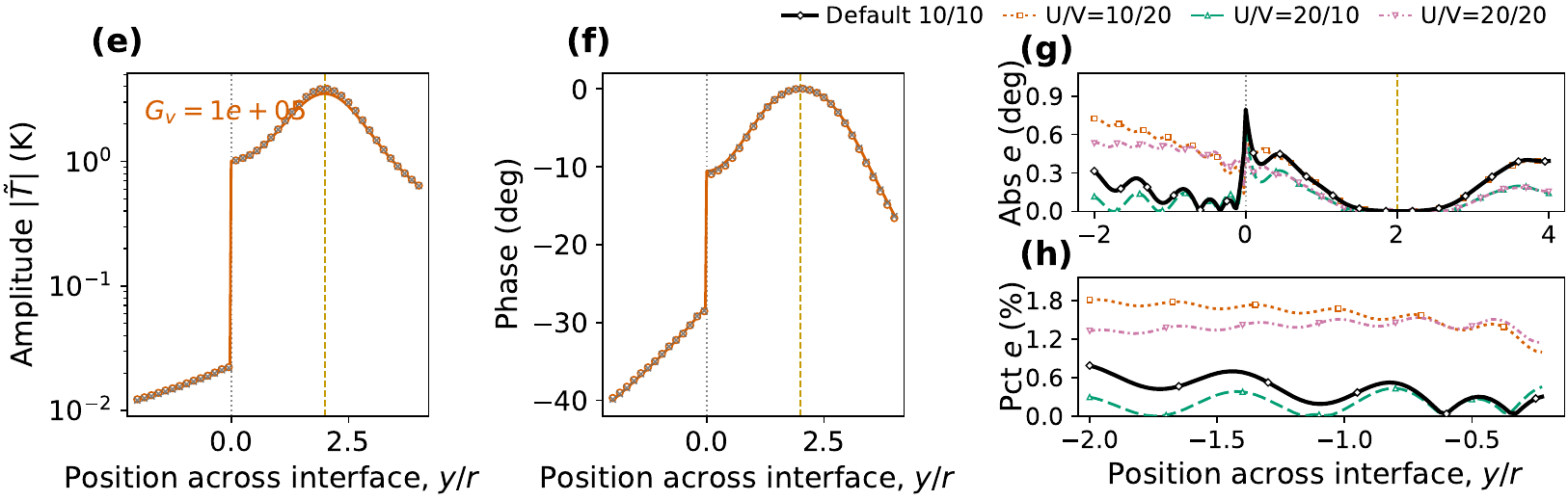}\par\vspace{-0.35em}
    \includegraphics[width=\linewidth]{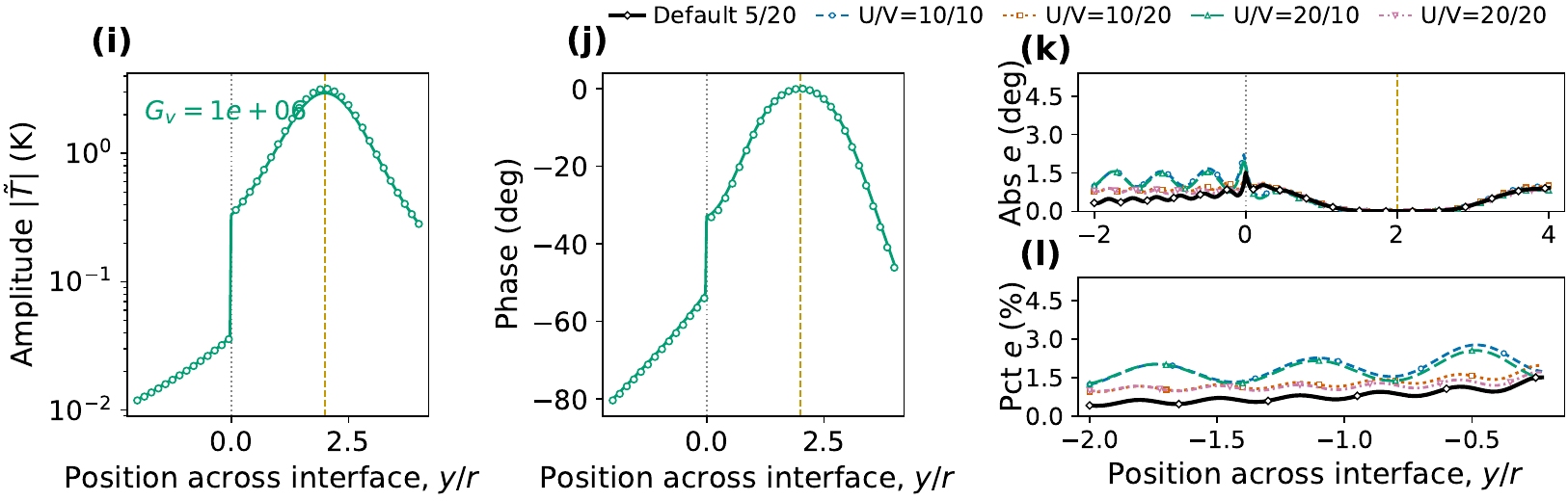}
    \caption{Featured cases 13--15, shown as panels (a)--(d), (e)--(h), and
    (i)--(l), respectively.  All use
    \(k_{\mathrm{un}}/k_{\mathrm{h}}=100/10\).  Case 13: Category III,
    \(rq_{\max}=0.752\), \(r=3~\si{\micro\metre}\),
    \(f=100~\si{\kilo\hertz}\), and
    \(G=10^6~\si{\watt\per\square\metre\per\kelvin}\); \(5/20\) gives
    \(0.524\%\), while \(10/20\) gives \(0.397\%\).  Case 14: Category II,
    \(rq_{\max}=0.251\), \(r=10~\si{\micro\metre}\), and
    \(f=1~\si{\kilo\hertz}\); \(10/10\) gives \(0.403\%\), while \(20/10\)
    gives \(0.198\%\).  Case 15: Category III, \(rq_{\max}=0.793\),
    \(r=10~\si{\micro\metre}\), \(f=10~\si{\kilo\hertz}\), and
    \(G=10^6~\si{\watt\per\square\metre\per\kelvin}\); \(5/20\) is best
    with \(E_i=0.781\%\).}
    \label{fig:supp_uv_cases13_15}
\end{figure}

\begin{figure}[p]
    \centering
    \includegraphics[width=\linewidth]{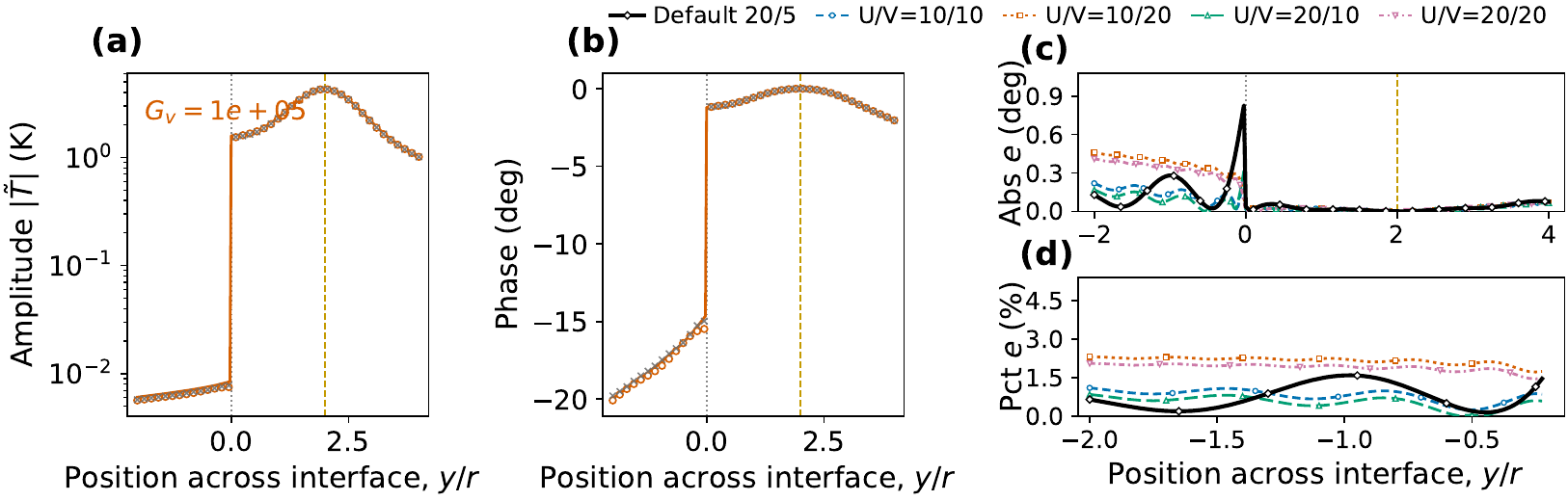}\par\vspace{-0.35em}
    \includegraphics[width=\linewidth]{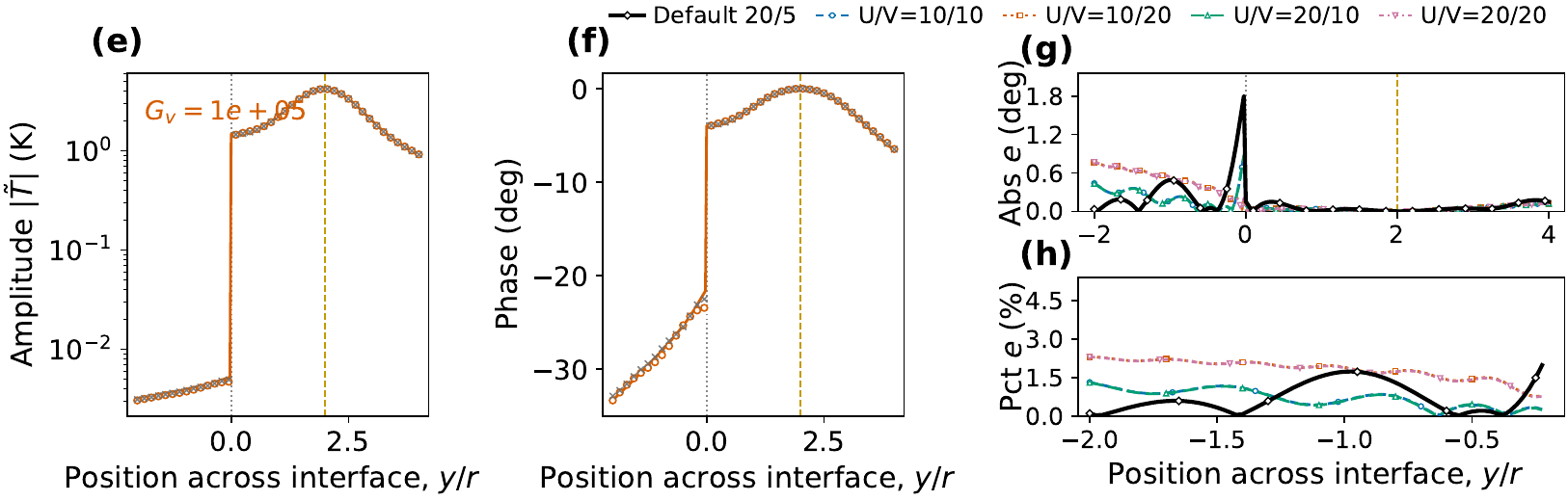}\par\vspace{-0.35em}
    \includegraphics[width=\linewidth]{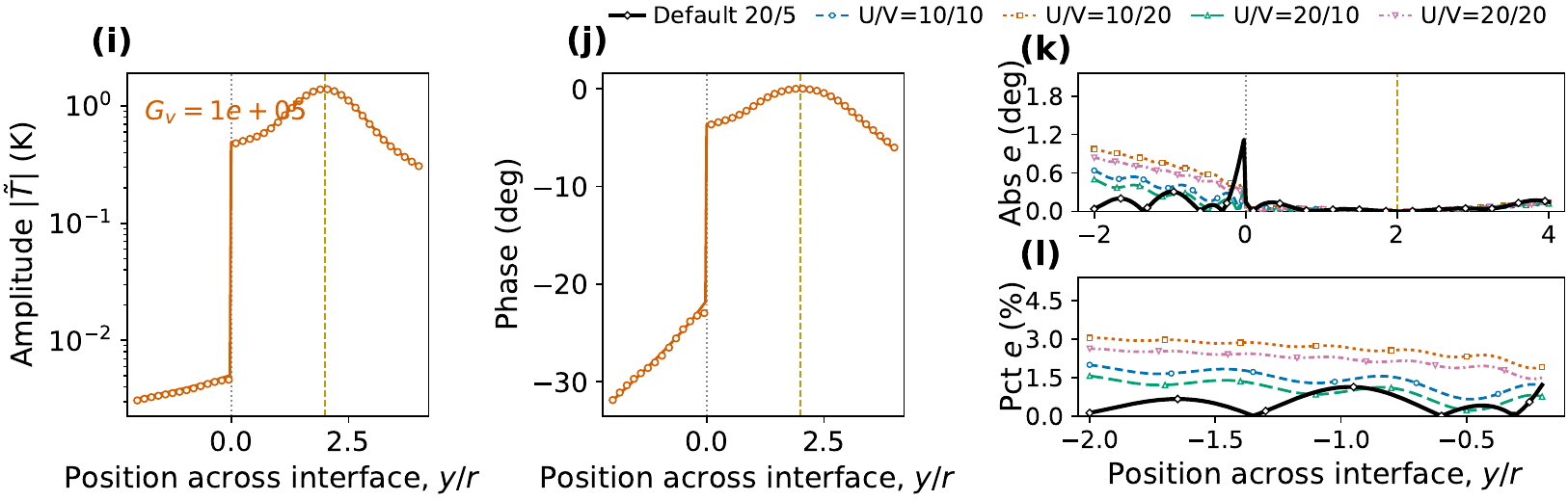}
    \caption{Featured cases 16--18, shown as panels (a)--(d), (e)--(h), and
    (i)--(l), respectively.  All are Category I
    with matched \(100/100~\si{\watt\per\metre\per\kelvin}\) materials.
    Case 16: \(rq_{\max}=0.025\), \(r=1~\si{\micro\metre}\), and
    \(f=10~\si{\kilo\hertz}\); \(20/5\) gives \(0.743\%\), while \(20/10\)
    gives \(0.533\%\).  Case 17: \(rq_{\max}=0.079\),
    \(r=1~\si{\micro\metre}\), and \(f=100~\si{\kilo\hertz}\); \(20/5\)
    gives \(0.710\%\), while \(20/10\) gives \(0.684\%\).  Case 18:
    \(rq_{\max}=0.075\), \(r=3~\si{\micro\metre}\), and
    \(f=10~\si{\kilo\hertz}\); \(20/5\) is best with \(E_i=0.530\%\).}
    \label{fig:supp_uv_cases16_18}
\end{figure}

\begin{figure}[htbp]
    \centering
    \includegraphics[width=\linewidth]{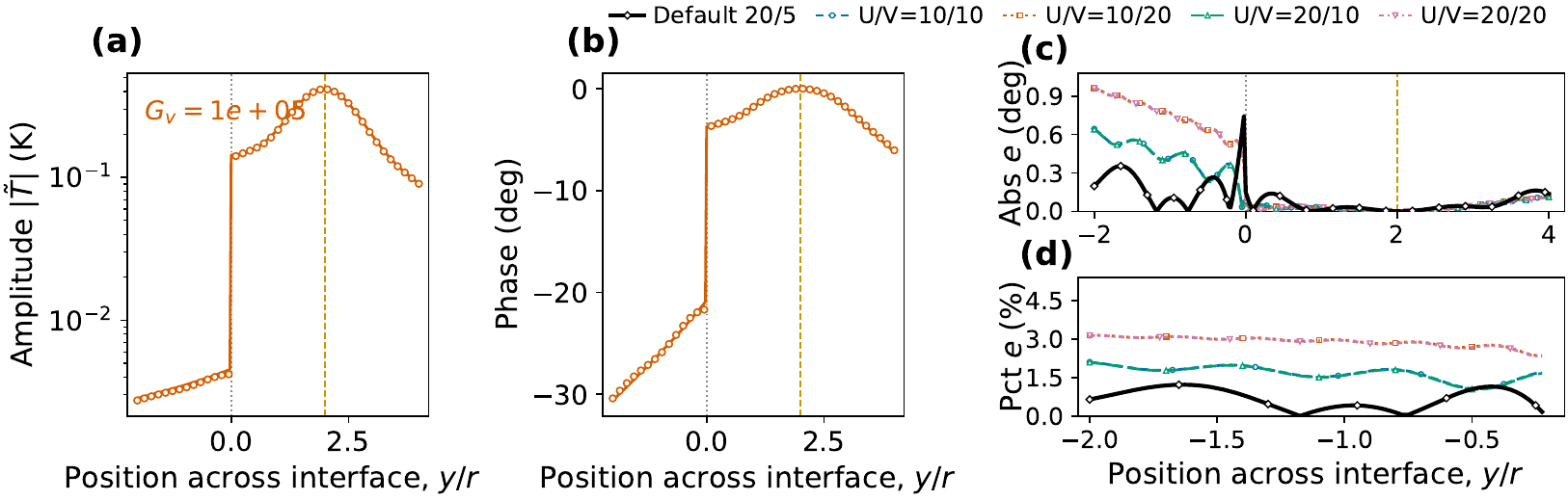}\par\vspace{-0.35em}
    \includegraphics[width=\linewidth]{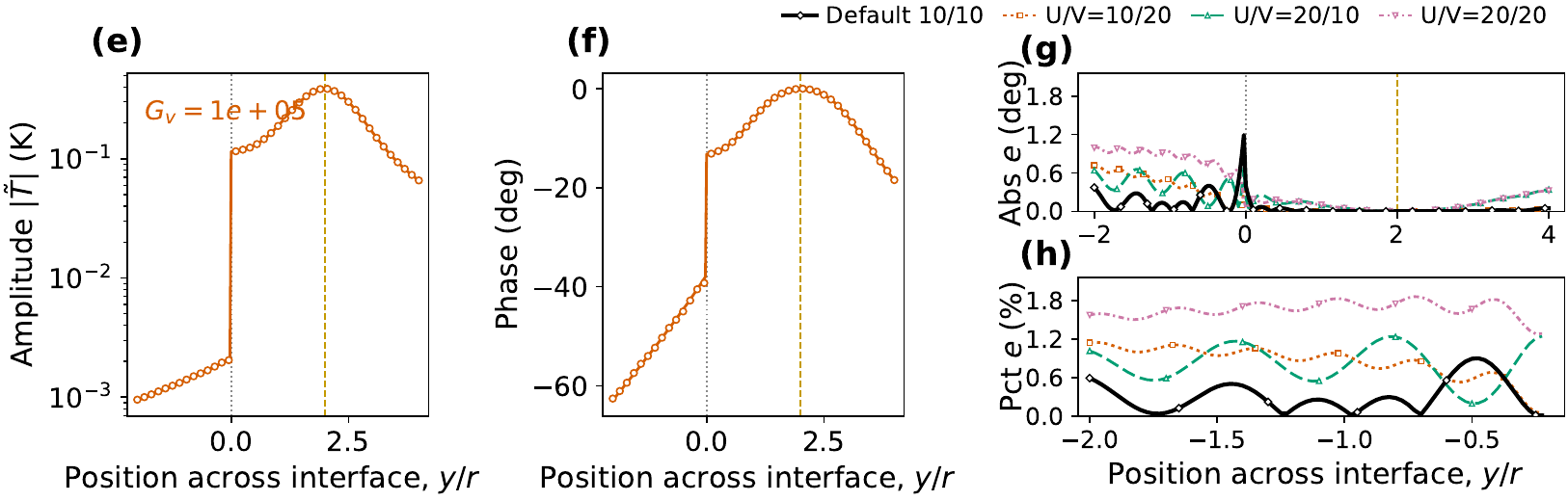}
    \caption{Featured cases 19 and 20, shown as panels (a)--(d) and
    (e)--(h), respectively, both with matched
    \(100/100~\si{\watt\per\metre\per\kelvin}\) materials and
    \(r=10~\si{\micro\metre}\).  Case 19: Category I,
    \(rq_{\max}=0.079\), \(f=1~\si{\kilo\hertz}\); \(20/5\) is best with
    \(E_i=0.683\%\).  Case 20: Category II, \(rq_{\max}=0.251\),
    \(f=10~\si{\kilo\hertz}\); \(10/10\) is best with
    \(E_i=0.310\%\).}
    \label{fig:supp_uv_cases19_20}
\end{figure}
\FloatBarrier

\subsection{Use for multilayer and anisotropic models}
\label{sec:supp_uv_multilayer_extension}

The fixed category rule was calibrated against homogeneous analytical
references and should not be transferred to a multilayer anisotropic model as
a universal optimum without additional validation. For an orthotropic stack, the
implementation evaluates \(q_x\), \(q_y\), and \(q_z\) over every layer on
both sides, uses the largest directional scales in
\cref{eq:supp_uv_combined_limits}, and treats the category rule as a starting
candidate.  All nine \((U,V)\) pairs are then compared with a more-resolved
Gear-5 GBIE solution of the \emph{same} physical stack, using
\cref{eq:supp_uv_phase_percentage} averaged over all requested interface
resistances.  A selected pair is accepted only when its score is below the
specified convergence tolerance.  This GBIE-to-GBIE comparison is a numerical
convergence test, not an independent validation of the physical model.

For the anisotropic multilayer COMSOL cohort used in the main-text
performance comparison, Gear~2 was held at
\((N_u,N_v,N_z)=(35,120,25)\) and \(U/V=10/10\), \(10/20\), and \(20/20\)
were compared directly.  The \(10/20\) pair gives the lowest mean phase RMSE
normalized by the COMSOL P95--P5 phase span, both over all five executions and
after the duplicated physical profile is counted once.  That independent
multilayer audit, including all profile plots, is reported in Supplementary
Section~S7.

For a practical first calculation, \(U/V=10/10\) is the most transferable
single starting pair in the homogeneous study: its mean pointwise phase
percentage error is below \(2\%\) in every one of the 20 featured cases and in
110 of the 126 structured Category I--III cases.  The category rules improve
the average score, but they are not universal optima for a new multilayer or
anisotropic class.  When the highest attainable accuracy is needed, one
domain-converged FEM calculation at the target or estimated property should
be used to select \(U/V\).  That calibrated pair can then be reused over the
intended nearby property range, with refinement checks at the range
boundaries.  A change in layer ordering, interface burial, strong anisotropy,
or the shape of \(G_v(z)\) defines a new numerical class and requires a new
calibration or GBIE-to-GBIE convergence check.

Additional multilayer scales include \(q_{z,j}h_j\), \(r/h_j\), conductivity
contrast, and interlayer resistance.  These quantities explain why
\(rq_{\max}\) alone cannot guarantee convergence outside the homogeneous
calibration set. The current implementation uses diagonal conductivity
tensors in the model coordinates. Nonzero off-diagonal components require
corresponding cross terms in the GBIE kernel and are not represented by
the present rule.

\section{Timing and Accuracy Against the Full Bulk Analytical Model}
\label{sec:supp_full_eq10_bie_timing}

The analytical validation in the main text compares the GBIE solution with the
published bulk vertical-interface solution for homogeneous isotropic
half-spaces.  To check the computational claim separately from the profile
comparison, we timed the same bulk cases using three evaluation routes:
(i) a direct adaptive evaluation of the full analytical source-convolution
integrals, (ii) the GBIE solver, and (iii) the optimized
vectorized full analytical implementation used here as the numerical
reference.  The full source-convolution model corresponds to the original
two-dimensional source integral, while the simplified analytical model denotes
the center-line reduction used only for \(x=0\).

The benchmark uses the three representative bulk cases from the main-text
analytical validation, \(k_{\mathrm{un}}/k_{\mathrm{h}}=10/10\), \(10/100\),
and \(100/10\), with \(G_v=10^6~\mathrm{W\,m^{-2}\,K^{-1}}\).  The matched
case uses \(r=100~\mu\mathrm{m}\) and \(f=100~\mathrm{Hz}\), while the two
thermal-contrast cases use \(r=10~\mu\mathrm{m}\) and
\(f=1~\mathrm{kHz}\).  For each case, the timing sweep used the same
\(121\)-point \(y\)-grid and one, three, and five lateral-offset sets:
\(x/r=1\), \(x/r=0,1,2\), and \(x/r=0,0.5,1,1.5,2\), respectively.  Thus,
the one-offset benchmark is off-axis rather than a center-line evaluation.  The
updated sweep was run on the same local workstation in one Python 3.11.9
process with NumPy 2.4.6 and SciPy 1.17.1 so that the offset-count runtimes
are internally consistent.
The additional point-count sweep used the same Python 3.11.9 environment. In
every paired timing, the full model and GBIE received the
identical \(x\)- and \(y\)-observation arrays.  This sweep varied both the
number of \(y\)-points in one off-axis line and the number of lines, as
detailed below.
Errors are reported relative to the vectorized full analytical solution,
excluding the grid point adjacent to the discontinuous interface.
The phase residual is computed after applying the same heat-center reference
used in the main validation figure,
\begin{equation}
    e_\phi(y;x)
    =
    \left[\phi_{\mathrm{GBIE}}(y;x)-\phi_{\mathrm{GBIE}}(d;x)\right]
    -
    \left[\phi_{\mathrm{full}}(y;x)-\phi_{\mathrm{full}}(d;x)\right],
    \label{eq:supp_full_analytical_phase_error}
\end{equation}
and this signed residual is shown in the detailed comparison below.  The
pointwise percentage-error panels use
\(100|e_\phi|/\max(|\phi_{\mathrm{full}}|,1^\circ)\) on the \(y<0\) side,
matching the convention used in the main-text isotropic validation.

The vectorized full analytical reference used
\((N_\delta,N_{x_0},N_{y_0})=(240,220,220)\), a source half-width of
\(4r\), and explicit source spillover across \(y=0\).  The direct adaptive
full analytical calculation used the same physical source model with
\((N_{x_0},N_{y_0})=(120,120)\) and complex adaptive quadrature over
\(\delta\) with relative and absolute tolerances \(10^{-5}\) and
\(10^{-10}\), respectively.  In the timing scripts this route is forced by
using \(\texttt{mode=eq10}\), so the center-line rows do not fall back to the
simplified expression.  The vectorized full reference is likewise forced by
disabling the homogeneous center-line shortcut and checking that the returned
path is the complete two-material Equation~10 evaluator. The GBIE settings
used for each case are listed in
Table~\ref{tab:supp_full_analytical_integration_specs}. Every GBIE run
uses the standard efficient setting
\((N_u,N_v,N_z)=(35,120,25)\), while the spectral source cutoffs are selected
after evaluating the physical inputs. This rule was obtained from an extensive
investigation comprising more than 3000 numerical GBIE evaluations.  Its
detailed definition and statistical assessment are given in Supplementary
Section~\ref{sec:supp_uv_source_factor}.

\begin{table}[htbp]
    \centering
    \caption{GBIE integration settings used in the full analytical timing
    benchmark. The standard efficient setting uses the source-cutoff procedure
    established by the numerical investigation in Supplementary
    Section~\ref{sec:supp_uv_source_factor}. Computational depth limits are
    given in dimensionless form using the beam radius \(r\).}
    \label{tab:supp_full_analytical_integration_specs}
    \begin{tabular}{lccccc}
        \toprule
        Case & \(N_u\) & \(N_v\) & \(N_z\) & Source cutoffs & \(z_{\mathrm{int,max}}/r\) \\
    \midrule
        \(10/10\) & 35 & 120 & 25 & input-selected & 10.0 \\
        \(10/100\) & 35 & 120 & 25 & input-selected & 10.0 \\
        \(100/10\) & 35 & 120 & 25 & input-selected & 10.0 \\
        \bottomrule
    \end{tabular}
\end{table}

To keep the depth-discretization workload controlled across thermal-property
cases, the timing run disabled material-dependent depth clustering and the automatic reuse
of side Green functions when the two materials are identical.  Consequently,
all cases use exactly 25 depth nodes and explicitly evaluate both sides of the
interface.  The spectral cutoffs vary only as prescribed by the adaptive rule;
these controls affect computational work, not the GBIE formulation.

\subsection{Full Analytical and GBIE Profiles at Multiple Lateral Offsets}
\label{sec:supp_multi_offset_profiles}

\begin{figure}[htbp]
    \centering
    \includegraphics[width=\linewidth]{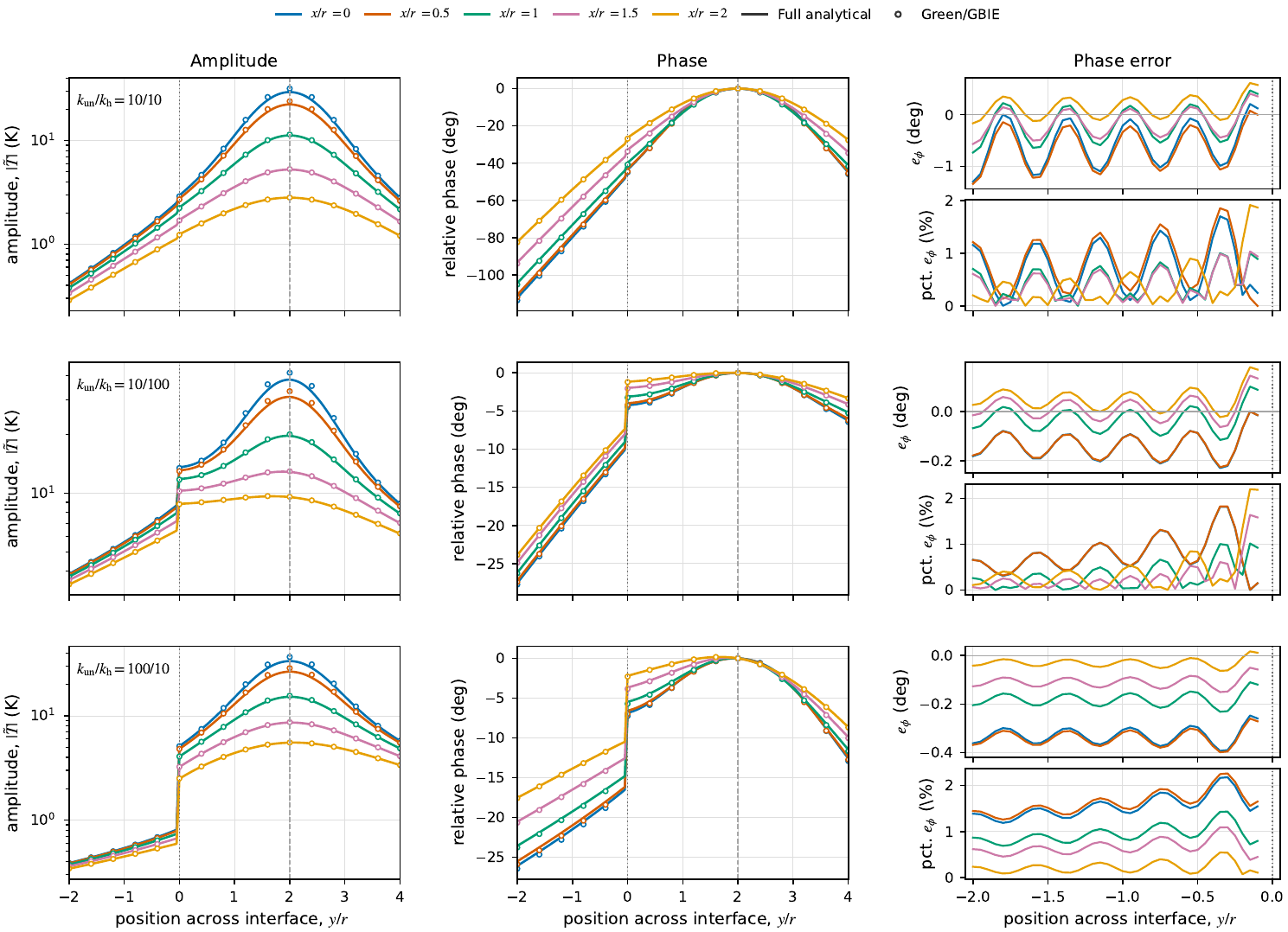}
    \caption{Amplitude and phase comparison between the full Equation~10
    analytical model (solid curves) and the GBIE solution (open circles)
    at five lateral offsets, \(x/r=0,0.5,1,1.5,2\).  Rows correspond to
    \(k_{\mathrm{un}}/k_{\mathrm{h}}=10/10\), \(10/100\), and \(100/10\).
    The third column shows the signed phase residual \(e_\phi\) and its
    pointwise percentage magnitude on the \(y/r<0\) side, with the
    interface-adjacent point masked.  Phase is referenced to the heat center at
    \(y/r=2\); the dotted and dashed vertical lines mark the interface and heat
    center, respectively.}
    \label{fig:supp_full_analytical_bie_five_offsets}
\end{figure}

Before comparing runtimes, we validate the off-axis response directly at five
lateral coordinates.  Figure~\ref{fig:supp_full_analytical_bie_five_offsets}
compares the vectorized full Equation~10 source-convolution model with the GBIE
for \(x/r=0,0.5,1,1.5,2\).  All analytical curves use the complete
two-dimensional source integral with explicit source spillover across the
interface; the simplified \(x=0\) reduction is not used.  The close overlap in
both amplitude and relative phase confirms that the GBIE reconstruction remains
consistent with the full analytical model away from the beam-center line.

\begin{figure}[htbp]
    \centering
    \includegraphics[width=\linewidth]{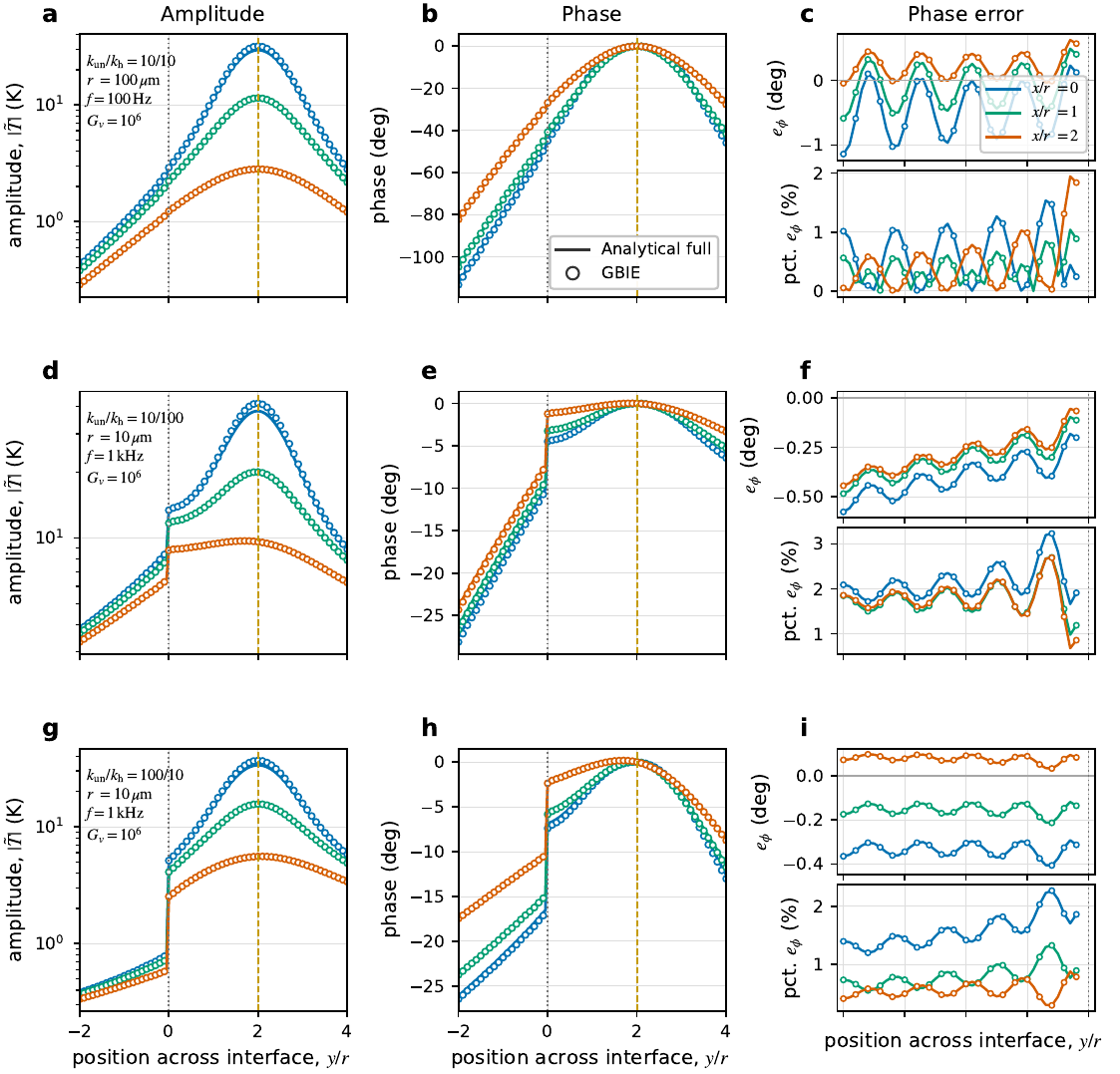}
    \caption{Direct comparison between the vectorized full analytical
    reference (solid lines) and the GBIE solution (open circles) for the
    same timing benchmark.  Rows correspond to
    \(k_{\mathrm{un}}/k_{\mathrm{h}}=10/10\), \(10/100\), and \(100/10\);
    colors correspond to \(x/r=0,1,2\).  Columns show amplitude, relative
    phase referenced to the heat center, and the signed phase residual together
    with its pointwise percentage error.  The crack-neighbor point is masked.}
    \label{fig:supp_full_analytical_bie_direct_comparison}
\end{figure}

\begin{figure}[htbp]
    \centering
    \includegraphics[width=\linewidth]{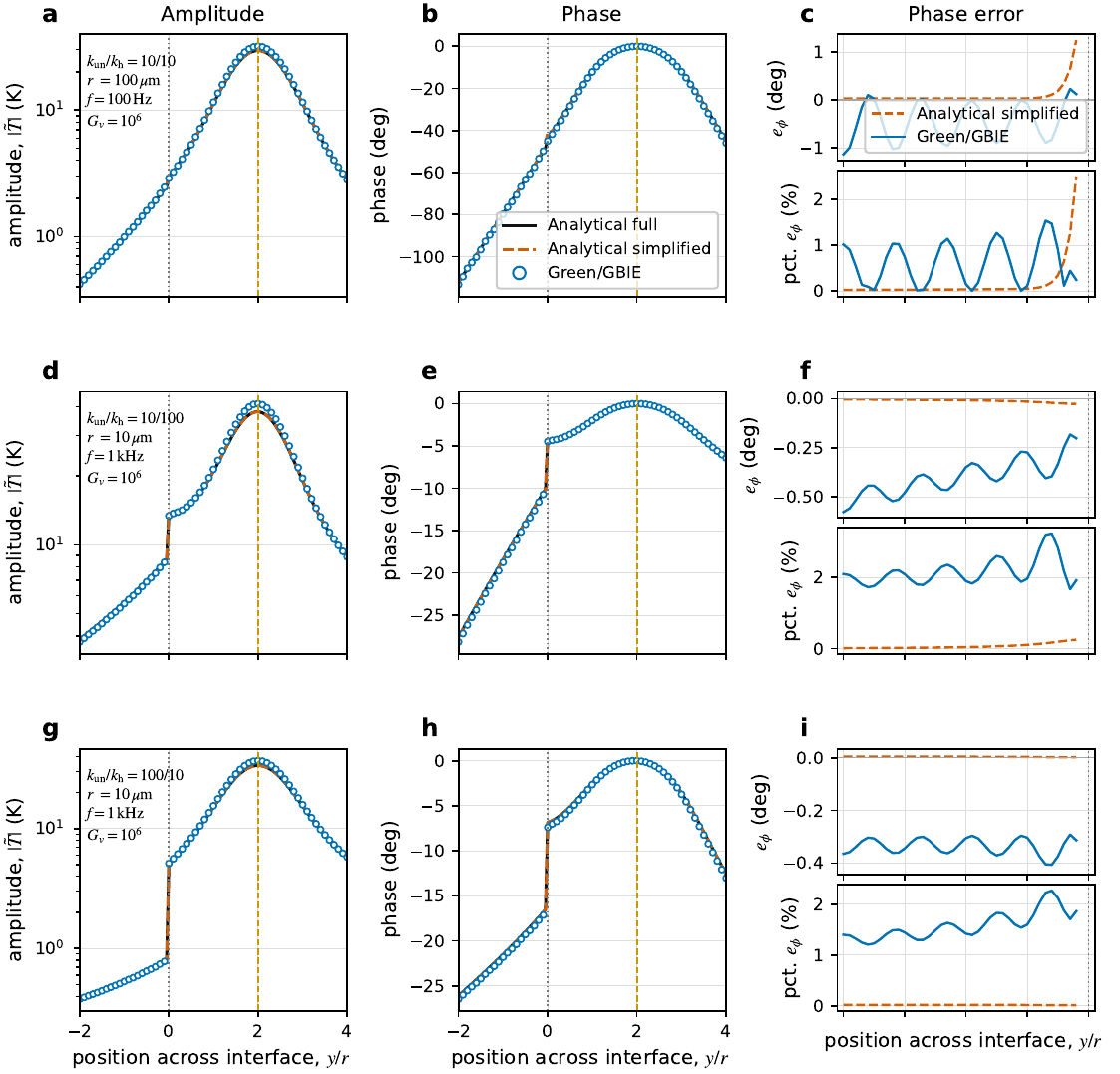}
    \caption{Zero-offset comparison among the full analytical
    source-convolution model, the simplified center-line analytical reduction,
    and the GBIE solution.  Rows correspond to
    \(k_{\mathrm{un}}/k_{\mathrm{h}}=10/10\), \(10/100\), and \(100/10\).
    The simplified curve is evaluated with the same analytical implementation
    as the full reference, but using the original center-line paper source
    convention without the split-source spillover correction.  The right column
    reports signed phase residuals and pointwise percentage errors relative to
    the full analytical reference.}
    \label{fig:supp_full_analytical_bie_x0_three_model}
\end{figure}

\clearpage
\subsection{Runtime Scaling with Number of Requested Points and Offsets}
\label{sec:supp_offset_count_timing}

\begin{table}[htbp]
    \centering
    \caption{Center-line runtimes at \(x/r=0\), including the simplified
    analytical reduction. The full analytical-model entries are forced
    evaluations of the complete Equation~10 source-convolution form, not the
    simplified center-line expression; one uses adaptive quadrature and the
    other uses vectorized quadrature.}
    \label{tab:supp_centerline_simplified_timing}
    \begin{tabular}{lcccc}
        \toprule
        Case & Simplified & \shortstack{Full analytical\\model} & GBIE & \shortstack{Vectorized\\analytical} \\
        \(k_{\mathrm{un}}/k_{\mathrm{h}}\) & \multicolumn{4}{c}{runtime for \(x/r=0\) (s)} \\
        \midrule
        \(10/10\) & 0.094 & 2.6 & 2.2 & 0.080 \\
        \(10/100\) & 0.198 & 2.6 & 2.2 & 0.077 \\
        \(100/10\) & 0.186 & 2.1 & 2.3 & 0.080 \\
        \bottomrule
    \end{tabular}
\end{table}

\begin{table}[htbp]
    \centering
    \caption{Paired point-count sweep for the direct adaptive full analytical
    model and the GBIE.  Both solvers receive exactly the same observation
    arrays in each row, so \(N_{\mathrm{obs}}=N_xN_y\) is identical.  The
    one-line cases use \(x/r=1\), the three-line case uses \(x/r=0,1,2\), and
    the five-line case uses \(x/r=0,0.5,1,1.5,2\).  A 241-point line has the
    same endpoints as the 121-point line and half its \(y\)-spacing.  Internal
    GBIE quadrature remains fixed at \((N_u,N_v,N_z)=(35,120,25)\), with
    adaptively selected source cutoffs.  Each entry is one paired timing in
    the Python 3.11.9 environment stated above.}
    \label{tab:supp_point_count_timing}
    \begingroup
    \scriptsize
    \setlength{\tabcolsep}{4pt}
    \begin{tabular}{llrrrr}
        \toprule
        Case & Layout & \(N_{\mathrm{obs}}\) & \shortstack{Full analytical\\model} & GBIE & Analytical/GBIE \\
        \(k_{\mathrm{un}}/k_{\mathrm{h}}\) & \(N_x\times N_y\) & & \multicolumn{2}{c}{runtime (s)} & ratio \\
        \midrule
        \(10/10\)  & \(1\times61\)  & 61  & 1.2 & 2.3 & \(0.5\times\) \\
                    & \(1\times121\) & 121 & 3.5 & 2.4 & \(1.5\times\) \\
                    & \(1\times241\) & 241 & 9.8 & 2.4 & \(4.1\times\) \\
                    & \(1\times363\) & 363 & 16.2 & 2.4 & \(6.6\times\) \\
                    & \(3\times121\) & 363 & 10.7 & 2.3 & \(4.6\times\) \\
                    & \(5\times121\) & 605 & 17.5 & 2.3 & \(7.7\times\) \\
        \addlinespace
        \(10/100\) & \(1\times61\)  & 61  & 1.4 & 2.4 & \(0.6\times\) \\
                    & \(1\times121\) & 121 & 2.9 & 2.4 & \(1.2\times\) \\
                    & \(1\times241\) & 241 & 5.8 & 2.4 & \(2.5\times\) \\
                    & \(1\times363\) & 363 & 8.7 & 2.4 & \(3.7\times\) \\
                    & \(3\times121\) & 363 & 8.7 & 2.4 & \(3.7\times\) \\
                    & \(5\times121\) & 605 & 14.6 & 2.4 & \(6.1\times\) \\
        \addlinespace
        \(100/10\) & \(1\times61\)  & 61  & 1.1 & 2.4 & \(0.4\times\) \\
                    & \(1\times121\) & 121 & 2.2 & 2.4 & \(0.9\times\) \\
                    & \(1\times241\) & 241 & 4.4 & 2.4 & \(1.9\times\) \\
                    & \(1\times363\) & 363 & 6.6 & 2.4 & \(2.8\times\) \\
                    & \(3\times121\) & 363 & 6.9 & 2.4 & \(2.9\times\) \\
                    & \(5\times121\) & 605 & 11.5 & 2.3 & \(5.0\times\) \\
        \bottomrule
    \end{tabular}
    \endgroup
\end{table}

Table~\ref{tab:supp_point_count_timing} demonstrates that ``single line'' is
not a computational-size definition.  With 61 requested points, direct
adaptive evaluation is \(1.7\)--\(2.2\times\) faster than the GBIE.  At the
current 121-point definition the two are comparable, whereas doubling the
sampling density to 241 points makes the GBIE \(1.9\)--\(4.1\times\) faster.
Interpolating the sign change in the single-line timing difference gives
case-dependent crossover estimates of approximately 90--131 requested
points.  This range is specific to the tested tolerances, materials, and
workstation rather than a universal constant.

Total point count is the main scaling variable for the direct calculation but
does not determine its runtime completely.  At the same
\(N_{\mathrm{obs}}=363\), the \(1\times363\) and \(3\times121\) layouts are
nearly equivalent for the contrast cases, but differ for the matched case
because the refined \(y\)-grid changes the automatic adaptive integration
cutoff near the interface.  The actual coordinates and grid spacing should
therefore accompany \(N_{\mathrm{obs}}\) in any timing claim.

\begin{table}[htbp]
    \centering
    \caption{Same-environment runtimes for one, three, and five lateral
    offsets. ``Full analytical model'' and ``Vectorized analytical'' denote
    direct adaptive-quadrature and optimized vectorized evaluations of the complete Equation~10
    source-convolution model.  The offset sets for \(N_x=1,3,5\) are
    \(x/r=1\), \(x/r=0,1,2\), and \(x/r=0,0.5,1,1.5,2\), respectively.  The
    speedup compares the direct adaptive runtime with the GBIE runtime for
    the same offset set.}
    \label{tab:supp_full_analytical_bie_timing}
    \begin{tabular}{lccccc}
        \toprule
        Case & \(N_x\) & \shortstack{Full analytical\\model} & GBIE & \shortstack{Vectorized\\analytical} & Analytical/GBIE \\
        & & \multicolumn{3}{c}{runtime (s)} & speedup \\
        \midrule
        \(10/10\)  & 1 & 3.4  & 2.2 & 0.078 & \(1.5\times\) \\
                     & 3 & 10.1 & 2.2 & 0.227 & \(4.5\times\) \\
                     & 5 & 16.8 & 2.2 & 0.378 & \(7.6\times\) \\
        \addlinespace
        \(10/100\) & 1 & 2.8  & 2.3 & 0.083 & \(1.2\times\) \\
                     & 3 & 8.7 & 2.3 & 0.228 & \(3.7\times\) \\
                     & 5 & 14.1 & 2.3 & 0.403 & \(6.1\times\) \\
        \addlinespace
        \(100/10\) & 1 & 2.1  & 2.3 & 0.080 & \(0.9\times\) \\
                     & 3 & 6.5 & 2.3 & 0.230 & \(2.9\times\) \\
                     & 5 & 11.1 & 2.2 & 0.394 & \(5.0\times\) \\
        \bottomrule
    \end{tabular}
\end{table}

Tables~\ref{tab:supp_centerline_simplified_timing} and
\ref{tab:supp_full_analytical_bie_timing} show the crossover clearly.  The
simplified center-line expression is the fastest \(x/r=0\) route, requiring
\(0.094\)--\(0.198~\mathrm{s}\), but it is not available off axis.  For one
off-axis offset at \(x/r=1\), the GBIE and direct adaptive full analytical
calculation are comparable, and either may be faster depending on the thermal
contrast.  For three offsets the GBIE is \(2.9\)--\(4.5\times\) faster, and
for five offsets it requires only \(2.2\)--\(2.3~\mathrm{s}\), compared with
\(11.1\)--\(16.8~\mathrm{s}\) for the adaptive full analytical evaluation,
a \(5.0\)--\(7.6\times\) speedup.  Under the adaptive source-cutoff rule, the
five-offset GBIE runtime varies by less than \(4\%\) among the three
thermal-property cases and remains
nearly unchanged as offsets are added because the interfacial solution is
common to all observation offsets; additional \(x\)-coordinates enter only
through the final cosine reconstruction.  By contrast, both full analytical implementations scale
approximately linearly with the number of offsets.

The optimized vectorized full analytical implementation is much faster than
the GBIE in this benchmark
(\(0.378\)--\(0.403~\mathrm{s}\) for the five-offset calculation) because it is
tailored specifically to the homogeneous isotropic bulk geometry and evaluates
the source convolution in large array operations.  This timing should
therefore be interpreted as a specialized reference implementation, not as the
cost of a general analytical model.  The relevant comparison for prior full
source-convolution semi-analytical evaluation is the direct adaptive route, for
which the GBIE is already competitive while retaining the ability to treat
anisotropy, multilayers, and vertical interfaces that may be surface-breaking
or buried and finite or semi-infinite in depth.

\section{Three-Dimensional COMSOL Model and Numerical Setup}
\label{sec:supp_comsol_setup}

\subsection{Computational Environment and Model Definition}

The finite-element reference calculations reported in
Section~3.2 of the main text were performed with COMSOL Multiphysics
6.2.0.415 using computational resources provided by the University of
Pittsburgh Center for Research Computing (CRC).  A fully
three-dimensional model was constructed with the \emph{Heat Transfer in
Solids} interface.  The film, the two anisotropic substrate regions, the
vertical interface, and the horizontal film--substrate interfaces were
represented explicitly.  Each solid region was assigned the thermal
conductivity tensor
\begin{equation}
    \mathbf{k}=\operatorname{diag}(k_x,k_y,k_z)
    \label{eq:supp_comsol_conductivity_tensor}
\end{equation}
and volumetric heat capacity \(C_v\) listed in the main text.  The vertical
and horizontal thermal boundary conductances were imposed as interfacial
thermal resistances satisfying
\begin{equation}
    q_n=G\left(T^{+}-T^{-}\right),
    \label{eq:supp_comsol_contact_law}
\end{equation}
with \(G=G_v\) at the vertical interface and \(G=G_h\) at each
film--substrate interface.

The pump was applied as a harmonically modulated Gaussian surface heat flux
with the same absorbed power, \(1/e^2\) radius \(r\), and center position as
in the GBIE calculation.  The remainder of the top surface was thermally
insulated.  COMSOL used an adiabatic rear surface placed beyond the
through-plane thermally active region, where it approximates the
semi-infinite decay condition used by the GBIE.  The retained summary does not specify the exact boundary-feature
assignment on all four lateral faces; their distance from
the pump was selected using the thermal-penetration-depth criterion described
below.  A \emph{Frequency Domain} study was used at angular frequency
\(\omega=2\pi f\), so that COMSOL solved directly for the complex harmonic
temperature amplitude.  The exported amplitude and phase were sampled along
the surface line through the pump center and normal to the vertical
interface, matching the line used for the GBIE comparison.

\subsection{Lateral Domain and Boundary-Effect Criterion}

For an anisotropic multilayer, the conservative lateral thermal penetration
depth was defined as the largest in-plane value among all layers and both
lateral directions,
\begin{equation}
    \delta_{\mathrm{lat}}
    =\max_{j,\,i\in\{x,y\}}
    \sqrt{\frac{k_{i,j}}{\pi C_{v,j} f}},
    \label{eq:supp_comsol_lateral_tpd}
\end{equation}
where \(j\) indexes the film and substrate regions. Let
\(y_{\min}^{\mathrm{data}}\) and \(y_{\max}^{\mathrm{data}}\) denote the ends
of the retained comparison interval and
\(L_e=y_{\max}^{\mathrm{data}}-y_{\min}^{\mathrm{data}}\). The two boundaries
in the scan direction were chosen to satisfy
\begin{equation}
    y_{\min}^{\mathrm{box}}
    \leq y_{\min}^{\mathrm{data}}-4\delta_{\mathrm{lat}},
    \qquad
    y_{\max}^{\mathrm{box}}
    \geq y_{\max}^{\mathrm{data}}+4\delta_{\mathrm{lat}},
    \qquad
    L_y\geq L_e+8\delta_{\mathrm{lat}}.
    \label{eq:supp_comsol_lateral_size}
\end{equation}
The production-domain design interval was
\(-3r\leq y\leq5r\), so \(L_e=8r\). Equation~
\eqref{eq:supp_comsol_lateral_size} places a
\(4\delta_{\mathrm{lat}}\) buffer beyond each end of that interval. If the
two-sided thermal width is denoted
\(\sigma_{\mathrm{lat}}=2\delta_{\mathrm{lat}}\), the same criterion is the
shorthand
\begin{equation}
    L_y\geq L_e+4\sigma_{\mathrm{lat}}.
    \label{eq:supp_comsol_lateral_padding}
\end{equation}
For a common heat capacity \(C_v\), the largest in-plane conductivity
compatible with a specified lateral width is therefore
\begin{equation}
    k_{\parallel,\max}^{\mathrm{allow}}
    =
    \pi C_v f
    \left(\frac{L_y-L_e}{8}\right)^2,
    \qquad
    f_{\min}
    =
    \frac{k_{\parallel,\max}}
    {\pi C_v[(L_y-L_e)/8]^2}.
    \label{eq:supp_comsol_lateral_inverse}
\end{equation}
For unequal heat capacities, the criterion is applied to the largest
directional diffusivity \(k_{i,j}/C_{v,j}\), rather than to conductivity
alone.
At a distance of four penetration depths, the local harmonic thermal-wave
amplitude is reduced by the factor \(e^{-4}\simeq 1.8\times10^{-2}\) before
the additional geometric attenuation is considered. The same four-depth
padding was applied around the active region in the orthogonal lateral
direction. Equation~
\eqref{eq:supp_comsol_lateral_size} was therefore used for the production
COMSOL models as a domain-design estimate. It does not by itself establish that the
truncating boundaries perturb the retained signal by less than the
reported numerical discrepancy.  When multiple
frequencies were considered, the domain was sized using the lowest frequency,
which gives the largest \(\delta_{\mathrm{lat}}\).  Separate low-frequency
domain-size tests and the consequences of smaller domains are reported in
Supplementary Section~\ref{sec:supp_edge_effects}.

\subsection{Adaptive Mesh and Solver Settings}

The mesh was refined most strongly in the region containing the large
temperature gradients.  A hemispherical refinement region of radius \(5r\)
was centered beneath the pump spot.  Within this region, the maximum element
size was varied from \(r/10\) in the finest mesh to \(r/3\) in the coarser
mesh-convergence tests.  The remaining domains used COMSOL's
physics-controlled \emph{Finer} mesh.  Adaptive refinement was then applied
to the frequency-domain solution, with refinement concentrated near the pump,
the thin film, and the vertical and horizontal interfaces.  The setup notes describe surface-profile checks after refinement, but
the retained summary does not supply numerical solver tolerances or a
per-case table of mesh-refinement changes. Consequently, these settings
alone do not certify that mesh error is smaller than the reported GBIE--FEM
discrepancy.

\begin{table}[htbp]
    \centering
    \small
    \caption{Summary of the three-dimensional COMSOL setup used for the
    numerical validation and expanded five-case performance audit.}
    \label{tab:supp_comsol_settings}
    \begin{tabular}{p{0.29\linewidth}p{0.63\linewidth}}
        \toprule
        Setting & Implementation \\
        \midrule
        Software and resource & COMSOL Multiphysics 6.2.0.415 on University of
        Pittsburgh CRC Slurm nodes with AMD EPYC 9374F processors; one CPU
        core used and a \(64~\mathrm{GB}\) job-memory allocation \\
        Physics and study & Heat Transfer in Solids, three dimensions;
        Frequency Domain at the prescribed modulation frequency \(f\) \\
        Material model & Orthotropic
        \(\mathbf{k}=\operatorname{diag}(k_x,k_y,k_z)\) and volumetric heat
        capacity \(C_v\) in every region \\
        Heat source & Gaussian surface heat flux with \(1/e^2\) radius \(r\),
        absorbed modulated power and center position matched to the GBIE model;
        the expanded five-case audit uses \(1~\mathrm{mW}\) \\
        Interface conditions & \(q_n=G_v\Delta T\) at the vertical interface
        and \(q_n=G_h\Delta T\) at horizontal film--substrate interfaces \\
        Outer boundaries & Unheated top surface; rear surface adiabatic
        unless stated otherwise; lateral truncation faces placed according
        to the domain-size criterion below \\
        Lateral extent & Four \(\delta_{\mathrm{lat}}\) of padding beyond each
        end of the retained data region; \(L_y\geq L_e+8\delta_{\mathrm{lat}}\) \\
        Near-field mesh & Hemispherical refinement region of radius \(5r\), with
        \(\Delta_{\mathrm{mesh,max}}=r/10\) to \(r/3\) in the mesh-convergence study \\
        Far-field mesh & Physics-controlled \emph{Finer} mesh with adaptive
        refinement \\
        Solver and convergence & Frequency-domain perturbation equations,
        quadratic Lagrange shape functions, and a stationary linear solution;
        surface profiles checked after mesh refinement \\
        Output & Complex surface temperature along the pump-center line;
        amplitude and phase exported at the same coordinates as the GBIE result \\
        \bottomrule
    \end{tabular}
\end{table}

\subsection{\texorpdfstring{Finite-Depth \(G_v(z)\) Interface on a
Connected Support}{Finite-Depth Gv(z) Interface on a Connected Support}}
\label{sec:supp_finite_depth_connected_support}

The finite-depth vertical interface with depth-dependent \(G_v(z)\) on a
connected support featured in the main text represents a local cross-section
through two laterally patterned device regions with different effective
thermal conductivities. The \(300\times300\times300~\mu\mathrm{m}^3\)
COMSOL model contains adjacent \(2~\mu\mathrm{m}\) top-layer sections,
dissimilar upper regions, and a connected support below
\(z_t=50~\mu\mathrm{m}\). The vertical boundary crosses the top-layer
junction and continues between the upper regions, then terminates where the
two regions merge into one connected support. Thus, there is no vertical
thermal-contact boundary for \(z>z_t\).
The \(300~\mu\mathrm{m}\) rear surface is held at the model reference
temperature. The \(1~\mathrm{mW}\) Gaussian source has \(r=1~\mu\mathrm{m}\),
is centered at \(y/r=2\), and is modulated at \(100~\mathrm{kHz}\).
The saved COMSOL model uses its \(x\) coordinate across the vertical
interface and its \(y\) coordinate along it, whereas the manuscript uses
\(x\) along and \(y\) across. Accordingly, the saved
\((k_{1x},k_{1y},k_{1z})=(60,75,50)\) and
\((k_{2x},k_{2y},k_{2z})=(90,55,30)\) map to the manuscript-coordinate
tensors listed below.

\begin{table}[htbp]
    \centering
    \small
    \caption{Physical inputs for the finite-depth,
    depth-dependent-\(G_v(z)\) benchmark on a connected support, as stored in the
    authoritative COMSOL model. All regions use the common normalized heat
    capacity.}
    \label{tab:supp_finite_depth_connected_support_inputs}
    \begin{tabular}{p{0.34\linewidth}p{0.56\linewidth}}
        \toprule
        Input & Value \\
        \midrule
        Surface sections &
        Adjacent top-layer sections, \(h_f=2~\mu\mathrm{m}\),
        \(k_x=k_y=k_z=100~\mathrm{W\,m^{-1}\,K^{-1}}\) \\
        Heated upper region &
        \((k_x,k_y,k_z)=(75,60,50)~
        \mathrm{W\,m^{-1}\,K^{-1}}\), \(2<z<50~\mu\mathrm{m}\) \\
        Unheated upper region &
        \((k_x,k_y,k_z)=(55,90,30)~
        \mathrm{W\,m^{-1}\,K^{-1}}\), \(2<z<50~\mu\mathrm{m}\) \\
        Connected support &
        \(k_x=k_y=k_z=60~
        \mathrm{W\,m^{-1}\,K^{-1}}\), \(50<z\leq300~\mu\mathrm{m}\) \\
        Volumetric heat capacities &
        \(C_v=10^6~\mathrm{J\,m^{-3}\,K^{-1}}\) in every region \\
        Horizontal conductances &
        \(G_h=10^8~\mathrm{W\,m^{-2}\,K^{-1}}\) at the
        film--upper-region and both upper-region--support contacts \\
        Vertical interface &
        \(z_{\min}=0\), \(z_t=50~\mu\mathrm{m}\); the two upper regions
        merge into the connected support for \(z>z_t\) \\
        Pump &
        \(P_0=1~\mathrm{mW}\), \(r=1~\mu\mathrm{m}\),
        \(y_0/r=2\), \(f=100~\mathrm{kHz}\) \\
        COMSOL domain &
        \(300\times300\times300~\mu\mathrm{m}^3\) \\
        GBIE discretization and cutoffs &
        \(N_u/N_v/N_z=35/120/25\), \(U/V=30/20\) \\
        \bottomrule
    \end{tabular}
\end{table}

The COMSOL vertical contact uses
\begin{equation}
G_v(z)=
\begin{cases}
10^8, & 0\le z<2~\mu\mathrm{m},\\
10^7, & 2\le z<5~\mu\mathrm{m},\\
2\times10^8, & 5\le z\le50~\mu\mathrm{m},
\end{cases}
\qquad
\left[\mathrm{W\,m^{-2}\,K^{-1}}\right].
    \label{eq:supp_finite_depth_conductance_profile}
\end{equation}
The GBIE inserts \(z=2\), \(5\), and \(50~\mu\mathrm{m}\) as protected
depth-grid boundaries. The third and final conductance,
\(2\times10^8~\mathrm{W\,m^{-2}\,K^{-1}}\), applies only over
\(5\leq z\leq50~\mu\mathrm{m}\). At \(z_t=50~\mu\mathrm{m}\), the
vertical contact terminates; for \(z>z_t\), the two upper regions are
replaced by one connected support, so no fourth physical contact conductance is defined. The GBIE nevertheless
retains an auxiliary coupling plane with $R_v=0$ in the support down to
$z_{\mathrm{int,max}}=192.73~\mu\mathrm{m}$ for this saved case. The \(250~\mu\mathrm{m}\)-thick support spans approximately
\(18\) support penetration depths at \(100~\mathrm{kHz}\), so the GBIE
semi-infinite continuation is expected to approximate the finite COMSOL rear closely on thermal-length
grounds. A dedicated fixed-input rear-depth sweep is needed to quantify
the remaining difference.

The comparison uses COMSOL export job 23547982. With
\(N_u/N_v/N_z=35/120/25\) and \(U/V=30/20\), the GBIE gives a
\(0.055^\circ\) P95 absolute phase residual, a \(1.5\%\) P95
floor-regularized pointwise phase-percentage residual, a \(0.17\%\)
robust-span-normalized phase RMSE, and a \(0.70\%\) amplitude
relative-\(L_2\) error. The cutoff screen in
\Cref{tab:supp_finite_depth_connected_support_cutoffs} selected \(30/20\)
because it gives the lowest P95 phase residual while retaining sub-percent
amplitude error.

\begin{table}[htbp]
    \centering
    \small
    \caption{Source-cutoff screen for the finite-depth connected-support
    benchmark at fixed \(N_u/N_v/N_z=35/120/25\), compared with the same
    COMSOL export (job 23547982).}
    \label{tab:supp_finite_depth_connected_support_cutoffs}
    \begin{tabular}{ccc}
        \toprule
        \(U/V\) & \(\mathrm{P95}\,|e_\phi|\) (deg) &
        Amplitude relative-\(L_2\) error (\%) \\
        \midrule
        \(10/20\) & 0.173057 & 0.394899 \\
        \(20/20\) & 0.103443 & 0.496691 \\
        \(20/50\) & 0.099488 & 0.617593 \\
        \(30/20\) & 0.055153 & 0.696067 \\
        \(30/50\) & 0.071713 & 0.747789 \\
        \(40/50\) & 0.081174 & 0.972879 \\
        \bottomrule
    \end{tabular}
\end{table}

COMSOL job 23547982 used a detached copy of the authoritative saved model.
For compatibility with the present three-region \(G_v(z)\) profile, only the
obsolete output-metadata field \texttt{G\_cond\_vert\_4} was removed. The
geometry, mesh, physics, study, solver, and all physical inputs were
unchanged.

\subsection{\texorpdfstring{Supplementary Rear-Surface Four-Level
\(G_v(z)\) Benchmark}{Supplementary Rear-Surface Four-Level Gv(z) Benchmark}}
\label{sec:supp_depth_dependent_gv}

The supplementary rear-surface comparison uses a
\(0.5~\mu\mathrm{m}\) transducer film, \(r=1~\mu\mathrm{m}\),
\(f=100~\mathrm{kHz}\), and a
\(300\times300\times300~\mu\mathrm{m}^3\) COMSOL domain. The
\(1~\mathrm{mW}\) Gaussian heat source is centered at \(y/r=2\).
All results in this supplementary benchmark use the same COMSOL export
(job 23523159); the cutoff study below changes only the GBIE source cutoffs.
\Cref{tab:supp_depth_dependent_inputs} lists the remaining physical and
numerical inputs. In the manuscript coordinate system, \(x\) is parallel
to the vertical interface, \(y\) is normal to it, and \(z\) is the
through-depth direction.

\begin{table}[htbp]
    \centering
    \small
    \caption{Physical and numerical inputs for the supplementary
    rear-surface four-level depth-dependent-\(G_v(z)\) comparison
    (COMSOL job 23523159,
    \(f=100~\mathrm{kHz}\)).}
    \label{tab:supp_depth_dependent_inputs}
    \begin{tabular}{p{0.34\linewidth}p{0.56\linewidth}}
        \toprule
        Input & Value \\
        \midrule
        Film &
        \(h_f=0.5~\mu\mathrm{m}\),
        \(k_x=k_y=k_z=35~\mathrm{W\,m^{-1}\,K^{-1}}\) \\
        Heated substrate &
        \((k_x,k_y,k_z)=(27,30,30)~
        \mathrm{W\,m^{-1}\,K^{-1}}\) \\
        Unheated substrate &
        \((k_x,k_y,k_z)=(25,30,30)~
        \mathrm{W\,m^{-1}\,K^{-1}}\) \\
        Volumetric heat capacities &
        \(C_{v,f}=C_{v,1}=C_{v,2}
        =10^6~\mathrm{J\,m^{-3}\,K^{-1}}\) \\
        Horizontal conductance &
        \(G_h=10^8~\mathrm{W\,m^{-2}\,K^{-1}}\) \\
        Pump &
        \(P_0=1~\mathrm{mW}\), \(r=1~\mu\mathrm{m}\),
        \(y_0/r=2\), \(f=100~\mathrm{kHz}\) \\
        COMSOL domain &
        \(300\times300\times300~\mu\mathrm{m}^3\) \\
        GBIE quadrature &
        \(N_u/N_v/N_z=35/120/25\) \\
        Source cutoffs &
        \(U/V=20/50\), giving
        \(\xi_{\max}=2.0\times10^7~\mathrm{m^{-1}}\) and
        \(\eta_{\max}=5.0\times10^7~\mathrm{m^{-1}}\) \\
        Interface integration depth &
        \(z_{\mathrm{int,max}}=52.8~\mu\mathrm{m}\) \\
        \bottomrule
    \end{tabular}
\end{table}

COMSOL evaluates the vertical-contact condition with the piecewise function
\begin{equation}
G_v(z)=
\begin{cases}
10^8, & 0\le z<0.5~\mu\mathrm{m},\\
10^7, & 0.5\le z<5~\mu\mathrm{m},\\
2\times10^8, & 5\le z<50~\mu\mathrm{m},\\
10^{10}, & 50\le z\le300~\mu\mathrm{m},
\end{cases}
\qquad
\left[\mathrm{W\,m^{-2}\,K^{-1}}\right].
    \label{eq:supp_rear_surface_conductance_profile}
\end{equation}
Thus the fourth conductance applies from the final transition at
\(50~\mu\mathrm{m}\) to the rear surface at \(z=300~\mu\mathrm{m}\).
The GBIE reads the same four conductances and three transition depths from the
COMSOL metadata, applies \(R_v(z)=1/G_v(z)\), and inserts every transition
as an exact boundary of the composite depth quadrature. Thus no quadrature
interval averages across a conductance discontinuity.

\begin{table}[htbp]
    \centering
    \small
    \caption{GBIE source-cutoff comparison against the same
    \(100~\mathrm{kHz}\) COMSOL export (job 23523159), at fixed
    \(N_u/N_v/N_z=35/120/25\). The selected \(20/50\) setting is used
    for this supplementary profile comparison.}
    \label{tab:supp_depth_dependent_cutoffs}
    \begin{tabular}{cccc}
        \toprule
        \(U/V\) & \(\mathrm{P95}\,|e_\phi|\) (deg) &
        \shortstack{P95 floor-regularized\\phase-percentage residual (\%)} &
        Amplitude \(L_2\) (\%) \\
        \midrule
        \(10/50\) & 0.157 & 1.145 & 0.366 \\
        \(20/20\) & 0.149 & 0.956 & 0.460 \\
        \(20/50\) & 0.066 & 0.807 & 0.474 \\
        \(30/50\) & 0.066 & 0.801 & 0.476 \\
        \bottomrule
    \end{tabular}
\end{table}

The lower residual at \(V=50\) is consistent with the interface-normal
\(v\)-integration requiring broader spectral support. Splitting the Gaussian
source at \(y=0\), together with the sharp field variation across the
vertical interface, produces a more demanding high-\(v\) tail than the full
unsplit Gaussian. At fixed \(N_v\), increasing \(V\) from 20 to 50 reduces
the observed COMSOL--GBIE \(\mathrm{P95}\,|e_\phi|\). This trend is consistent
with improved representation of the high-\(v\) tail, but it does not by itself
establish asymptotic convergence in \(V\). At \(V=50\), increasing \(U\) from
20 to 30 changes \(\mathrm{P95}\,|e_\phi|\) by less than
\(0.001^\circ\), showing insensitivity to the tested increase in \(U\). This cutoff
effect is distinct from resolving the discontinuities in \(G_v(z)\), which
is handled by the protected composite depth quadrature described above.
Because a larger cutoff does not increase quadrature order, the adopted
\(U/V\) pair is a case-specific comparison rather than a rule that
larger cutoffs are always more accurate.

\begin{figure}[tbp]
    \centering
    \includegraphics[width=\linewidth]{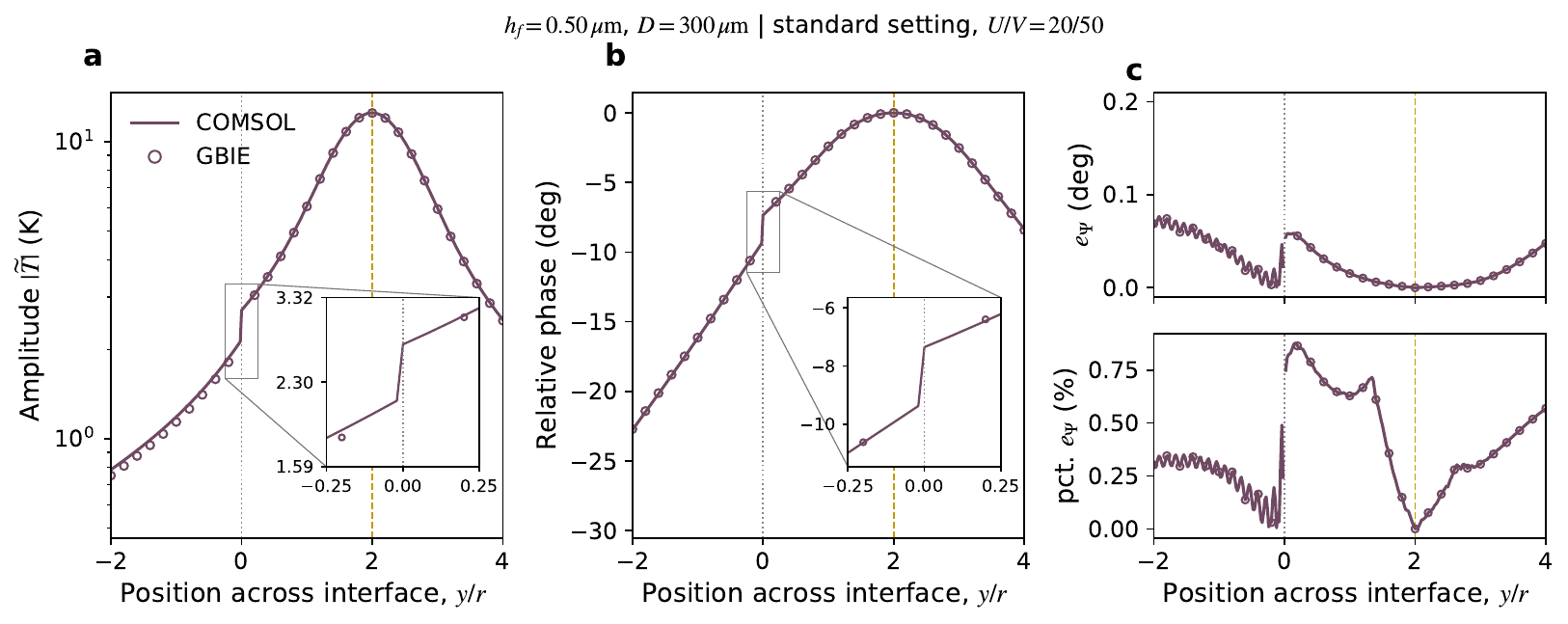}
    \caption{Supplementary COMSOL--GBIE comparison for the rear-surface
    four-level depth-dependent-\(G_v(z)\) benchmark
    (COMSOL job 23523159). The conductances
    are \(10^8\), \(10^7\), \(2\times10^8\), and
    \(10^{10}~\mathrm{W\,m^{-2}\,K^{-1}}\), with transitions at
    \(z=0.5\), \(5\), and \(50~\mu\mathrm{m}\); the fourth value extends
    from \(50~\mu\mathrm{m}\) to the \(300~\mu\mathrm{m}\) rear surface.
    Solid curves are COMSOL and open circles are GBIE results.}
    \label{fig:supp_rear_surface_four_level_gv}
\end{figure}

\section{COMSOL--GBIE Accuracy and Computational-Cost Audit}
\label{sec:supp_comsol_bie_five_case}

This section reports the expanded comparison used for the performance
statements in the main text. Five COMSOL executions are denoted Jobs~1--5.
Jobs~1 and 2 use identical physical inputs and exported profiles and are
retained as repeat runs for evaluating workflow-time stability. The study
therefore contains five executions and four unique physical profiles.
These five source-cutoff-cohort executions use \(f=10~\si{\kilo\hertz}\), a pump radius of
\(r=1~\si{\micro\metre}\), a \(1~\si{\micro\metre}\) surface film,
\(C_v=10^6~\si{\joule\per\cubic\metre\per\kelvin}\) in every region, and a
pump center \(2~\si{\micro\metre}\) from the vertical interface. The
absorbed pump power is \(1~\si{\milli\watt}\), and
\(G_h=10^8~\si{\watt\per\square\metre\per\kelvin}\) in every job.
Jobs~1--2 use \(G_v=10^7~\si{\watt\per\square\metre\per\kelvin}\), whereas
Jobs~3--5 use \(G_v=10^8~\si{\watt\per\square\metre\per\kelvin}\). The GBIE
uses the standard efficient setting,
\begin{equation}
    (N_u,N_v,N_z)=(35,120,25),
    \label{eq:supp_comsol_audit_quadrature}
\end{equation}
and each COMSOL and GBIE execution is restricted to one CPU core.
An independent exact-node series with \(500\), \(200\), and
\(100~\si{\nano\metre}\) films is reported separately below. It was evaluated
only at the adopted \(U/V=10/20\) setting and is excluded from the five-job
source-cutoff ranking, aggregate statistics, and duplicate-aware means.

\subsection{Source-cutoff sensitivity and accuracy}
\label{sec:supp_comsol_bie_five_case_accuracy}

The source-cutoff pairs \(U/V=10/10\), \(10/20\), and \(20/20\) were evaluated
for every COMSOL profile.  Before comparison, the phase of each solution was
referenced to zero at the pump center.  Let \(\mathcal{M}\) denote the sampled
points in \(-2\leq y/r\leq4\), excluding the point immediately on the
discontinuous vertical interface.  The primary phase metric is the root-mean-
square phase residual normalized by the robust COMSOL phase span,
\begin{equation}
  E_{\phi}
  =
  100\,
  \frac{
    \left[
      |\mathcal{M}|^{-1}
      \sum_{i\in\mathcal{M}}
      \left(\phi_{\mathrm{GBIE},i}-\phi_{\mathrm{COMSOL},i}\right)^2
    \right]^{1/2}
  }{
    P_{95}\!\left(\phi_{\mathrm{COMSOL}}\right)
    -
    P_{5}\!\left(\phi_{\mathrm{COMSOL}}\right)
  } .
  \label{eq:supp_comsol_bie_phase_nrmse}
\end{equation}
This normalization remains meaningful when the phase crosses zero and avoids
allowing a single point with a small denominator to dominate the score.  The
amplitude metric is the relative \(L_2\) error
\begin{equation}
  E_A
  =
  100\,
  \frac{
    \left\|\mathbf{A}_{\mathrm{GBIE}}-\mathbf{A}_{\mathrm{COMSOL}}\right\|_2
  }{
    \left\|\mathbf{A}_{\mathrm{COMSOL}}\right\|_2
  } .
  \label{eq:supp_comsol_bie_amplitude_l2}
\end{equation}
Here $\operatorname{RMSE}$ denotes the root-mean-square error, $P_p$ the
$p$th percentile, $\|\cdot\|_2$ the Euclidean norm, and $\mathbf A$ the
vector of amplitudes at the comparison points. The robust span $P_{95}-P_5$
scales $E_\phi$ by the phase variation in the reference profile.

\begin{table}[htbp]
  \centering
  \caption{Aggregate source-cutoff sensitivity over the five COMSOL jobs.
  The unique-profile mean counts the two identical \(250~\si{\micro\metre}\)
  profiles once.}
  \label{tab:supp_comsol_bie_uv_aggregate}
  \small
  \begin{tabular}{lccccc}
    \toprule
    \(U/V\) &
    mean \(E_\phi\) &
    median \(E_\phi\) &
    unique mean \(E_\phi\) &
    mean \(E_A\) &
    unique mean \(E_A\) \\
    & \multicolumn{5}{c}{(\%)} \\
    \midrule
    \(10/10\) & 0.644 & 0.778 & 0.708 & \textbf{0.473} & \textbf{0.493} \\
    \(10/20\) & \textbf{0.479} & 0.576 & \textbf{0.536} & 0.564 & 0.592 \\
    \(20/20\) & 0.512 & \textbf{0.570} & 0.581 & 0.643 & 0.690 \\
    \bottomrule
  \end{tabular}
\end{table}

At the adopted \(U/V=10/20\) setting, the detailed per-job accuracy values are
listed in Table~\ref{tab:supp_comsol_bie_adopted_accuracy}. The two
\(250~\si{\micro\metre}\) rows are identical because Jobs~1 and 2 use the same
physical inputs and exported profile.

\begin{table}[htbp]
  \centering
  \caption{Per-job accuracy for the \(1~\si{\micro\metre}\)-film
  domain/property audit at the adopted \(U/V=10/20\) setting.}
  \label{tab:supp_comsol_bie_adopted_accuracy}
  \small
  \begin{tabular}{lrrrr}
    \toprule
    Job & Width & \(E_\phi\) & \(E_A\) &
    P95 \(\lvert e_\phi\rvert\) \\
    & (\(\si{\micro\metre}\)) & (\%) & (\%) & (deg) \\
    \midrule
    1 & 250 & 0.253 & 0.454 & 0.043 \\
    2 & 250 & 0.253 & 0.454 & 0.043 \\
    3 & 300 & 0.576 & 0.538 & 0.070 \\
    4 & 400 & 0.594 & 0.648 & 0.047 \\
    5 & 500 & 0.721 & 0.728 & 0.049 \\
    \bottomrule
  \end{tabular}
\end{table}

The \(10/20\) pair gives the lowest mean phase error both across all five jobs
and after duplicate-profile removal, and is therefore used as the default for
the computational-cost comparison.  The result is an average-based choice,
not a claim that \(10/20\) is the pointwise optimum for every profile:
\(20/20\) gives the smallest phase error for the repeated \(250~\si{\micro
\metre}\) profile and the \(300~\si{\micro\metre}\) profile, whereas \(10/20\)
wins for the \(400\) and \(500~\si{\micro\metre}\) profiles.  The \(10/10\)
pair gives the lowest mean amplitude error but a substantially larger mean
phase error.  These multilayer results complement, rather than replace, the
broader homogeneous-case cutoff investigation in Supplementary Section~S4.

\begin{figure}[htbp]
  \centering
  \includegraphics[width=\linewidth]{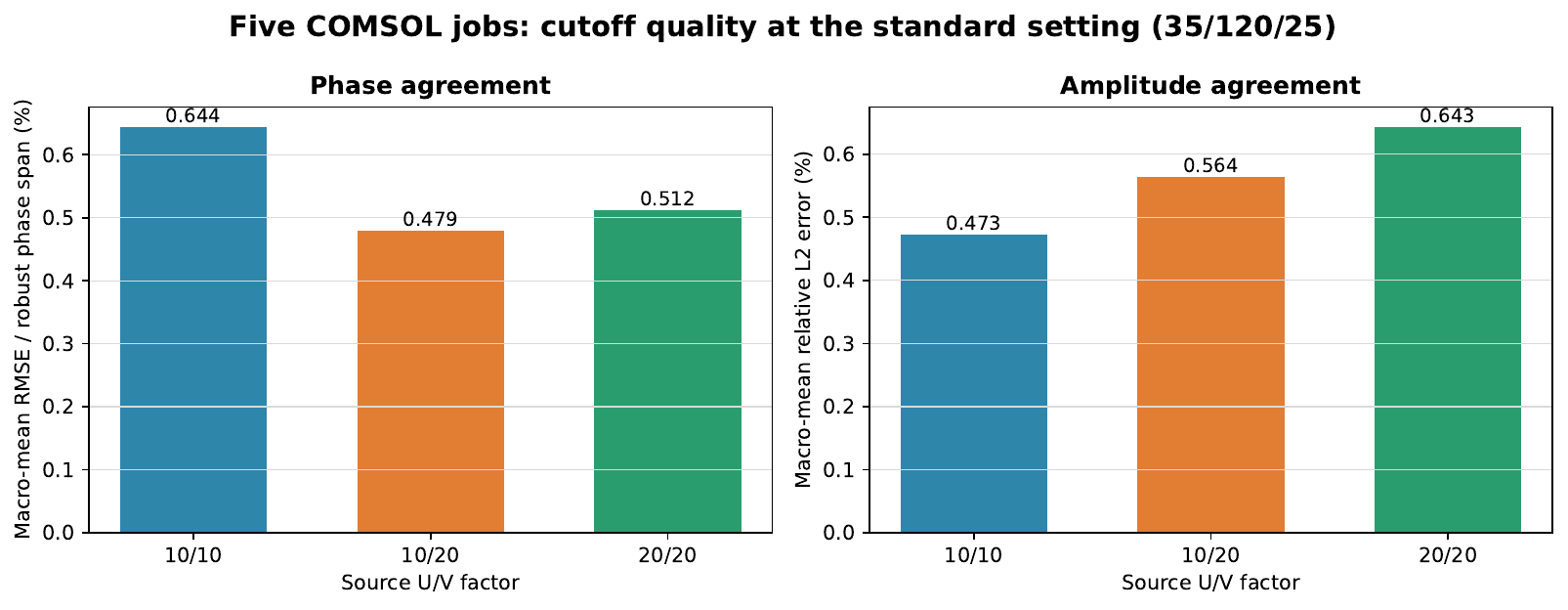}
  \caption{Aggregate GBIE source-cutoff comparison against the five COMSOL
  jobs.  Phase accuracy is ranked using
  \cref{eq:supp_comsol_bie_phase_nrmse}; amplitude accuracy uses
  \cref{eq:supp_comsol_bie_amplitude_l2}.  The duplicate-aware result confirms
  that the \(10/20\) phase recommendation is not caused by counting the
  repeated \(250~\si{\micro\metre}\) profile twice.}
  \label{fig:supp_comsol_bie_uv_aggregate}
\end{figure}

\subsection{Runtime and resource scaling}
\label{sec:supp_comsol_bie_five_case_efficiency}

For COMSOL, ``complete time'' is the logged elapsed time for opening the
model, running it, and saving the result.  ``Model run'' contains equation
compilation and mesh preparation, dependent-variable setup, the stationary
linear solve, and result evaluation; ``solver'' is the stationary solve alone.
For the GBIE, ``complete time'' is measured outside a fresh Python process and
therefore includes interpreter startup, input parsing, GBIE assembly and
solution, profile reconstruction, metric evaluation, and output writing.
The COMSOL values are its logged physical-memory peaks, whereas GBIE
values are the Python process maximum resident set size. Both are converted
to GiB. These reporters have different scopes; a common process-tree memory
measurement was not used for the published ratios.  The one-thread
environment and CPU-time/wall-time ratios confirm that the GBIE calculation
used one core.

\begin{table}[htbp]
  \centering
  \caption{Per-job one-core computational cost at the adopted
  \(U/V=10/20\). Both workflows ran in the University of Pittsburgh CRC
  \texttt{smp} partition on one core of an AMD EPYC 9374F processor. COMSOL
  memory values were converted from reported decimal GB to GiB before forming
  the ratios. Jobs~1 and 2 provide the repeat-run timing check.}
  \label{tab:supp_comsol_bie_performance}
  \begingroup
  \scriptsize
  \setlength{\tabcolsep}{1.8pt}
  \begin{tabular}{@{}lrrrrrrrrr@{}}
    \toprule
    Job & Width & Solved+internal DOFs &
    \multicolumn{3}{c}{COMSOL time (s)} &
    GBIE complete & Time ratio &
    COMSOL/GBIE RSS & RSS ratio \\
    & (\(\si{\micro\metre}\)) & (\(10^6\)) &
    Complete & Model run & Solver & (s) &
    & (GiB) & \\
    \midrule
    1 & 250 & \(4.33+1.15\) & 264 & 214 & 174 & 8.64 & 31 & \(9.31/0.0775\) & 120 \\
    2 & 250 & \(4.33+1.15\) & 250 & 217 & 176 & 8.75 & 29 & \(9.41/0.0742\) & 127 \\
    3 & 300 & \(5.73+1.58\) & 432 & 368 & 236 & 7.20 & 60 & \(12.85/0.0742\) & 173 \\
    4 & 400 & \(9.20+2.70\) & 681 & 603 & 388 & 7.90 & 86 & \(20.30/0.0747\) & 272 \\
    5 & 500 & \(13.71+4.13\) & 1022 & 940 & 621 & 6.92 & 148 & \(30.44/0.0742\) & 410 \\
    \bottomrule
  \end{tabular}
  \endgroup
\end{table}

Across the domain-width sweep, COMSOL complete time grows from \(250\)--\(264\)
to \(1022~\si{\second}\), while GBIE complete time remains between \(6.92\) and
\(8.75~\si{\second}\).  The corresponding processor-class-controlled one-core
time ratio grows from \(29\)--\(31\) to \(148\).  The internal GBIE
pipeline time, which excludes Python startup and external timing overhead, is
\(4.25\)--\(4.34~\si{\second}\).  COMSOL peak physical memory grows from
\(9.31\)--\(9.41\) to \(30.44~\mathrm{GiB}\), whereas the isolated GBIE
process remains between \(0.074\) and \(0.078~\mathrm{GiB}\), giving memory
ratios of \(120\)--\(410\).  This contrasting scaling follows from the numerical
representations: increasing the finite domain enlarges the three-dimensional
COMSOL mesh, but it does not change the standard GBIE quadrature dimensions.

\begin{figure}[htbp]
  \centering
  \includegraphics[width=\linewidth]{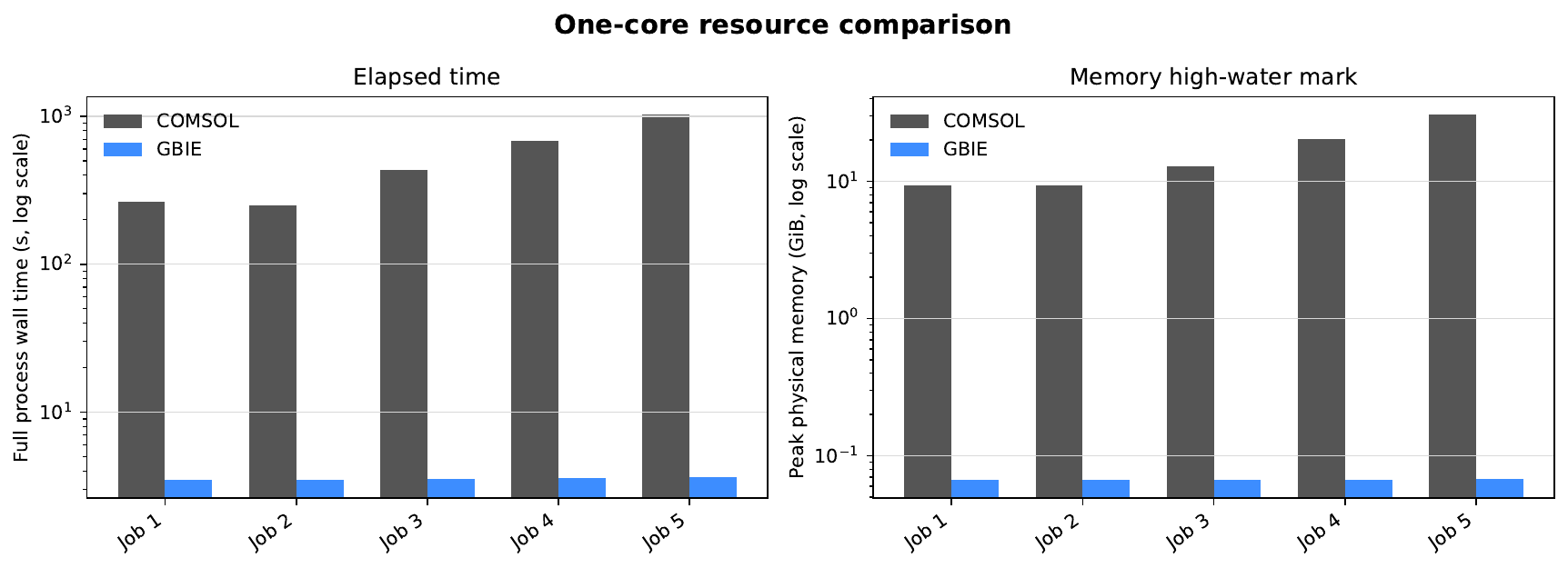}
  \caption{Complete-process runtime, peak physical memory, and observed
  COMSOL/GBIE ratios for the five one-core jobs. The ratios show the measured
  workflow scaling with processor model, partition, and core count controlled.}
  \label{fig:supp_comsol_bie_runtime_resources}
\end{figure}

\subsection{Independent exact-node thin-film series}
\label{sec:supp_comsol_bie_500nm_check}

The three additional calculations differ only in surface-film thickness:
\(500\), \(200\), or \(100~\si{\nano\metre}\). All source, domain, material,
and interface properties listed below are common to the three thicknesses.
They are
\(f=10~\si{\kilo\hertz}\), \(r=1~\si{\micro\metre}\), and a
\(300\times300\times300~\si{\micro\metre}^3\) domain. The absorbed pump
power is \(1~\si{\milli\watt}\), and its center is
\(2~\si{\micro\metre}\) from the vertical interface. In \((x,y,z)\) order,
the conductivity vectors are
\(\boldsymbol{k}_a=(30,27,30)\),
\(\boldsymbol{k}_b=(30,25,30)\), and
\(\boldsymbol{k}_f=(35,35,35)~
\si{\watt\per\metre\per\kelvin}\).
Every volumetric heat capacity is
\(10^6~\si{\joule\per\cubic\metre\per\kelvin}\), and
\(G_v=G_h=10^8~\si{\watt\per\square\metre\per\kelvin}\).
Each COMSOL/GBIE pair used one-core allocations on the same physical AMD EPYC
9374F node. All three GBIE calculations use
\((N_u,N_v,N_z)=(35,120,25)\), and only \(U/V=10/20\) was evaluated, so the
series checks the adopted setting but is excluded from the five-job cutoff
ranking and does not define a new source-cutoff rule. Its common anisotropic
property set is distinct from that of the \(1~\si{\micro\metre}\)-film Job~3.
The film-relative composite depth grid places six, six, and five of the
requested 25 nodes inside the \(500\), \(200\), and
\(100~\si{\nano\metre}\) films, respectively. Every film/substrate boundary
is an exact quadrature breakpoint.

\begin{table}[htbp]
  \centering
  \caption{Accuracy, numerical size, and exact-node resource use for the
  independent thin-film series at \(U/V=10/20\). Slash-separated COMSOL times
  are opening/running/saving \(=\) complete time; GBIE times are
  complete/internal-pipeline/solve. COMSOL memory is its reported peak
  physical memory converted from decimal GB to GiB; GBIE memory is the
  GNU-\texttt{time} peak resident set. Time/RSS entries are the corresponding
  COMSOL/GBIE ratios.}
  \label{tab:supp_comsol_bie_500nm_check}
  \begingroup
  \small
  \setlength{\tabcolsep}{4pt}
  \begin{tabular}{@{}rrrrrrr@{}}
    \toprule
    \multicolumn{7}{c}{\textit{Accuracy and timing}} \\
    \cmidrule(lr){1-7}
    Film &
    \(E_\phi\) &
    \(E_A\) &
    P95 \(\lvert e_\phi\rvert\) &
    COMSOL time &
    Solver &
    GBIE time \\
    (\(\si{\nano\metre}\)) &
    (\%) &
    (\%) &
    (deg) &
    (s) &
    (s) &
    (s) \\
    \midrule
    500 & 0.60 & 0.51 & 0.067 &
      \(32/1000/67=1099\) & 648 & \(6.05/4.27/4.19\) \\
    200 & 0.57 & 0.51 & 0.069 &
      \(35/1122/44=1201\) & 683 & \(5.90/4.26/4.21\) \\
    100 & 0.58 & 0.48 & 0.071 &
      \(16/1155/72=1244\) & 691 & \(5.85/4.25/4.21\) \\
    \bottomrule
  \end{tabular}

  \vspace{0.8em}

  \begin{tabular}{@{}rrrrr@{}}
    \toprule
    \multicolumn{5}{c}{\textit{Numerical size and resources}} \\
    \cmidrule(lr){1-5}
    Film &
    RSS &
    Time/RSS &
    Elements &
    Solved+internal DOFs \\
    (\(\si{\nano\metre}\)) &
    (GiB) &
    ratios &
    &
    \\
    \midrule
    500 & \(32.94/0.076\) & \(182/435\) &
      \(9\,519\,839\) & \(14\,906\,865+4\,478\,996\) \\
    200 & \(33.29/0.075\) & \(204/445\) &
      \(9\,521\,011\) & \(14\,912\,977+4\,489\,968\) \\
    100 & \(33.07/0.075\) & \(213/439\) &
      \(9\,609\,009\) & \(15\,076\,751+4\,597\,108\) \\
    \bottomrule
  \end{tabular}
  \endgroup
\end{table}

Across the series, the complete COMSOL application time is
\(1099\)--\(1244~\si{\second}\), whereas the plot-free GBIE process requires
\(6.05\), \(5.90\), and \(5.85~\si{\second}\) for the \(500\), \(200\), and
\(100~\si{\nano\metre}\) films, respectively. The resulting complete-time
ratios are \(182\), \(204\), and \(213\), while the peak-memory ratios remain
between \(435\) and \(445\). For the \(100~\si{\nano\metre}\) mesh,
\(q_{\min}=0.007744\); COMSOL issued a low-minimum-element-quality warning but
completed the solve successfully. The plotted and timed GBIE runs were
separated so that PDF rendering does not enter the reported workflow time.

\begin{figure}[htbp]
  \centering
  \includegraphics[width=0.82\linewidth]{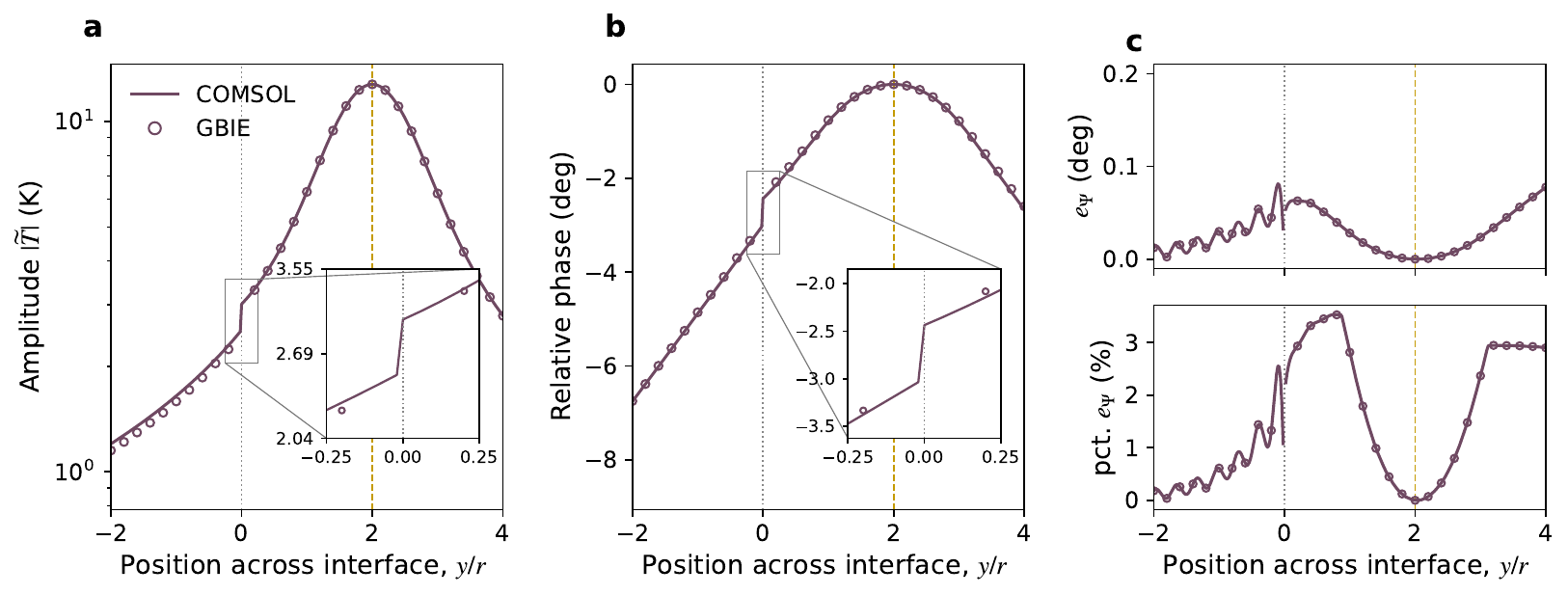}\par\vspace{1em}
  \includegraphics[width=0.82\linewidth]{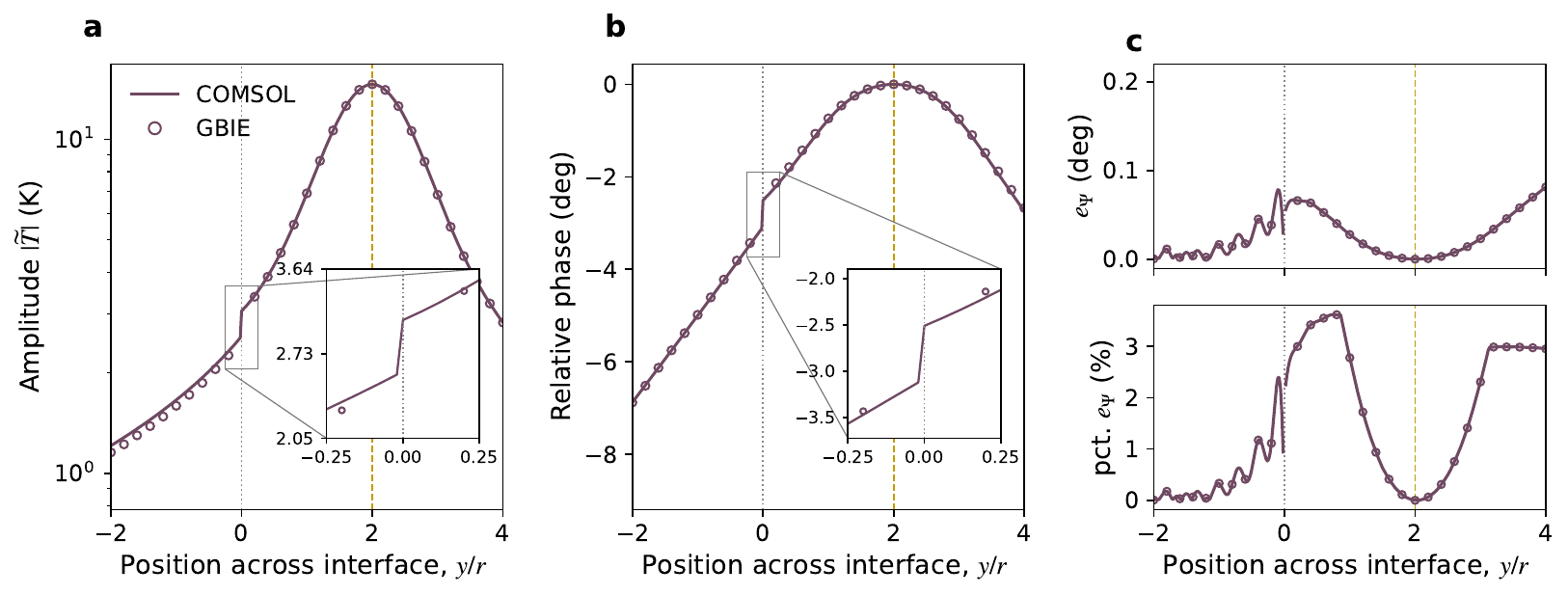}\par\vspace{1em}
  \includegraphics[width=0.82\linewidth]{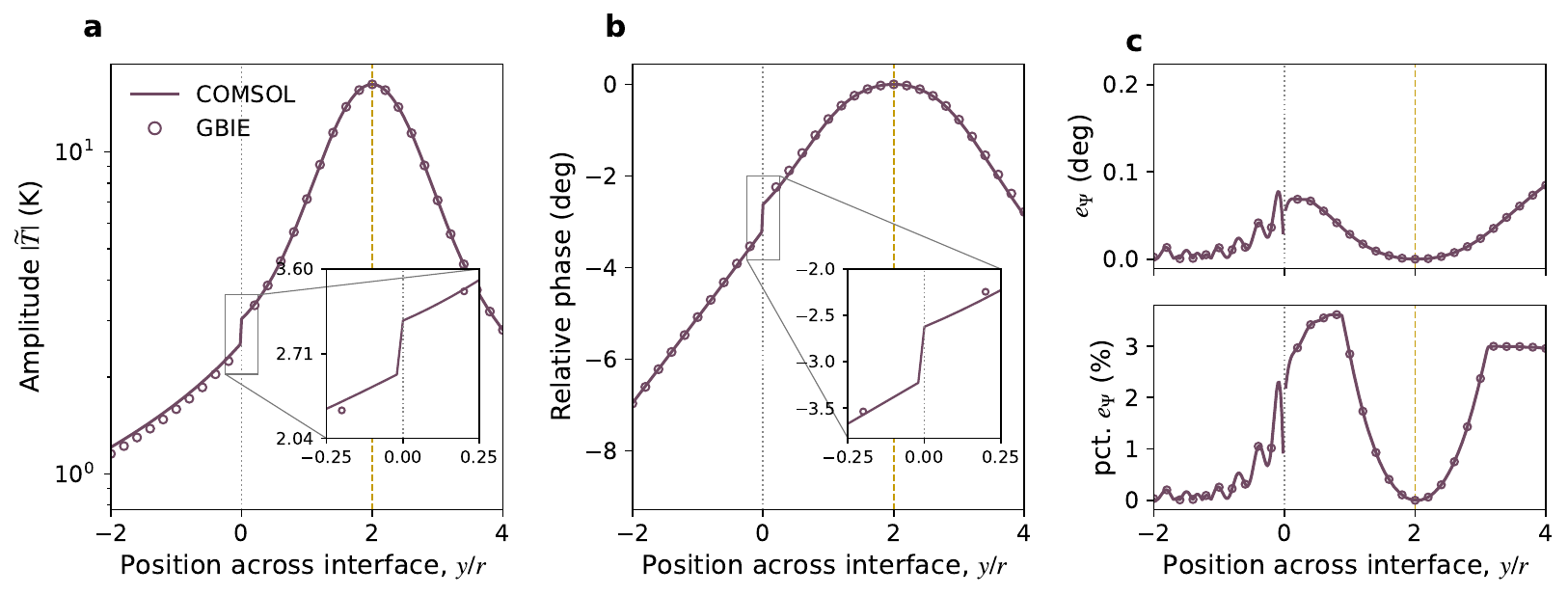}
  \caption{Independent exact-node thin-film comparisons in the
  \(300~\si{\micro\metre}\) domain at \(U/V=10/20\), ordered from top to
  bottom as \(500\), \(200\), and \(100~\si{\nano\metre}\). COMSOL (solid)
  and GBIE (open circles) agree in amplitude and relative phase; the right
  panels show the absolute and percentage phase residuals.}
  \label{fig:supp_comsol_bie_500nm_check}
\end{figure}

\FloatBarrier
\subsection{Detailed profile comparisons}
\label{sec:supp_comsol_bie_five_case_profiles}

\begin{figure}[htbp]
  \centering
  \includegraphics[width=\linewidth]{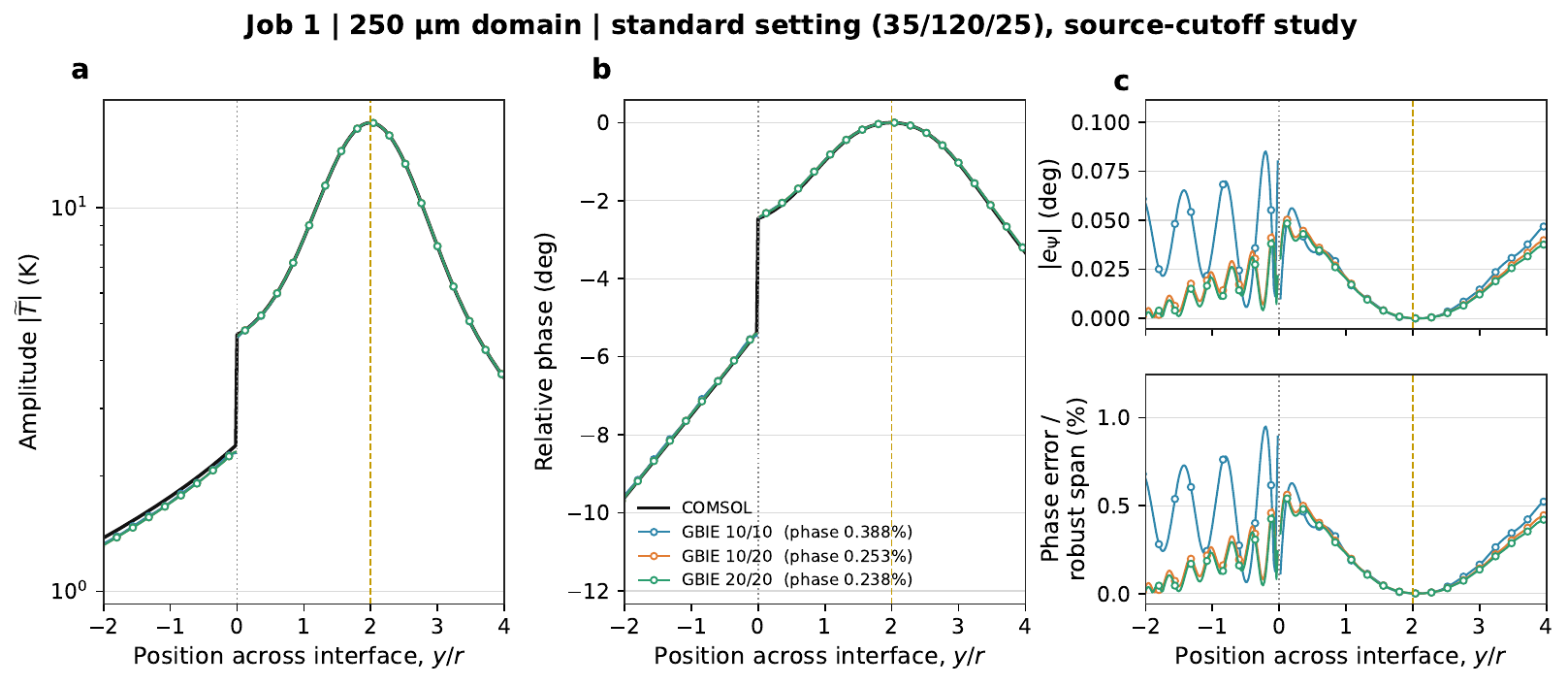}
  \caption{Detailed comparison for Job~1, the first
  \(250~\si{\micro\metre}\) execution. At \(U/V=10/20\),
  \(E_\phi=0.253\%\) and \(E_A=0.454\%\).}
  \label{fig:supp_comsol_bie_job_23511498}
\end{figure}

\begin{figure}[htbp]
  \centering
  \includegraphics[width=\linewidth]{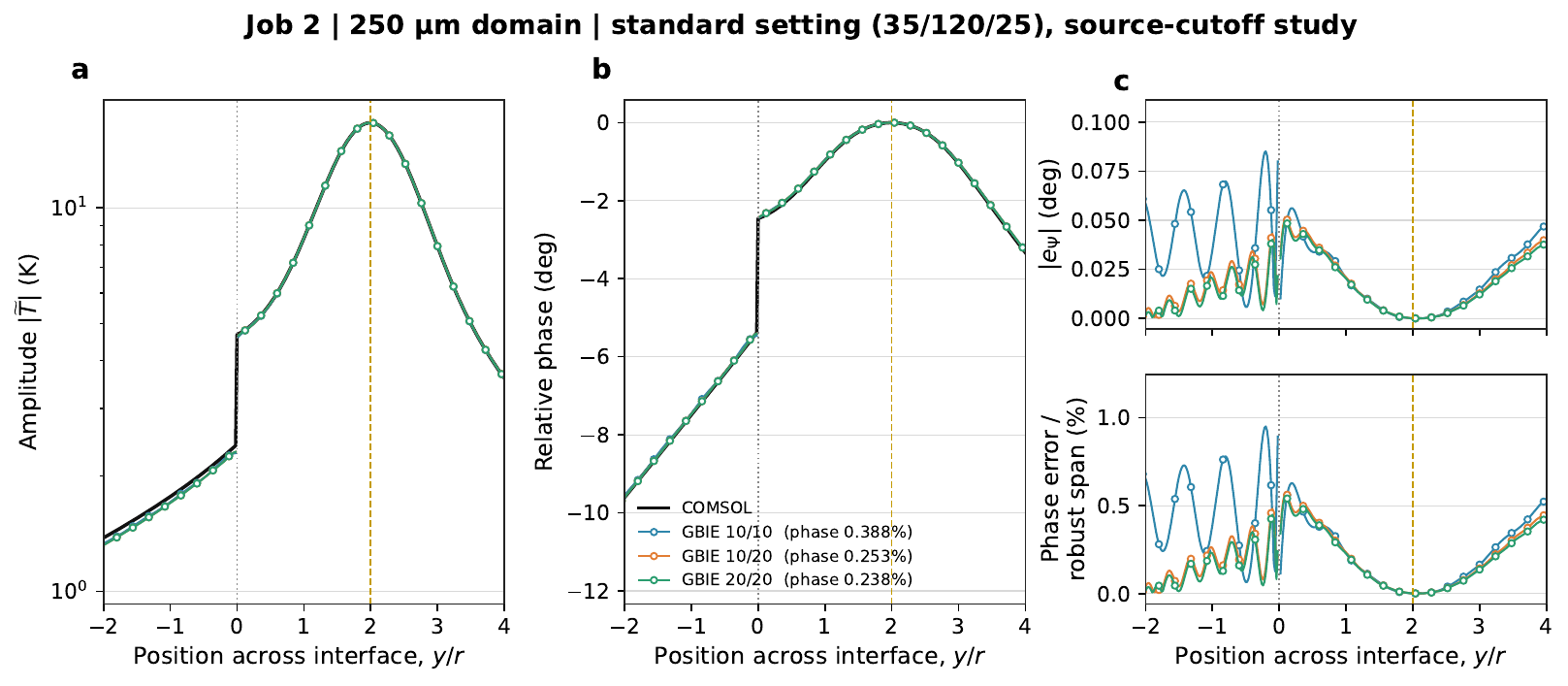}
  \caption{Detailed comparison for Job~2, the repeated
  \(250~\si{\micro\metre}\) execution. Its physical inputs and exported
  profile match Job~1; the pair is retained to evaluate workflow-time
  stability.}
  \label{fig:supp_comsol_bie_job_23511619}
\end{figure}

\begin{figure}[htbp]
  \centering
  \includegraphics[width=\linewidth]{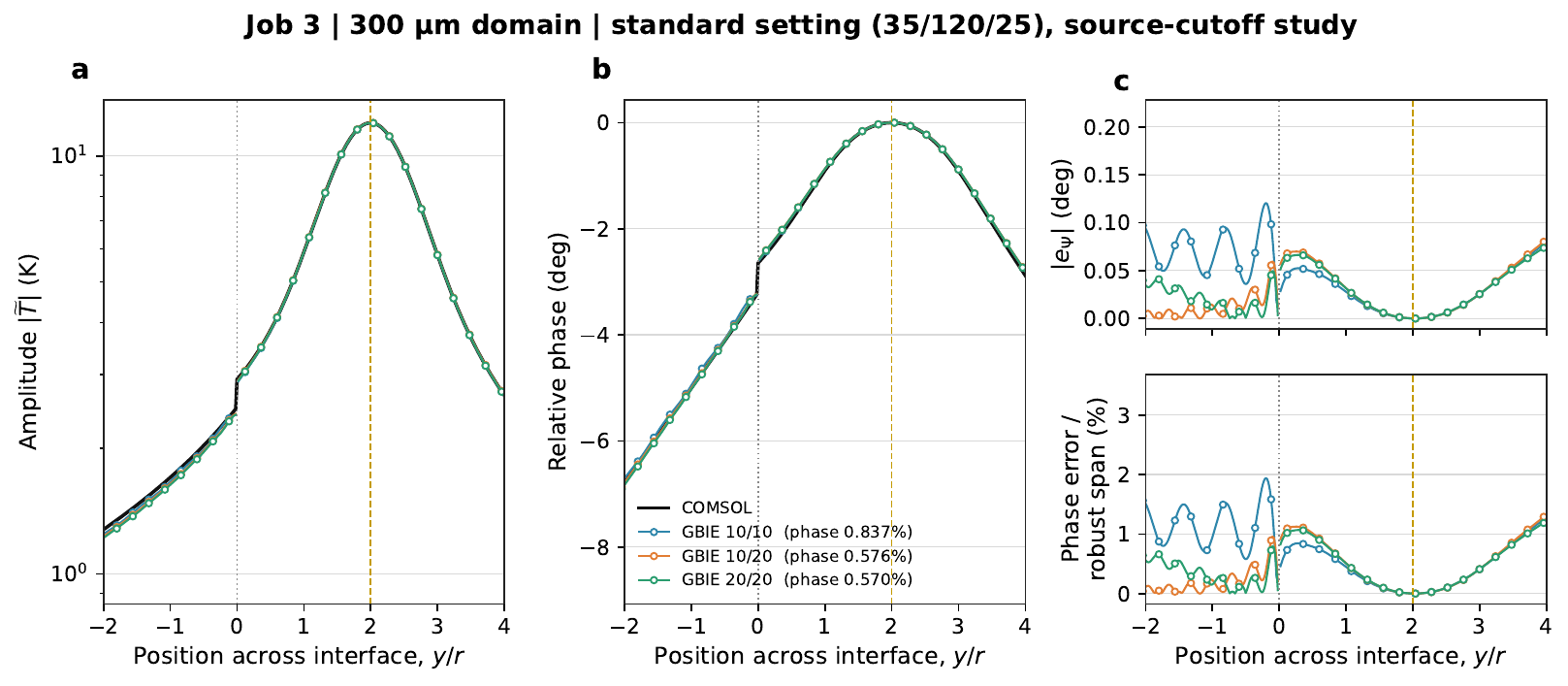}
  \caption{Detailed comparison for the \(300~\si{\micro\metre}\) COMSOL
  Job~3. At \(U/V=10/20\), \(E_\phi=0.576\%\) and
  \(E_A=0.538\%\).}
  \label{fig:supp_comsol_bie_job_23511733}
\end{figure}

\begin{figure}[htbp]
  \centering
  \includegraphics[width=\linewidth]{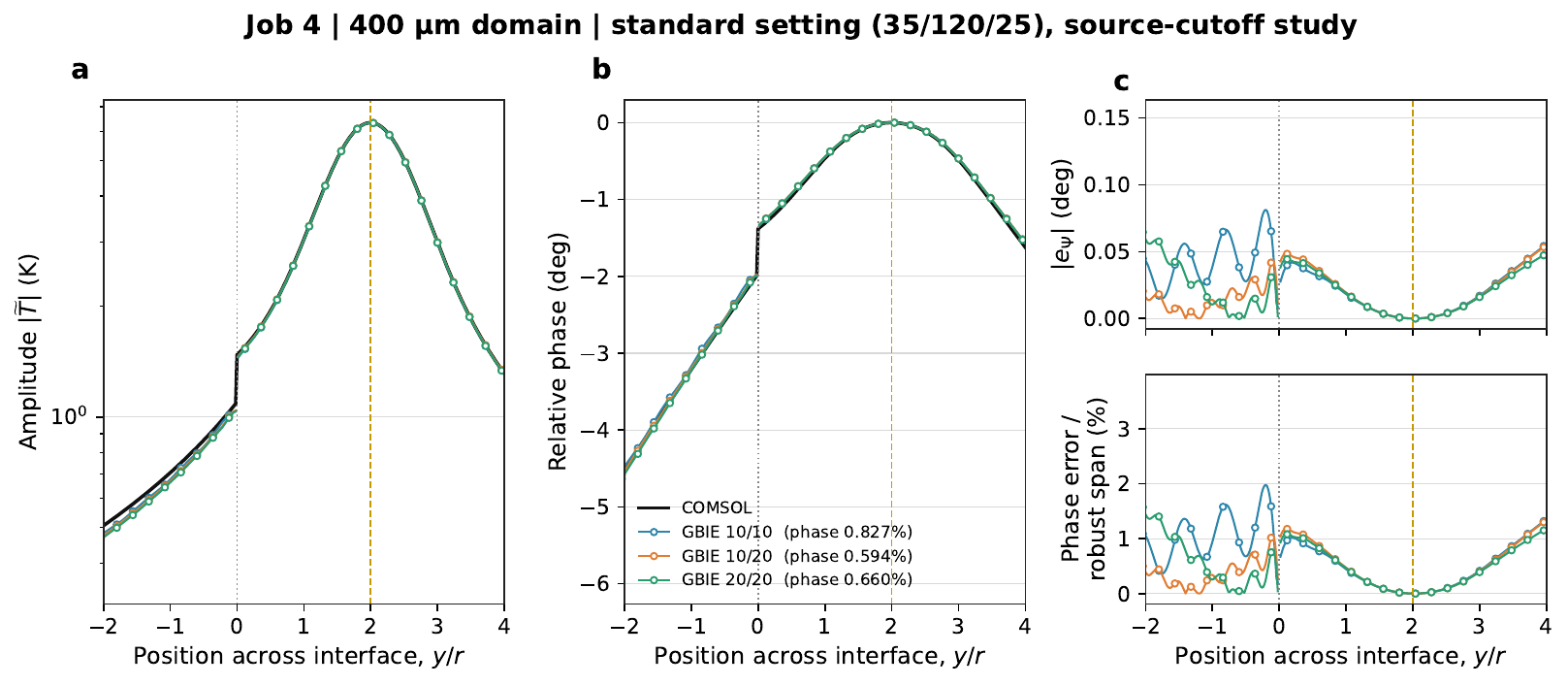}
  \caption{Detailed comparison for the \(400~\si{\micro\metre}\) COMSOL
  Job~4. At \(U/V=10/20\), \(E_\phi=0.594\%\) and
  \(E_A=0.648\%\).}
  \label{fig:supp_comsol_bie_job_23511801}
\end{figure}

\begin{figure}[htbp]
  \centering
  \includegraphics[width=\linewidth]{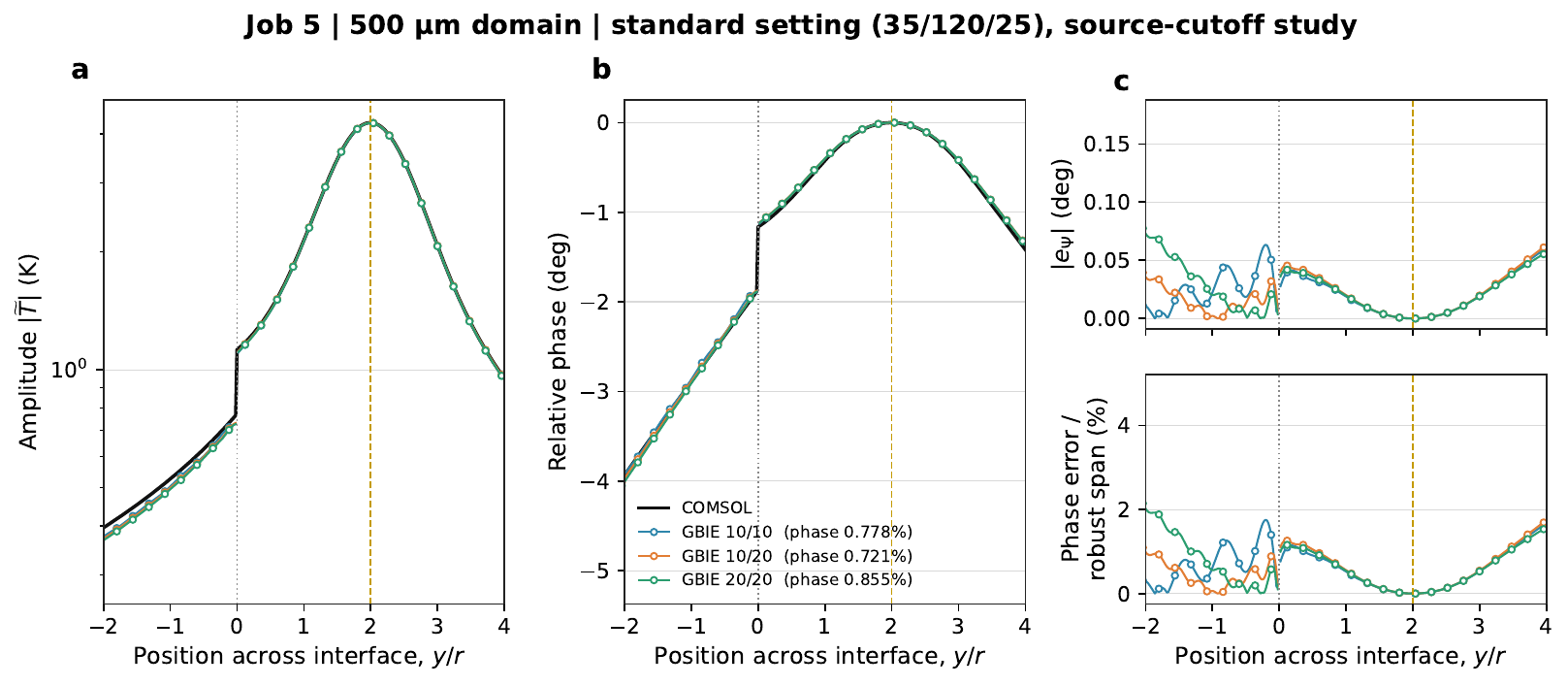}
  \caption{Detailed comparison for the \(500~\si{\micro\metre}\) COMSOL
  Job~5. At \(U/V=10/20\), \(E_\phi=0.721\%\) and
  \(E_A=0.728\%\).}
  \label{fig:supp_comsol_bie_job_23511820}
\end{figure}

Figures~\ref{fig:supp_comsol_bie_job_23511498}--
\ref{fig:supp_comsol_bie_job_23511820} show the profile-level evidence behind
the aggregate metrics.  Each figure compares COMSOL with all three GBIE
source-cutoff pairs and reports both phase and amplitude residuals.

\FloatBarrier

\section{Finite-Domain Size Effects in the COMSOL Reference}
\label{sec:supp_edge_effects}

The lateral domain used in the volumetric COMSOL reference must be large
enough that the truncating side boundaries do not perturb the temperature field
near the heated spot and vertical interface.  The requirement is most severe at
low frequency because the lateral thermal penetration depth
\begin{equation}
    \mu_i=\sqrt{\frac{k_i}{\pi C_v f}},\qquad i=x,y,
    \label{eq:supp_directional_penetration_depth}
\end{equation}
increases as \(f^{-1/2}\) and as \(\sqrt{k_i/C_v}\).  For the edge-effect
tests below, \(f=1~\mathrm{kHz}\), \(C_v=10^6~\mathrm{J\,m^{-3}\,K^{-1}}\),
and the in-plane conductivities \(k=60\)--\(120~\mathrm{W\,m^{-1}\,K^{-1}}\)
give \(\mu=138\)--\(195~\mu\mathrm{m}\). Thus, \(0.3\)--\(0.35~\mathrm{mm}\)
domains place the side boundaries only of order one thermal penetration depth
from the active region, whereas a \(0.5~\mathrm{mm}\) domain is the first size
tested that begins to separate the boundaries from the thermal wave.  In all
comparisons, the substrate thickness \(h_N\) was matched to the corresponding
COMSOL box thickness.

\subsection{Production-domain property envelope}

The production COMSOL audit used \(f=10~\mathrm{kHz}\),
\(C_v=10^6~\mathrm{J\,m^{-3}\,K^{-1}}\), \(r=1~\mu\mathrm{m}\), and the
design interval \(-3r\leq y\leq5r\), so \(L_e=8~\mu\mathrm{m}\). Applying
Eq.~\eqref{eq:supp_comsol_lateral_inverse} gives the largest allowable
in-plane conductivity for each lateral box. This is a coupled
property--domain audit, not a fixed-property domain-size sweep: the
\(250\), \(300\), \(400\), and \(500~\mu\mathrm{m}\) boxes use different
anisotropic conductivity sets selected to satisfy the corresponding bound.
Table~
\ref{tab:supp_production_domain_properties} lists both this design limit and
the complete conductivity tensors used in the five executions. Jobs~1 and 2
use the same material properties and form the repeat-run timing check.

\begin{table}[htbp]
    \centering
    \scriptsize
    \caption{Thermal properties and domain-size envelope for the production
    COMSOL audit. Conductivity tuples are
    \((k_x,k_y,k_z)\) in \(\mathrm{W\,m^{-1}\,K^{-1}}\). The film is
    isotropic with conductivity \(k_f\). The allowable \(k_{\parallel}\) is
    obtained from Eq.~\eqref{eq:supp_comsol_lateral_inverse}; \(L_y^{\rm req}\)
    is the width required by the actual maximum in-plane conductivity.}
    \label{tab:supp_production_domain_properties}
    \resizebox{\linewidth}{!}{%
    \begin{tabular}{lrrrrrrr}
        \toprule
        Job & \(L_y\) & \((k_{1x},k_{1y},k_{1z})\) &
        \((k_{2x},k_{2y},k_{2z})\) & \(k_f\) &
        \(k_{\parallel,\max}\) & \(k_{\parallel}^{\rm allow}\) &
        \(L_y^{\rm req}\) \\
        & (\(\mu\mathrm{m}\)) & & & &
        \multicolumn{2}{c}{(\(\mathrm{W\,m^{-1}\,K^{-1}}\))} &
        (\(\mu\mathrm{m}\)) \\
        \midrule
        1--2 & 250 & \((22,20,25)\)  & \((20,25,10)\) & 25  & 25  & 29  & 234 \\
        3    & 300 & \((30,27,30)\)  & \((30,25,20)\) & 35  & 35  & 42  & 275 \\
        4    & 400 & \((70,65,75)\)  & \((60,75,80)\) & 65  & 75  & 75  & 399 \\
        5    & 500 & \((100,95,110)\)& \((90,70,80)\) & 100 & 100 & 119 & 459 \\
        \bottomrule
    \end{tabular}%
    }
\end{table}

\subsection{Low-frequency finite-domain test}

The dedicated \(1~\mathrm{kHz}\) size-effect cases use an isotropic film with
\(k_f=100~\mathrm{W\,m^{-1}\,K^{-1}}\), a heated-side substrate with
\((k_x,k_y,k_z)=(60,80,90)~\mathrm{W\,m^{-1}\,K^{-1}}\), and an
unheated-side substrate with
\((k_x,k_y,k_z)=(120,100,110)~\mathrm{W\,m^{-1}\,K^{-1}}\). All three
regions use \(C_v=10^6~\mathrm{J\,m^{-3}\,K^{-1}}\); the pump radius is
\(10~\mu\mathrm{m}\), \(G_v=5\times10^7~\mathrm{W\,m^{-2}\,K^{-1}}\),
and \(G_h=10^8~\mathrm{W\,m^{-2}\,K^{-1}}\).

\begin{table}[htbp]
    \centering
    \small
    \caption{Geometry and phase accuracy for the COMSOL finite-domain
    comparison.  The P95 phase difference is evaluated against the
    corresponding laterally unbounded GBIE solution on the same
    \(501\)-point center-line scan.}
    \label{tab:supp_edge_effect_cases}
    \begin{tabular}{cccc>{\raggedright\arraybackslash}p{4.1cm}}
        \toprule
        \(h_f\) & Lateral box & \(h_N\) & P95 \(|e_\phi|\) &
        Finite-size assessment \\
        \((\mu\mathrm{m})\) & \((\mathrm{mm}^2)\) & \((\mathrm{mm})\) &
        \((\mathrm{deg})\) & \\
        \midrule
        1 & \(0.3\times0.3\) & 0.3 & 5.12 &
        Strong lateral-boundary bias \\
        1 & \(0.35\times0.35\) & 0.35 & 3.07 &
        Reduced but still appreciable boundary bias \\
        5 & \(0.5\times0.5\) & 0.5 & 0.14 &
        Close agreement with the unbounded reference \\
        \bottomrule
    \end{tabular}
\end{table}

\begin{figure}[htbp]
    \centering
    \includegraphics[width=\linewidth]{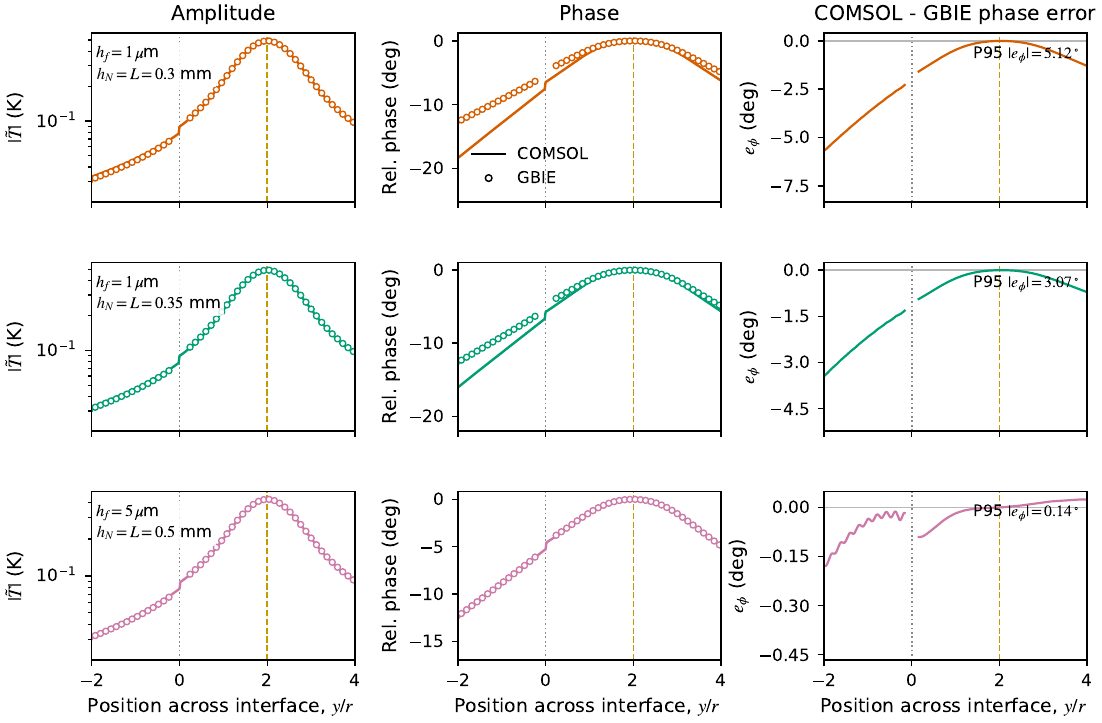}
    \caption{COMSOL--GBIE comparison for the lateral-domain tests at
    \(f=1~\mathrm{kHz}\), \(r=10~\mu\mathrm{m}\), and
    \(C_v=10^6~\mathrm{J\,m^{-3}\,K^{-1}}\).  The substrate thickness
    \(h_N\) is matched to the COMSOL box thickness in each row.  The vertical
    gray dotted line marks the material interface and the gold dashed line
    marks the pump center.  For the \(1~\mu\mathrm{m}\) film, enlarging the
    domain from \(0.3\times0.3\) to \(0.35\times0.35~\mathrm{mm}^2\) reduces
    but does not eliminate the multi-degree phase error.  For the
    \(5~\mu\mathrm{m}\) film, the \(0.5\times0.5~\mathrm{mm}^2\) domain
    agrees with the GBIE to \(0.14^\circ\) at the P95 level.}
    \label{fig:supp_edge_effect_comsol_bie}
\end{figure}

Figure~\ref{fig:supp_edge_effect_comsol_bie} compares the COMSOL center-line
solutions with the GBIE solution for the same material stack.  For the
\(h_f=1~\mu\mathrm{m}\) film, the \(0.3\times0.3~\mathrm{mm}^2\) case
shows a clear edge effect: its phase differs from the GBIE by
\(5.12^\circ\) at the P95 level.  The largest lateral domain that produced a
\(1~\mu\mathrm{m}\)-film COMSOL solution included in this comparison is
\(0.35\times0.35~\mathrm{mm}^2\).  Enlarging the box from \(0.3\) to
\(0.35~\mathrm{mm}\) lowers the P95 phase difference from \(5.12^\circ\) to
\(3.07^\circ\), but the remaining multi-degree discrepancy shows that the
lateral boundary is still too close to the thermally active region.  For the
\(h_f=5~\mu\mathrm{m}\) comparison, the \(0.5\times0.5~\mathrm{mm}^2\)
domain agrees with the GBIE to \(0.14^\circ\) P95 phase error.  The comparison
shows that phase is the more sensitive indicator of an insufficient lateral
domain: an amplitude curve can look acceptable while the side boundaries
still bias the phase.

\begin{figure}[htbp]
    \centering
    \includegraphics[width=0.9\linewidth]{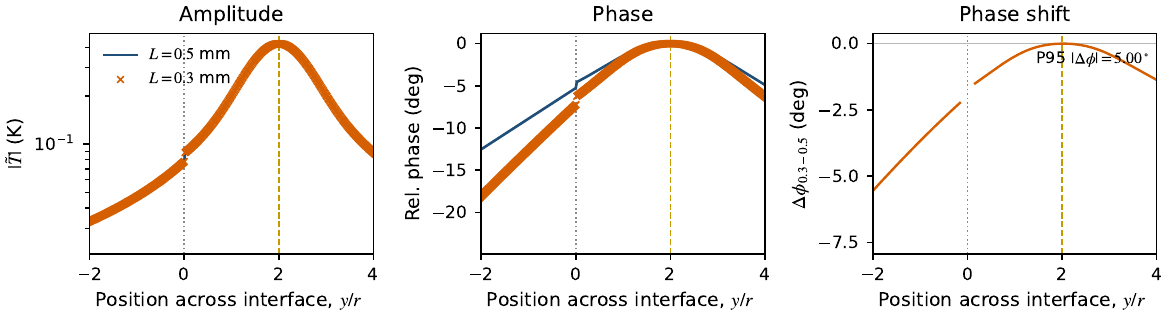}
    \caption{Direct COMSOL comparison between the
    \(0.5\times0.5~\mathrm{mm}^2\) and \(0.3\times0.3~\mathrm{mm}^2\) domains
    for the same \(h_f=5~\mu\mathrm{m}\) material stack.  The small domain
    produces a multi-degree phase shift while leaving the amplitude largely
    unchanged, confirming the finite-domain edge effect.}
    \label{fig:supp_edge_effect_domain_check}
\end{figure}

To isolate the edge effect from changes in material properties, the
\(h_f=5~\mu\mathrm{m}\) stack was also solved with the same material parameters
but two lateral domains,
\(0.5\times0.5~\mathrm{mm}^2\) and \(0.3\times0.3~\mathrm{mm}^2\).
Figure~\ref{fig:supp_edge_effect_domain_check} directly compares these two
COMSOL data sets.  The two solutions have nearly the same amplitude, with a
P95 amplitude difference of \(1.78\%\), but their relative phase differs by
\(5.00^\circ\) at the P95 level.  This direct COMSOL--COMSOL comparison
confirms that the phase discrepancy is introduced by the finite lateral
boundary rather than by the GBIE reference.

These tests illustrate the need for a domain-size check for the volumetric reference.
The lateral box must extend several thermal penetration depths beyond the
thermally active region, and convergence should be assessed from phase as well
as amplitude.  At \(1~\mathrm{kHz}\), the \(0.3\)--\(0.35~\mathrm{mm}\)
boxes retain multi-degree phase shifts even when their amplitude profiles
appear similar.  Increasing the lateral extent to \(0.5~\mathrm{mm}\) reduces
the discrepancy to \(0.14^\circ\) for the \(5~\mu\mathrm{m}\)-film test.
Because the GBIE represents the laterally unbounded multilayer directly, it
introduces no finite lateral boundary and helps identify this finite-size
bias.

\FloatBarrier

\section{Thin-Film Limit for a Semi-Infinite Vertical Interface}
\label{sec:supp_thin_film_thickness_sweep}

This section tests the thin-film limit of the multilayer GBIE
formulation for a surface-breaking, semi-infinite vertical interface.  For
every nonzero film thickness, the interface starts at the surface,
\(z_{\min}=0\), and extends through both the surface film and the substrate.
The purpose is to check whether the finite-film GBIE solution approaches the
no-film bulk-substrate solution as the film thickness vanishes.

The same anisotropic material properties as the main multilayer validation
were used, with \(G_v=10^8~\mathrm{W\,m^{-2}\,K^{-1}}\),
\(r=1~\mu\mathrm{m}\), \(f=100~\mathrm{kHz}\), and source position
\(d/r=2\).  The film/substrate conductance was set to
\(G_h\to\infty\), so that the film is perfectly coupled to the substrate and
the sweep isolates the geometric effect of a finite surface film.  The
\(h_f=0\) calculation removes both the film and the film/substrate interface
and is used as the GBIE bulk reference.

This is a GBIE sensitivity study rather than an additional FEM comparison.  We
computed \(h_f=0\), \(1~\mathrm{nm}\), \(10~\mathrm{nm}\),
\(50~\mathrm{nm}\), \(100~\mathrm{nm}\), \(500~\mathrm{nm}\), and
\(1~\mu\mathrm{m}\).  For visual clarity, the figure shows the representative
subset \(h_f=0\), \(1~\mathrm{nm}\), \(10~\mathrm{nm}\),
\(100~\mathrm{nm}\), and \(1~\mu\mathrm{m}\).  All calculations used the same
refined quadrature setting,
\begin{equation}
    (N_u,N_v,N_z)=(50,200,50),\qquad
    u_{\max}r=15,\qquad v_{\max}r=50 .
    \label{eq:supp_thin_film_refined_quadrature}
\end{equation}
The depth grid uses the layer-relative breakpoints described in
Supplementary Section~S3.  Thus even the \(1~\mathrm{nm}\) film contains five
protected interface-flux collocation intervals, while the transfer matrix
still represents propagation inside each layer analytically.
With the film/substrate resistance removed, the finite-film solutions should
approach the \(h_f=0\) bulk response as the film thickness vanishes.

\begin{figure}[htbp]
    \centering
    \includegraphics[width=0.80\linewidth]{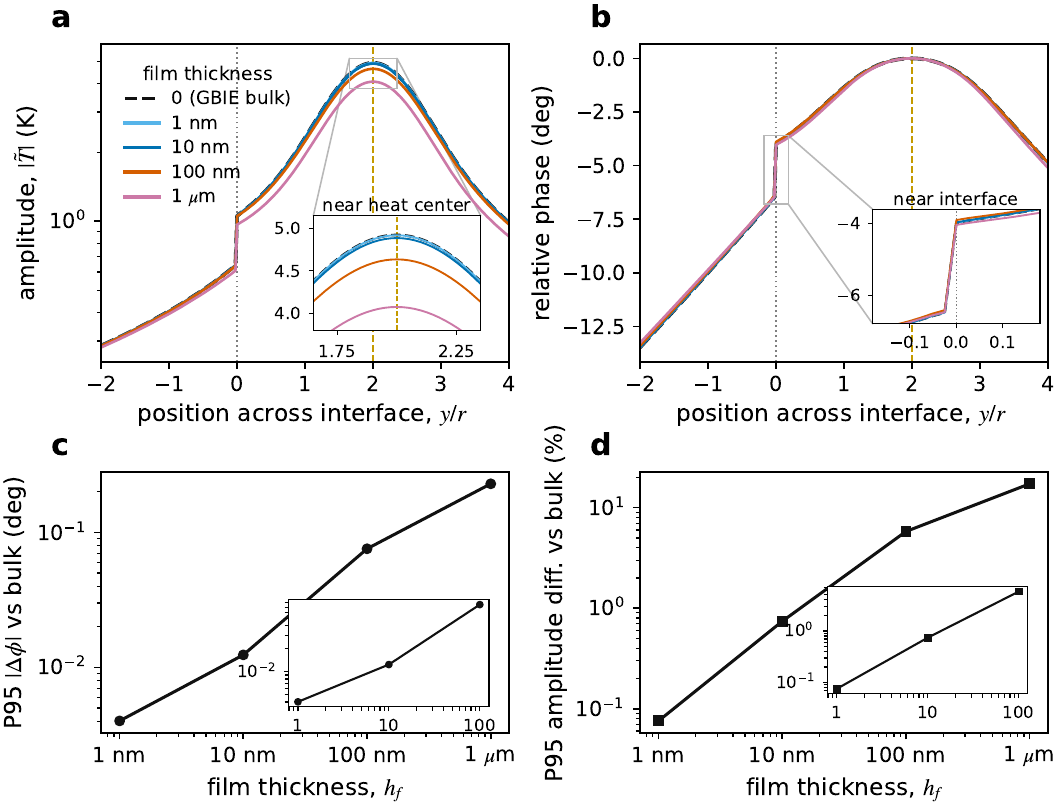}
    \caption{GBIE thin-film-limit test for a surface-breaking, semi-infinite
    vertical interface.  The \(h_f=0\) curve is the bulk-substrate reference
    with no film and no film/substrate interface; the finite-film cases use
    \(G_h\to\infty\).  Panels show (a) amplitude, (b) relative phase, (c) P95
    phase difference relative to the bulk reference, and (d) P95 relative
    amplitude difference relative to the bulk reference.  Insets in (a) and
    (b) zoom the heat-center amplitude and near-interface phase, respectively.
    Panels (c,d) use log--log axes; their insets zoom the
    \(1\)--\(100~\mathrm{nm}\) range where convergence to the bulk reference is
    most apparent.}
    \label{fig:supp_thin_film_thickness_sweep}
\end{figure}

Figure~\ref{fig:supp_thin_film_thickness_sweep} shows the expected convergence
to the \(h_f=0\) bulk limit as the film becomes thinner.  Relative to the bulk
reference, the P95 phase differences are \(0.0040^\circ\), \(0.012^\circ\),
\(0.044^\circ\), \(0.076^\circ\), \(0.188^\circ\), and \(0.229^\circ\) for
\(h_f=1~\mathrm{nm}\), \(10~\mathrm{nm}\), \(50~\mathrm{nm}\),
\(100~\mathrm{nm}\), \(500~\mathrm{nm}\), and \(1~\mu\mathrm{m}\),
respectively.  The corresponding P95 relative amplitude differences are
\(0.076\%\), \(0.74\%\), \(3.30\%\), \(5.79\%\), \(14.4\%\), and \(17.3\%\).
The monotonic decrease toward the \(h_f=0\) reference confirms that the
multilayer Green's function recovers the correct bulk limit for
nanometer-scale films.  Relative to the earlier beam-scale-only grid, adding
the layer-relative breaks changes the complete \(1~\mathrm{nm}\) phase
profile by at most \(2.4\times10^{-4}\) degrees and its amplitude by at most
\(2.0\times10^{-6}~\mathrm{K}\); the bulk solution is unchanged.  The
refinement therefore removes the accidental dependence on \(h_f=r\) without
altering the thin-film limiting behavior.

\FloatBarrier
\section{Single-Node Shared-Memory Scaling of COMSOL and GBIE}
\label{sec:supp_parallel_scaling}

\subsection{Exact \(100~\si{\nano\metre}\) thin-film benchmark}
\label{sec:supp_parallel_exact_100nm}

The one-core comparisons in Supplementary
Section~\ref{sec:supp_comsol_bie_500nm_check} isolate the numerical cost of
COMSOL and the GBIE at matched core count. To determine whether the reported
advantage persists under normal multicore COMSOL use, we repeated the exact
\(100~\si{\nano\metre}\)-film case in
Table~\ref{tab:supp_comsol_bie_500nm_check} on 1, 2, 4, 8, and 16 physical
cores. This is the same \(10~\si{\kilo\hertz}\),
\(300\times300\times300~\si{\micro\metre}^3\) calculation, with
\(\boldsymbol{k}_a=(30,27,30)\),
\(\boldsymbol{k}_b=(30,25,30)\), and
\(\boldsymbol{k}_f=(35,35,35)~
\si{\watt\per\metre\per\kelvin}\);
\(C_v=10^6~\si{\joule\per\cubic\metre\per\kelvin}\); and
\(G_v=G_h=10^8~\si{\watt\per\square\metre\per\kelvin}\).
The mesh and algebraic problem were locked at \(9{,}609{,}009\) elements and
\(15{,}076{,}751\) solved plus \(4{,}597{,}108\) internal degrees of
freedom. Only the physical-core count was varied.

The benchmark follows the
\href{https://crc-pages.pitt.edu/user-manual/slurm/batch-jobs/}
{University of Pittsburgh CRC guidance for multithreaded Slurm jobs}: one task
was allocated on a single
\href{https://crc-pages.pitt.edu/user-manual/hardware_profiles/smp/}
{AMD EPYC 9374F Genoa node}, and its CPU allocation was set with
\texttt{--cpus-per-task}. COMSOL's shared-memory parallelism was controlled
with \texttt{-np}, following the
\href{https://www.comsol.com/support/knowledgebase/1096}
{COMSOL multicore guidance}. Simultaneous multithreading was disabled, and
each fresh process was restricted to its assigned physical cores with local
CPU and memory binding. The runtime settings were
\texttt{-mpmode turnaround}, \texttt{-blas auto},
\texttt{-alloc native}, and \texttt{-numasets 1}; these were selected by the
lower-cost screen described in
Sec.~\ref{sec:supp_comsol_runtime_screen}, rather than assumed to be
universally optimal.

All realizations ran serially within one 16-CPU allocation and therefore did
not compete with one another. The node was not exclusive, so unrelated jobs
could occupy its remaining resources. We recorded the host, CPU affinity, and
node load, alternated ascending and descending core-count blocks, and report
the full repeat range together with the median. One complete four-core run was
discarded as a first-use warm-up. Each measured point contains three fresh
COMSOL processes, with no model solution reused between repetitions.

COMSOL wrote temporary files and the completed output model to node-local
Slurm scratch. For each realization, \(T_{\mathrm{proc}}\) is the externally
measured fresh-process time through output-model closure, and
\(T_{\mathrm{arc}}\) is the separately measured time to copy the approximately
\(2.65~\si{\giga\byte}\) output model to durable project storage. The copied
model was hashed and then deleted; hashing, numerical validation, and cleanup
are outside the timing interval. The primary durable-workflow metric is
\begin{equation}
    T_{\mathrm{dur}}=T_{\mathrm{proc}}+T_{\mathrm{arc}}.
    \label{eq:supp_parallel_durable_time}
\end{equation}
Each reported \(T_{\mathrm{dur}}\) is the median of the per-realization sums.
Because the component columns are medianed separately, their displayed
medians need not sum exactly to the displayed durable median. Relative to the
one-core median,
\begin{equation}
    S_p=\frac{\widetilde T_{\mathrm{dur},1}}
              {\widetilde T_{\mathrm{dur},p}},
    \qquad
    \eta_p=\frac{S_p}{p},
    \qquad
    C_p=p\,\widetilde T_{\mathrm{dur},p},
    \label{eq:supp_parallel_metrics}
\end{equation}
where \(S_p\), \(\eta_p\), and \(C_p\) are speedup, parallel efficiency, and
allocated core-seconds, respectively.

\begin{table}[htbp]
    \centering
    \caption{Single-node shared-memory scaling of the exact
    \(100~\si{\nano\metre}\) COMSOL case (benchmark job 23549615).
    Times and COMSOL-reported peak physical memory are medians over three
    fresh processes. Parenthetical values are the observed
    minimum--maximum ranges of \(T_{\mathrm{dur}}\).}
    \label{tab:supp_parallel_scaling}
    \begingroup
    \scriptsize
    \setlength{\tabcolsep}{2.4pt}
    \resizebox{\linewidth}{!}{%
    \begin{tabular}{@{}rrrrrrrrrr@{}}
        \toprule
        \(p\) & Repeats &
        \(T_{\mathrm{solve}}\) &
        \(T_{\mathrm{proc}}\) &
        \(T_{\mathrm{arc}}\) &
        \(T_{\mathrm{dur}}\) (range) &
        \(S_p\) &
        \(\eta_p\) (\%) &
        \(C_p\) (core-s) &
        Memory (GiB) \\
        \midrule
        1  & 3 & 643 & 1104.46 & 28.31 &
        1132.47 (1121.82--1135.16) & 1.00 & 100.0 & 1132 & 33.94 \\
        2  & 3 & 402 & 706.07 & 32.37 &
        738.24 (726.43--741.38) & 1.53 & 76.7 & 1476 & 32.80 \\
        4  & 3 & 272 & 503.53 & 33.42 &
        533.65 (527.42--557.98) & 2.12 & 53.1 & 2135 & 34.79 \\
        8  & 3 & 255 & 497.68 & 30.99 &
        528.67 (505.37--541.98) & 2.14 & 26.8 & 4229 & 36.90 \\
        16 & 3 & 237 & 489.01 & 35.51 &
        521.63 (499.98--582.82) & 2.17 & 13.6 & 8346 & 36.16 \\
        \bottomrule
    \end{tabular}%
    }
    \endgroup
\end{table}

\begin{figure}[htbp]
    \centering
    \includegraphics[width=\linewidth]
    {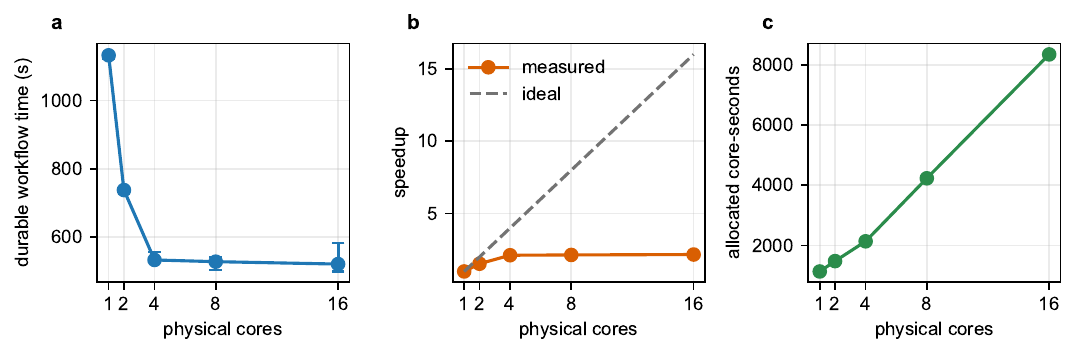}
    \caption{Shared-memory scaling of the exact
    \(100~\si{\nano\metre}\) COMSOL case. Points show repeat medians, and
    error bars in (a) span the observed minimum and maximum. Panels show
    (a) durable-workflow time, (b) speedup relative to the one-core median
    with ideal linear speedup for reference, and (c) allocated
    durable-workflow core-seconds.}
    \label{fig:supp_parallel_scaling}
\end{figure}

COMSOL's median solution time decreases from \(640\) to \(240~\si{\second}\)
between one and 16 cores, a \(2.7\times\) solver-stage speedup. The complete
durable workflow decreases from \(1.1\times10^3\) to
\(5.2\times10^2~\si{\second}\), a
\(2.2\times\) speedup with \(14\%\) parallel efficiency. The lowest
measured median occurs at 16 cores, but the 4-, 8-, and 16-core repeat ranges
overlap. Four cores therefore offer a favorable turnaround/resource
compromise for this case: their median is only \(2.3\%\) above the 16-core
median, while their allocation is approximately one quarter of the 16-core
core-seconds. Model opening, equation compilation, mesh and result
processing, saving, and archival scale less effectively than the solver and
produce this plateau.

For context, the independent one-core GBIE execution of the identical case in
Table~\ref{tab:supp_comsol_bie_500nm_check} required
\(5.9~\si{\second}\). Even the lowest measured COMSOL median is therefore
about \(89\times\) longer. This comparison does not imply that the exact GBIE
case was rerun at every core count; its shared-memory behavior is assessed
separately below.

All 15 timed COMSOL exports passed locked metadata, observation-coordinate,
algebraic-problem-size, core-count, and numerical-equivalence checks. Relative
to the original one-core export, the largest scaled numerical difference was
\(7.91\times10^{-6}\), within the prescribed
\(10^{-5}\) relative and \(10^{-8}\) absolute tolerances; the largest
absolute difference in an exported numeric column was
\(4.66\times10^{-5}\). The \(1244~\si{\second}\) entry in
Table~\ref{tab:supp_comsol_bie_500nm_check} remains the independent,
single-execution sum of COMSOL's reported opening, running, and saving
components. It is not replaced by the repeated
\(1132.47~\si{\second}\) median here because the timing boundaries and run
context differ.

\subsection{Supporting runtime-setting and GBIE threading screens}
\label{sec:supp_comsol_runtime_screen}

Before running the expensive exact-case sweep, we screened COMSOL runtime
settings on a lower-cost, fixed \(300\times300\times300~
\si{\micro\metre}^3\) finite-depth connected-support model
(job 23529103). That auxiliary model contains \(2{,}296{,}036\) solved plus
\(524{,}741\) internal degrees of freedom. It was used only to select a
reasonable production configuration and to test GBIE native-thread
sensitivity; conclusions about the \(100~\si{\nano\metre}\) case come from
the exact benchmark above.

COMSOL recommends measuring the core count for each model because using every
available core can slow smaller problems. It also recommends testing the
memory allocator and available BLAS libraries. COMSOL~6.2 includes a custom
AOCL~4.1.1 build that
\href{https://www.comsol.com/support/knowledgebase/1311}
{COMSOL identifies as beneficial for many AMD Zen~4 models}. We therefore
screened \texttt{-blas \{auto,aocl\}} and
\texttt{-alloc \{native,scalable\}} at 16 physical cores while retaining
\texttt{-mpmode turnaround} and \texttt{-numasets 1}. Each setting used one
discarded warm-up and three fresh-process repetitions in balanced,
interleaved order.

\begin{table}[htbp]
    \centering
    \caption{Supporting COMSOL BLAS and memory-allocator screen on the
    lower-cost model at 16 physical cores. Times are medians in seconds;
    ranges are the observed minimum--maximum durable-workflow times.}
    \label{tab:supp_comsol_runtime_screen}
    \small
    \begin{tabular}{@{}llrrrrr@{}}
        \toprule
        \texttt{-blas} & \texttt{-alloc} &
        \(T_{\mathrm{solve}}\) & \(T_{\mathrm{proc}}\) &
        \(T_{\mathrm{arc}}\) & \(T_{\mathrm{dur}}\) (range) &
        Relative time \\
        \midrule
        auto & native & 21 & 60.67 & 5.02 &
        65.95 (65.69--69.31) & 1.000 \\
        aocl & native & 22 & 60.92 & 5.43 &
        66.51 (64.05--76.70) & 1.008 \\
        auto & scalable & 22 & 61.97 & 6.57 &
        68.54 (67.80--68.62) & 1.039 \\
        aocl & scalable & 23 & 62.55 & 6.56 &
        69.19 (66.93--70.29) & 1.049 \\
        \bottomrule
    \end{tabular}
\end{table}

All 12 exports passed the same numerical-equivalence checks. For this
auxiliary model, COMSOL~6.2 build~415, and Genoa node, the predefined
\texttt{auto/native} combination has the lowest median. AOCL with the native
allocator is \(0.8\%\) slower in the durable metric, and the repeat ranges
overlap. The scalable allocator is approximately \(4\%\) slower than the
native allocator at either BLAS choice. The AOCL recommendation is therefore
worth screening on Zen~4 hardware, but it is not a universal improvement.

We screened \texttt{-mpmode} separately at the selected
\texttt{auto/native} combination. COMSOL describes \texttt{owner} as the
highest-performance option in many cases, \texttt{turnaround} as suitable
when COMSOL is the only active process, and \texttt{throughput} as suitable
when other processes are active. The same three-repeat interleaved protocol
was used.

\begin{table}[htbp]
    \centering
    \caption{Supporting COMSOL multiprocessing-mode screen on the lower-cost
    model at 16 physical cores with \texttt{-blas auto} and
    \texttt{-alloc native}. Times are medians in seconds; ranges are the
    observed minimum--maximum durable-workflow times.}
    \label{tab:supp_comsol_mpmode_screen}
    \small
    \begin{tabular}{@{}lrrrrr@{}}
        \toprule
        \texttt{-mpmode} & \(T_{\mathrm{solve}}\) &
        \(T_{\mathrm{proc}}\) & \(T_{\mathrm{arc}}\) &
        \(T_{\mathrm{dur}}\) (range) & Relative time \\
        \midrule
        turnaround & 21 & 59.99 & 4.66 &
        63.90 (63.32--66.20) & 1.000 \\
        throughput & 21 & 59.90 & 4.76 &
        64.66 (64.24--73.87) & 1.012 \\
        owner & 21 & 61.74 & 5.92 &
        67.30 (66.91--70.68) & 1.053 \\
        \bottomrule
    \end{tabular}
\end{table}

All nine mode-screen exports passed numerical-equivalence checks.
\texttt{turnaround} has the shortest median, \(63.90~\si{\second}\);
\texttt{throughput} is \(1.2\%\) slower with an overlapping range, and
\texttt{owner} is \(5.3\%\) slower. The solution time is
\(21~\si{\second}\) for every mode, so the observed differences occur outside
the solver stage. We consequently adopted
\texttt{auto/native/turnaround} for the exact-case core-count sweep without
claiming that it is a universal COMSOL optimum.

The same auxiliary model was also used to check native-thread sensitivity of
the present GBIE implementation. Across eight fresh-process repetitions at
each of 1, 2, 4, 8, and 16 physical cores, the workflow medians were
\(5.08\), \(5.04\), \(5.04\), \(5.07\), and
\(5.07~\si{\second}\), respectively. The spread is smaller than the repeat
ranges, so one core is the practical setting: it minimizes allocated
processor use without a meaningful latency penalty. This supporting result
does not substitute for the independent \(5.85~\si{\second}\) one-core
measurement of the exact \(100~\si{\nano\metre}\) case.

\subsection{Interpretation and scope}

COMSOL parallelizes substantial matrix assembly and iterative/multigrid
solution work, but these operations are only part of its complete workflow;
sparse-memory traffic, synchronization, and NUMA effects further reduce
efficiency as the core count increases. In the present GBIE implementation,
the dominant source-quadrature traversal is largely serial and its native
linear-algebra calls operate on many small depth systems, so additional
threads do not reduce wall time. This is an implementation-specific result,
not an intrinsic limit of Green's-function or boundary-integral methods.

Distributed-memory COMSOL execution was not tested because the exact
\(100~\si{\nano\metre}\) model used only \(33\)--\(37~\si{\gibi\byte}\) of
peak physical memory and fit on one CRC SMP node. Absolute times and the
scaling plateau are model-, implementation-, filesystem-, and
hardware-dependent; the robust result is that multicore execution reduces
COMSOL turnaround without removing the large same-case advantage of the
one-core GBIE workflow.

\clearpage
\section{Additional bulk timings and thin-film profiles}
\label{sec:supp_moved_results}
The detailed bulk timing table and thin-film-limit profile are retained here
to keep the main validation section focused. These are the original saved
benchmark results; moving them does not constitute a new numerical run.
\begin{table}[htbp]
    \centering
    \caption{Runtime comparison with direct adaptive evaluation of the
    complete source-convolution form of the established semi-analytical bulk
    reference. Each line contains 121 \(y\)-points, and both methods receive
    identical observation coordinates. The first column lists the ordered
    conductivity pair \((k_{\mathrm{un}},k_{\mathrm{h}})\) in
    \(\mathrm{W\,m^{-1}\,K^{-1}}\).
    Numerical settings and cutoff calibration are detailed in Supplementary
    Sections~S2 and S4.}
    \label{tab:bulk_analytical_timing}
    \begingroup
    \scriptsize
    \setlength{\tabcolsep}{5pt}
    \begin{tabular}{@{}llrccc@{}}
        \toprule
        Case & Offset set & \(N_{\mathrm{obs}}\) & \shortstack{Full source-\\convolution} & GBIE & Reference/GBIE \\
        \((k_{\mathrm{un}},k_{\mathrm{h}})\) & & & \multicolumn{2}{c}{runtime (s)} & ratio \\
        \midrule
        \((10,10)\)  & \(x/r=0\) & 121 & 2.6 & 2.2 & \(1.2\times\) \\
                    & \(x/r=1\) & 121 & 3.4 & 2.2 & \(1.5\times\) \\
                    & \(x/r=0,1,2\) & 363 & 10.1 & 2.2 & \(4.5\times\) \\
                    & \(x/r=0,0.5,1,1.5,2\) & 605 & 16.8 & 2.2 & \(7.6\times\) \\
        \addlinespace
        \((10,100)\) & \(x/r=0\) & 121 & 2.6 & 2.2 & \(1.1\times\) \\
                    & \(x/r=1\) & 121 & 2.8 & 2.3 & \(1.2\times\) \\
                    & \(x/r=0,1,2\) & 363 & 8.7 & 2.3 & \(3.7\times\) \\
                    & \(x/r=0,0.5,1,1.5,2\) & 605 & 14.1 & 2.3 & \(6.1\times\) \\
        \addlinespace
        \((100,10)\) & \(x/r=0\) & 121 & 2.1 & 2.3 & \(0.9\times\) \\
                    & \(x/r=1\) & 121 & 2.1 & 2.3 & \(0.9\times\) \\
                    & \(x/r=0,1,2\) & 363 & 6.5 & 2.3 & \(2.9\times\) \\
                    & \(x/r=0,0.5,1,1.5,2\) & 605 & 11.1 & 2.2 & \(5.0\times\) \\
        \bottomrule
    \end{tabular}
    \endgroup
\end{table}
\begin{figure}[htbp]
    \centering
    \includegraphics[width=\linewidth]{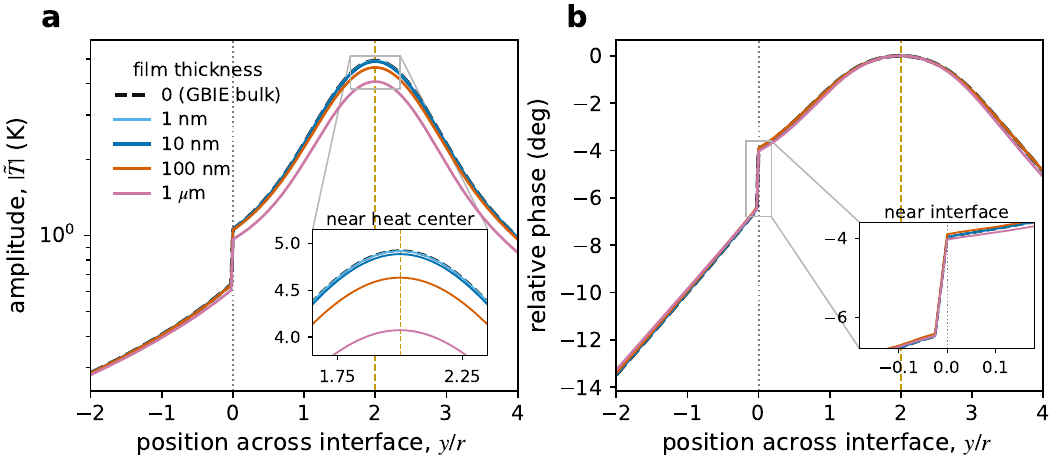}
    \caption{GBIE thin-film-limit test for a surface-breaking, semi-infinite
    vertical interface in the anisotropic multilayer stack. The black dashed
    \(h_f=0\) curve is the GBIE bulk-substrate reference; finite-film colors
    denote \(1~\mathrm{nm}\) (light blue), \(10~\mathrm{nm}\) (blue),
    \(100~\mathrm{nm}\) (vermilion), and \(1~\mu\mathrm{m}\) (pink). All
    finite-film cases use \(G_h\to\infty\) to isolate the geometric
    film-thickness effect. Panels
    show (a) amplitude and (b) relative phase; the insets enlarge the
    heat-center amplitude and near-interface phase. The full four-panel
    diagnostic, including the P95 phase and amplitude differences, is given
    in Supplementary Section~S9.}
    \label{fig:thin_film_thickness_sweep}
\end{figure}
\clearpage
\section{Additional Field Metrics and Numerical Evidence Limits}
\label{sec:review_numerical_audit}

\subsection{Direct complex-temperature and absolute-error measures}

The archived surface profiles allow an additional comparison that retains
both amplitude and the original phase relative to the applied heating.
For the retained set of sample indices $\mathcal I$, define
\begin{align}
 E_{\Theta}&=100\frac{\left[\sum_{i\in\mathcal I}
 |\Theta_{\mathrm{GBIE},i}-\Theta_{\mathrm{FEM},i}|^2\right]^{1/2}}
 {\left[\sum_{i\in\mathcal I}|\Theta_{\mathrm{FEM},i}|^2\right]^{1/2}},
 \label{eq:review_complex_l2}\\
 e_{A,{\rm rms}}\equiv\mathrm{RMSE}_{A}&=\left[\frac{1}{|\mathcal I|}
 \sum_{i\in\mathcal I}(|\Theta_{\mathrm{GBIE},i}|-
 |\Theta_{\mathrm{FEM},i}|)^2\right]^{1/2},
 \label{eq:review_absolute_amplitude_rmse}\\
 e_{\phi,i}^{\mathrm{raw}}&=\operatorname{wrap}_{[-180^\circ,180^\circ)}
 \!\left(\arg\Theta_{\mathrm{FEM},i}-\arg\Theta_{\mathrm{GBIE},i}\right).
 \label{eq:review_raw_phase_error}
\end{align}
Here the reference is FEM, $N=|\mathcal I|$ is the number of comparison
points, and $A_i=|\Theta_i|$ is the temperature amplitude. The amplitude
RMSE is reported in kelvin. The absolute phase error is
$|\arg(\Theta_{\mathrm{GBIE},i}\Theta_{\mathrm{FEM},i}^{*})|
=|e_{\phi,i}^{\mathrm{raw}}|$ in degrees, with $*$ denoting complex
conjugation. These comparisons use the original heating reference without
fitted complex rescaling. No phase floor or phase-span normalization is applied.
The primary sample mask is the original retained window,
nominally $-2\le y/r\le4$, excluding the sampled vertical-interface point.
There are 299 or 300 retained samples, according to the archived coordinate
and interface masks. The quadrature is not used to reweight these uniformly
sampled surface-line comparisons.

The GBIE files store real and imaginary temperature. Although the processed
FEM files store beam-referenced phase, they also retain the original
FEM-minus-GBIE phase difference at the beam. If $i_b$ is that saved beam
index, $\phi_{\mathrm{FEM},i}^{\mathrm{rel}}$ the archived relative phase,
and $\Delta\phi_b^{\mathrm{raw}}$ the saved raw phase difference, the
original FEM phase is recovered, modulo $360^\circ$, by
\begin{equation}
 \phi_{\mathrm{FEM},i}^{\mathrm{raw}}=
 \phi_{\mathrm{FEM},i}^{\mathrm{rel}}+
 \arg\Theta_{\mathrm{GBIE},i_b}+\Delta\phi_b^{\mathrm{raw}}.
 \label{eq:review_restore_fem_phase}
\end{equation}
This identity reverses the stored phase referencing; it does not align the
two predictions. The reconstructed coordinates, amplitudes, and raw phases
were checked against the original COMSOL text exports for all ten
executions, including the three thin-film cases.

\begin{table}[htbp]
\centering
\small
\caption{Direct complex-temperature and absolute-error measures recomputed
from archived GBIE and COMSOL surface profiles. The five multilayer rows use
$U/V=10/20$; rows 1 and 2 are repeat executions with identical profiles.
The thin-film rows use $U/V=10/20$, the finite-depth support uses $30/20$,
and the four-level conductance uses $20/50$. All use requested
$(N_u,N_v,N_z)=(35,120,25)$. Phase errors retain the original heating reference.}
\label{tab:review_direct_complex_metrics}
\begin{tabular}{@{}lrrrr@{}}
\toprule
Case & $E_{\Theta}$ & $\mathrm{RMSE}_{A}$ &
 P95 $|e_{\phi}^{\mathrm{raw}}|$ & $\max|e_{\phi}^{\mathrm{raw}}|$\\
 & (\%) & (K) & (deg) & (deg)\\
\midrule
Multilayer 1 & 0.456 & 0.0383 & 0.058 & 0.066 \\
Multilayer 2 (repeat) & 0.456 & 0.0383 & 0.058 & 0.066 \\
Multilayer 3 & 0.543 & 0.0328 & 0.095 & 0.108 \\
Multilayer 4 & 0.650 & 0.0206 & 0.063 & 0.070 \\
Multilayer 5 & 0.730 & 0.0159 & 0.069 & 0.082 \\
Film, 500 nm & 0.519 & 0.0334 & 0.092 & 0.106 \\
Film, 200 nm & 0.510 & 0.0369 & 0.091 & 0.103 \\
Film, 100 nm & 0.485 & 0.0379 & 0.091 & 0.105 \\
Finite-depth support & 0.699 & 0.0148 & 0.081 & 0.091 \\
Four-level $G_v(z)$ & 0.475 & 0.0298 & 0.074 & 0.085 \\
\bottomrule
\end{tabular}
\end{table}

The ten executions correspond to nine unique physical inputs. The complex
relative-$L_2$ errors range from $0.456\%$ to $0.730\%$ on the original
retained masks. Including the exported interface sample gives
$0.458\%$--$0.732\%$; using the full exported line while excluding that
sample gives $0.518\%$--$0.838\%$. These additional masks quantify the
sensitivity of the reported surface errors to window selection. The side
represented by an exported point exactly on a discontinuity is not used to
infer a contact jump. The accompanying CSV includes every mask, raw phase
RMSE, amplitude relative-$L_2$ and maximum absolute error, and the source
profile paths. These surface comparisons do not evaluate an independently
reconstructed contact-law residual, two-sided normal-flux continuity, or a
whole-domain harmonic energy balance.

\subsection{Exact depth grids and finite-depth continuation}

The three-level finite-depth benchmark uses protected breakpoints at
$2$, $5$, and $50~\mu\mathrm m$, with the physical resistive interface
ending at $z_t=50~\mu\mathrm m$. The flux equation is solved on an auxiliary
plane extending to $192.7299~\mu\mathrm m$. Below $z_t$, the two stack
representations have the same support properties and $R_v=0$, so the
auxiliary equation enforces temperature continuity in the connected support.
It retains the flux across that plane. It does not impose zero flux at the
physical termination. The resistance and flux discretizations therefore
require a distinction between the physical contact interval and the
auxiliary integration interval.

The saved grid has 25 nodes distributed over 24 intervals. Near the
conductance transitions, the nearest nodes below/above the protected
breakpoints are listed in Table~\ref{tab:review_depth_jump_nodes}.
Independent Gauss--Legendre rules on the protected intervals prevent a
quadrature panel from crossing a prescribed resistance jump. This property
alone does not establish local convergence of the flux within each interval.

\begin{table}[htbp]
\centering
\small
\caption{Archived nodes adjacent to conductance transitions and to the
finite-depth endpoint. Values are in $\mu\mathrm m$. The full node and
weight lists are supplied as CSV files.}
\label{tab:review_depth_jump_nodes}
\begin{tabular}{@{}lrrr@{}}
\toprule
Case & Protected depth & Shallower node & Deeper node\\
\midrule
Finite depth & 2 & 1.8921 & 2.2215 \\
Finite depth & 5 & 4.9430 & 6.9603 \\
Finite depth & 50 & 44.5441 & 60.6825 \\
Four level & 0.5 & 0.3750 & 0.7386 \\
Four level & 5 & 4.9430 & 5.1388 \\
Four level & 50 & 46.1100 & 51.3876 \\
\bottomrule
\end{tabular}
\end{table}

For the finite-depth case, the last node above the support is
$44.5441~\mu\mathrm m$ and the first support node is
$60.6825~\mu\mathrm m$. Thus the saved grid provides an exact endpoint
breakpoint but does not resolve a local endpoint flux profile. No nodal
flux trace, local endpoint-refinement sequence, or independent endpoint
contact-law residual is retained in these archives. The local behavior at
the junction of the vertical and horizontal contacts is consequently not
established by the surface-temperature comparison. The four-level case has
25 nodes and a numerical integration depth of $52.7751~\mu\mathrm m$;
its fourth resistance is applied for $z>50~\mu\mathrm m$ through
the remaining integration interval.

\subsection{Runtime, arithmetic, and memory provenance}

The production GBIE workflow uses Python, NumPy, and SciPy, with
64-bit real and 128-bit complex arrays. The CRC records identify Python
3.11.5. The source traverses the spectral nodes serially, solves each
interface matrix with a general dense linear solver, and caches transverse
reconstruction weights within the calculation. Identical side stacks can
share Green-function evaluations. The OMP, OpenBLAS, MKL, VECLIB, NUMEXPR,
and BLIS thread-count environment variables were all set to one before
importing the numerical libraries. The archived runtime records do not
identify the resolved BLAS library or exact NumPy/SciPy versions; those
cannot be inferred from the environment-variable names.

For the five-case one-core benchmark, an external timer surrounds a fresh,
plot-free Python process. It includes interpreter startup and imports,
input reading, grid construction, Green-function and matrix assembly,
solution, surface reconstruction, metric evaluation, and result writing.
The separately saved forward-solver interval includes assembly, solution,
and reconstruction together. These three stages were not separately timed.
The complete/internal-pipeline/forward times for the five executions are,
in seconds,
$8.64/4.34/4.15$, $8.75/4.33/4.22$, $7.20/4.28/4.18$,
$7.90/4.31/4.18$, and $6.92/4.25/4.19$. Fresh processes do not reuse
previously solved kernels. The COMSOL complete interval includes model
opening, execution, and saving; its separately logged stationary-solver time
excludes other workflow stages. The multicore durable-workflow interval
in Section~\ref{sec:supp_parallel_scaling} additionally includes the
specified output-model archival step and should be interpreted with that
study's timing definition.

The GBIE memory entries use the external GNU-\texttt{time} maximum resident
set for the complete Python process, including interpreter, imported
libraries, input data, and working arrays. The internal
\texttt{getrusage} record can precede the final process peak and is not
substituted for it. COMSOL memory entries use the application's reported
peak physical memory, with decimal GB converted to GiB. Loaded model and
solution data contribute to that value, but the archives do not establish
that COMSOL and GNU-\texttt{time} use identical accounting for every runtime
component. These are application-level measured peaks, not allocated memory,
virtual memory, or a theoretical storage bound. The runtime and memory
ratios are specific to these workflows and hardware.

An archived 121-point, $100~\mathrm{kHz}$ anisotropic multilayer sensitivity
run provides a limited empirical order-scaling check. Varying
$N_u=16,24,32,48$ at $(N_v,N_z)=(160,48)$ gives
$3.64,5.16,6.86,10.29~\mathrm s$; varying
$N_v=48,72,120,160$ at $(N_u,N_z)=(48,48)$ gives
$3.04,4.60,7.62,10.13~\mathrm s$. The actual depth orders
$N_z=23,30,48,64$ at $(N_u,N_v)=(48,160)$ give
$5.29,6.56,10.08,13.23~\mathrm s$. The first depth order was raised from a
requested 16 to 23 to preserve all protected intervals. These are individual
archived timings, not repeated estimates of asymptotic scaling. Observation
count and layout must also be distinguished: additional $x$ rows reuse the
spectral and $y$-reconstruction work, as shown in the observation-count
benchmarks. Counting the dense operations gives kernel assembly proportional
to $N_uN_vN_z^2$ and dense solves proportional to $N_uN_z^3$; these are
operation-count estimates, not fitted timing laws.

\subsection{Numerical checks not established by the archive}

The existing depth-order comparisons use finite transverse cutoffs. They
therefore do not demonstrate convergence to the infinite-cutoff weakly
singular operator. The production kernel uses the finite-cutoff diagonal
without analytic singularity subtraction or integrated self-cell weights.
The dense depth-grid agreement and the selected source-cutoff comparisons
must be interpreted within this limitation.

Condition-number diagnostics are implemented but were disabled in the
principal production runs: their stored condition-number fields are null.
Consequently, no conditioning bound follows from those archived outputs.
The additional large-conductance tests in Section~S13 now supply matrix
condition numbers, backward errors, and an independent perfect-contact
FEM comparison for the stated benchmarks. The zero-resistance portions of a buried or finite-depth auxiliary plane also
prevent a uniform second-kind claim based on a strictly positive local
resistance. Independent reconstruction of the interface temperatures and
normal fluxes, a local endpoint refinement study, and a whole-domain energy
balance require additional calculations or exports. In the convention
$e^{i\omega t}$, the latter compares applied complex power with
$i\omega\int C_v\Theta\,dV$ plus outward conductive flux through the
external boundary. The stored surface lines contain neither the required
volume integral nor all external-boundary flux integrals.

The finite-domain COMSOL calculations and semi-infinite GBIE calculations
share sources, local material parameters, and interface laws, but their
remote boundaries differ. The archived low-frequency tests demonstrate
finite-domain bias when the boundary is too close. They do not quantify a
smaller-than-discrepancy domain error for each production case. Exact
production mesh-convergence tolerances, model-specific linear-solver
tolerances, and fixed-property lateral/rear-domain expansions are not
established by the retained setup descriptions and surface exports.

\clearpage
\section{Large-conductance stability and the perfect-contact limit}
\label{sec:supp_large_conductance}

\subsection{Conductance and discretization tests}
The additional calculations test both exact thermal continuity and its
approach from finite resistance. The GBIE is assembled directly in $R_v$,
so setting $R_v=0$ removes only the local resistance term while retaining
the Green-function coupling and unknown interfacial flux. This imposes
perfect contact between dissimilar stacks and applies equally to the
connected portions of buried and finite-depth interfaces. Finite-conductance
tests use $R_v=1/G_v$; in particular,
$G_v=10^{14}~\mathrm{W\,m^{-2}K^{-1}}$ corresponds to
$R_v=10^{-14}~\mathrm{m^2K\,W^{-1}}$.
The first benchmark retains the 100~nm film case: both films have
$k=35~\mathrm{W\,m^{-1}K^{-1}}$, the substrate conductivity components
$(k_x,k_y,k_z)$ are $(25,30,30)$ and $(27,30,30)~\mathrm{W\,m^{-1}K^{-1}}$,
and $C_v=10^6~\mathrm{J\,m^{-3}K^{-1}}$ throughout.
Each horizontal film/substrate contact retains
$G_h=10^8~\mathrm{W\,m^{-2}K^{-1}}$.
The Gaussian source has $P_0=1$~mW, $r=1~\mu\mathrm{m}$,
$d=2~\mu\mathrm{m}$ and $f=10$~kHz.
The depth integration extends to $166.89~\mu\mathrm{m}$ with a
semi-infinite terminal substrate. All surface comparisons use the same
501 points at $x=0$, $-3\le y\le7~\mu\mathrm{m}$, without phase referencing.

For each angular wavenumber $\xi$, we record the unscaled matrix
condition number $\kappa_2(A)$ and the normalized backward error
\begin{equation}
 \beta_\infty=
 \frac{\|A\mathbf q-\mathbf b\|_\infty}
 {\|A\|_\infty\|\mathbf q\|_\infty+\|\mathbf b\|_\infty}.
 \label{eq:supp_large_G_backward}
\end{equation}
The values below are maxima over all $\xi$ nodes. The original
complex-double-precision GBIE implementation uses a direct dense solve
without special preconditioning. The surface difference is
$E_T=100\|\boldsymbol\Theta-\boldsymbol\Theta_{\rm ref}\|_2/
\|\boldsymbol\Theta_{\rm ref}\|_2$.

\begin{table}[htbp]
\centering\small
\caption{Conductance sweep for the 100~nm multilayer benchmark at
$(N_u,N_v,N_z)=(35,120,25)$. Angular cutoffs are
$(\xi_{\max},\eta_{\max})=(1,2)\times10^7~\mathrm{rad\,m^{-1}}$.
The reference for $E_T$ is the GBIE solution with exactly $R_v=0$ on the
same grid. The $G_v=\infty$ row is evaluated by setting $R_v=0$ directly.}
\label{tab:supp_large_G_sweep}
\begin{tabular}{@{}cccc@{}}
\toprule
$G_v$ & $\max\kappa_2(A)$ & $\max\beta_\infty$ & $E_T$ (\%) \\
\midrule
$10^{8}$ & $4.84\times10^{2}$ & $3.17\times10^{-17}$ & $1.34\times10^{0}$ \\
$10^{10}$ & $4.09\times10^{4}$ & $1.86\times10^{-17}$ & $1.65\times10^{-2}$ \\
$10^{11}$ & $2.53\times10^{5}$ & $1.75\times10^{-17}$ & $1.64\times10^{-3}$ \\
$10^{12}$ & $5.26\times10^{5}$ & $1.88\times10^{-17}$ & $1.63\times10^{-4}$ \\
$10^{14}$ & $5.96\times10^{5}$ & $1.41\times10^{-17}$ & $1.63\times10^{-6}$ \\
$\infty$ & $5.97\times10^{5}$ & $1.87\times10^{-17}$ & $0$ \\
\bottomrule
\end{tabular}
\end{table}

For the tested multilayer conditions, $G_v=10^{11}~\mathrm{W\,m^{-2}K^{-1}}$
effectively represents perfect contact: its relative complex surface-temperature
difference is $0.00164\%$ on the baseline grid and $0.00152\%$ on the
$(35,240,100)$ grid with $\eta_{\max}=4\times10^7~\mathrm{rad\,m^{-1}}$.
At $G_v=10^{14}$, these differences decrease to
$1.63\times10^{-6}\%$ and $1.53\times10^{-6}\%$, respectively.
Thus $10^{11}$ is already sufficient for the stated surface-response
comparison, while $10^{14}$ provides a more extreme small-resistance test.

\begin{table}[htbp]
\centering\small
\caption{Discretization tests at $G_v=10^{14}~\mathrm{W\,m^{-2}K^{-1}}$
for the same multilayer. The cutoff pair is expressed in
$10^7~\mathrm{rad\,m^{-1}}$. The last column compares each surface solution
with the first row, rather than with an exact continuum solution.}
\label{tab:supp_large_G_grids}
\begin{tabular}{@{}cccc@{}}
\toprule
$N_u/N_v/N_z$ & $\xi_{\max}/\eta_{\max}$ & $\max\kappa_2(A)$ & $E_T$ (\%) \\
\midrule
35/120/25 & 1/2 & $5.96\times10^{5}$ & 0.0000 \\
35/120/50 & 1/2 & $2.89\times10^{5}$ & 0.0482 \\
35/120/100 & 1/2 & $1.54\times10^{7}$ & 0.4776 \\
35/240/25 & 1/4 & $3.26\times10^{5}$ & 0.2484 \\
35/480/25 & 1/8 & $1.78\times10^{5}$ & 0.4580 \\
70/120/25 & 2/2 & $5.97\times10^{5}$ & 0.2156 \\
35/240/100 & 1/4 & $8.05\times10^{6}$ & 0.3962 \\
\bottomrule
\end{tabular}
\end{table}

All tested multilayer systems, including the corresponding perfect-contact
systems, return finite solutions without solver warnings, with
$\beta_\infty<8\times10^{-17}$. The largest condition number is
$1.64\times10^7$, for the perfect-contact system with $N_z=100$ at the
baseline cutoffs. These diagnostics support reliable algebraic solution of
the tested systems through $G_v=10^{14}$ and at perfect contact.
Changes under grid and cutoff variation are larger than the difference
between $G_v=10^{11}$ and perfect contact; the depth trend is not monotonic.
The tests therefore distinguish algebraic stability from discretization
accuracy and do not imply a uniform conditioning bound or a general
cutoff-independent convergence result.

\subsection{Independent perfect-contact FEM comparison}
A separate bulk benchmark uses the two substrate tensors above without
films or horizontal interfaces; the source and observation points are
unchanged. The FEM reference Fourier-transforms $x$ and discretizes the
$y$--$z$ plane with bilinear quadrilateral elements. Shared degrees of
freedom at $y=0$ impose exact temperature continuity, while normal-flux
continuity follows from the weak form. No large-conductance penalty is used.
This is an independent Fourier--FEM reference, distinct from the
three-dimensional FEM benchmarks in the main text.

The finite domain is $|y|\le8\mu$ and $0\le z\le8\mu$, where
$\mu=\sqrt{2k_z/(\omega C_v)}=30.90~\mu\mathrm{m}$.
The remote lateral and bottom boundaries have zero oscillating temperature;
the top carries the Gaussian heat flux. The final mesh contains 65,296
nodes (64,491 unconstrained unknowns), with $0.03125~\mu\mathrm{m}$
lateral spacing near the source and interface and graded spacing farther
away. Independent Gauss quadrature integrates the surface load and
96 positive angular-wavenumber nodes reconstruct the $x=0$ field.
The numerical source integral equals the prescribed absorbed power.

Halving the near-source spacing from $0.25$ to $0.125~\mu\mathrm{m}$,
and then to $0.0625~\mu\mathrm{m}$, changes the complex surface profile by
$0.459\%$ and $0.119\%$, respectively. Refinement to the final mesh,
including finer far-field grading, changes it by $0.0534\%$.
Increasing the Fourier quadrature from 64 to 96 nodes changes the profile
by $2.12\times10^{-8}\%$; extending the domain from $6\mu$ to $8\mu$
changes it by $8.62\times10^{-6}\%$. These are successive numerical
checks, rather than certified bounds on the exact solution.

\begin{figure}[htbp]
\centering
\includegraphics[width=\linewidth]{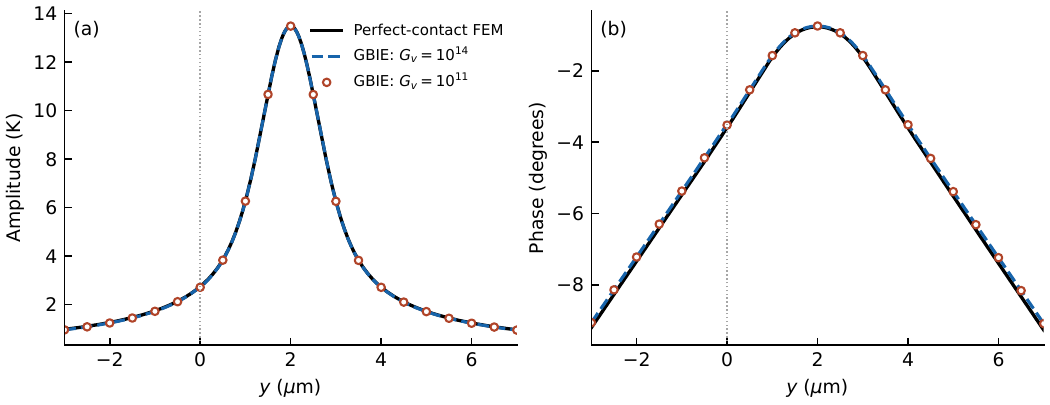}
\caption{Bulk perfect-contact comparison: (a) surface amplitude and
(b) unreferenced phase. FEM uses exact continuity; GBIE uses
$G_v=10^{11}$ or $10^{14}~\mathrm{W\,m^{-2}K^{-1}}$ with
$(N_u,N_v,N_z)=(35,240,100)$ and angular cutoffs
$(\xi_{\max},\eta_{\max})=(1,4)\times10^7~\mathrm{rad\,m^{-1}}$.
The two GBIE curves nearly overlap. Their relative complex $L_2$
differences from FEM are $0.2414\%$ and $0.2413\%$, respectively.}
\label{fig:supp_perfect_contact_fem}
\end{figure}

At $G_v=10^{14}$, the largest unreferenced phase difference from FEM
is $0.183^\circ$. The GBIE solution with exactly $R_v=0$ gives the same
$0.2413\%$ complex discrepancy to the reported precision.
This independent comparison supports the surface response at both exact
thermal continuity and large finite conductance for the stated benchmark. The supplementary files retain the
inputs, source code, complex profiles, per-wavenumber diagnostics, and
FEM refinement results for reproduction.

\clearpage
\section{Local JAX implementation comparison}
\label{sec:supp_jax_cpu}

To assess the cost of using more quadrature nodes, we compare the original
NumPy/SciPy GBIE solver with a just-in-time compiled JAX implementation
on the same macOS ARM CPU host (14 logical CPUs; Python~3.11.9,
JAX~0.10.2, NumPy~2.4.6, and SciPy~1.17.1). Both use
double-precision complex arithmetic (\texttt{complex128}). The test has
two identical semi-infinite media with
$k_x=k_y=k_z=100~\mathrm{W\,m^{-1}\,K^{-1}}$,
$C_v=10^6~\mathrm{J\,m^{-3}\,K^{-1}}$, and
$R_v=10^{-7}~\mathrm{m^2\,K\,W^{-1}}$.
The Gaussian source has $P_0=1~\mathrm{mW}$, $f=100~\mathrm{kHz}$,
$r=1~\mu\mathrm{m}$, and $d=2~\mu\mathrm{m}$, with heating on both
sides retained. Temperature is evaluated at 181 equally spaced surface
points with $x=0$ and $-2\le y/r\le4$.
Both solvers use identical composite Gauss--Legendre quadratures,
clustered depth nodes, and angular-wavenumber cutoffs
$\xi_{\max}=\eta_{\max}=10^7~\mathrm{rad\,m^{-1}}$
($\xi=2\pi u$, $\eta=2\pi v$).
The interface integral is truncated at
$z_{\mathrm{int,max}}=142.7299~\mu\mathrm{m}$, with the semi-infinite
decay admittance retained in the Green functions. The analytical bulk
shortcut is disabled.

Each timing includes host preparation, Green-function evaluation,
integral assembly, the dense solve, reconstruction, synchronization, and
host-visible output; imports, file export, plotting, and diagnostics are
excluded. The JAX first call includes tracing and compilation after
clearing its in-memory cache, but excludes process startup. Repeated
timings are medians of three further evaluations with alternating paired
solver order. Each evaluation recomputes the GBIE solution while reusing
compiled code. Core affinity was unconstrained, so these timings do not
constitute a matched-core COMSOL comparison.

\begin{table}[htbp]
\centering
\small
\caption{Local CPU timings in seconds. Speedup is the ratio of the NumPy
and repeated JAX medians. The relative complex $L_2$ error compares JAX
with NumPy over all observation points at identical quadrature.}
\label{tab:supp_jax_local}
\begin{tabular}{@{}lrrrrr@{}}
\toprule
$(N_u,N_v,N_z)$ & \shortstack{NumPy\\median}
& \shortstack{JAX\\first call} & \shortstack{JAX\\median}
& Speedup & \shortstack{Relative\\$L_2$ error} \\
\midrule
$(16,36,24)$  & 0.32089 & 0.43102 & 0.016602 & $19.33\times$ & $1.82\times10^{-16}$ \\
$(32,72,48)$  & 2.33597 & 0.55718 & 0.087535 & $26.69\times$ & $1.32\times10^{-16}$ \\
$(48,120,64)$ & 7.77868 & 0.74018 & 0.381674 & $20.38\times$ & $1.51\times10^{-16}$ \\
\bottomrule
\end{tabular}
\end{table}

After compilation, JAX accelerates repeated evaluations by
$19.3$--$26.7\times$ across the tested grids, with relative complex
$L_2$ differences below $2\times10^{-16}$ from NumPy. These results
measure implementation performance at identical quadrature; they do
not establish continuum convergence.
The coarse and medium NumPy grids differ from the finest tested grid
by $0.9656\%$ and $0.1946\%$, respectively. All three use fixed cutoffs
and integration depth, so these differences do not assess truncation
error. Inputs, raw timing samples, saved complex fields, software
versions, and source hashes accompany the comparison in
\texttt{evidence/jax\_local\_cpu}.

\clearpage
\section{JAX startup and repeated evaluation for the 100 nm benchmark}
\label{sec:supp_jax_crc_timings}

This comparison uses the archived $100~\mathrm{nm}$ film case from
COMSOL job 23517175 and original GBIE run 23517757: a
$300~\mu\mathrm{m}$ domain, $10~\mathrm{kHz}$ heating, a
$1~\mu\mathrm{m}$ pump radius, $2~\mu\mathrm{m}$ offset, and
501 surface observation points. The JAX implementation retains the
archived $(N_u,N_v,N_z)=(35,120,25)$ quadrature, source cutoffs,
material inputs, and \texttt{complex128} precision, with the analytical
bulk shortcut disabled. CRC job 24042690 ran on the same physical
AMD EPYC 9374F node as the archived COMSOL calculations. Exact affinity
checks enforced one logical CPU per physical core and the stated core
budgets for the process and its surviving helper threads. The runs
were independent series with different dates, software environments,
and background loads. The recorded JAX environment used Python~3.12.8,
JAX/JAXLIB~0.11.1, NumPy~2.5.3, and SciPy~1.18.1.

\paragraph{Fresh-process timing.}
An external timer surrounds each fresh Python process, including
startup, imports, device initialization, input preparation,
tracing/compilation, the synchronized forward calculation, numerical
validation, durable profile/metadata output, and shutdown. Persistent
JAX compilation caching is disabled. Each reported time is the median
of three processes; scheduler and step-launch latency are excluded.
The historical one-core COMSOL application total is $1244.0~\mathrm{s}$.
The optimized four- and sixteen-core COMSOL durable-workflow medians
are $533.65$ and $521.63~\mathrm{s}$ (Section~\ref{sec:supp_parallel_scaling}).
The latter include solved-model archival; JAX exports a complex surface
profile, arrays, and metadata. The resulting fresh-workflow ratios
compare these different saved outputs and are not pure solver ratios.
The separate optimized one-core COMSOL median of $1132.47~\mathrm{s}$
in Section~\ref{sec:supp_parallel_scaling} does not replace the historical
$1244.0~\mathrm{s}$ baseline in main-text Table~3.

\paragraph{Repeated evaluation.}
At each core count, three separate processes first compile the model
and then execute five calls to the full public forward function.
Each call reconstructs quadrature and device inputs, Green functions,
kernel matrices, the linear solve, and the surface temperature, ending
with synchronized host-visible output. No prepared operator,
factorization, or solved temperature is reused. Only the compiled
program is retained. Imports, compilation, file output, and plotting
are excluded from these repeated intervals. We first take the median
of five calls in each process, then the median across three processes.

\begin{table}[htbp]
\centering
\small
\caption{JAX timings for the archived $100~\mathrm{nm}$ case.
The fresh-process interval includes startup and compilation; the
repeated interval times a full forward evaluation after compilation.
Peak RSS is the maximum complete-process peak over the three fresh runs.}
\label{tab:supp_jax_crc_timings}
\begin{tabular}{@{}rrrr@{}}
\toprule
Physical cores & \shortstack{Fresh process\\(s)} &
\shortstack{Repeated evaluation\\(s)} &
\shortstack{Fresh peak RSS\\(GiB)} \\
\midrule
1 & 3.04 & 0.161996 & 0.399 \\
2 & 2.32 & 0.113585 & 0.416 \\
4 & 2.02 & 0.146243 & 0.381 \\
8 & 1.81 & 0.135373 & 0.368 \\
16 & 1.80 & 0.127704 & 0.408 \\
\bottomrule
\end{tabular}
\end{table}

All 30 cold and warm process records passed the archived quadrature,
source/input provenance, affinity, and numerical parity checks.
The largest relative complex $L_2$ difference from the original GBIE
was $2.07\times10^{-16}$ (rounded upward). This verifies the port at
the archived quadrature; it does not establish continuum convergence.
JAX memory is the external GNU-\texttt{time} process peak RSS, including
imports and compilation. The optimized COMSOL entries are medians of
COMSOL-reported physical-memory peaks, whereas JAX uses the maximum
peak over three fresh runs. The reporters and aggregation differ;
these ratios are not common-process-tree storage bounds. No warm-only
per-evaluation peak memory was measured.

\paragraph{Relevance to inverse fitting.}
An isolated calculation pays startup and compilation once. Parameter
sweeps and inverse fitting can instead retain a live process and
amortize that cost over many forward calls. Material-property values
enter the JAX kernel as dynamic inputs, so changing their values can
reuse the compiled program when the actual array shapes, precision,
and static options remain unchanged. Automatic quadrature construction
can change array sizes when properties change, requiring another
compilation; changing a scalar vertical resistance with the same
geometry preserves those sizes. New properties must still be prepared
and the forward solution recomputed. The archived repeated timings
use fixed inputs and do not constitute an inverse-fitting benchmark.
For the compiled rows in main-text Table~3, we retain the archived
COMSOL times of $1244.0$, $533.65$, and $521.63~\mathrm{s}$ as fixed
workflow references at one, four, and sixteen cores. The same reference
is used for the startup and compiled JAX rows within each group.
The reference ratio is defined as
\begin{equation}
    R_{\mathrm{ref}}=
    \frac{T_{\mathrm{COMSOL,workflow}}}
         {T_{\mathrm{JAX,compiled}}}.
    \label{eq:supp_jax_reference_ratio}
\end{equation}
At sixteen cores, the synchronized full forward evaluation takes
$0.128~\mathrm{s}$ after compilation, excluding JAX startup,
compilation, and file output. Compared with the $521.6~\mathrm{s}$
COMSOL workflow, which includes startup and solved-model archival,
this gives a reference ratio of approximately $4100\times$ using the
unrounded archived values. The startup-inclusive JAX process takes
$1.80~\mathrm{s}$, including compilation, validation, and file output,
giving approximately $290\times$ relative to the same COMSOL workflow.
These ratios describe the stated timing intervals; they do not compare
repeated solves by two initialized solvers. No corresponding
repeated COMSOL timing with model/mesh reuse is available, so retaining
its archived time as a reference does not establish that its repeated
execution time is unchanged. A complete inverse-fitting speedup was
not measured.

The input archive, source hashes, allocation and affinity records,
individual time measurements, complex profiles, and analysis tables
are retained under
\path{evidence/jax_crc_optimized_2026-09-10/results/slurm_24042690/}. These CRC results are separate from the
local macOS study in Section~\ref{sec:supp_jax_cpu}.